\documentclass[aps,prx,twocolumn,superscriptaddress]{revtex4-1}

\pdfoutput=1

\usepackage{amsmath,breqn,mathtools,amsfonts} \DeclareMathOperator{\Tr}{Tr}
\usepackage{hyperref}
\allowdisplaybreaks
\usepackage{graphicx}
\usepackage{tikz}
\usepackage[caption=false]{subfig}
\usepackage{xcolor}

\newcommand{\FF}[1]{{\color[rgb]{0,0,0}#1}}
\newcommand{\FFnew}[1]{{\color[rgb]{0,0,0}#1}}

\makeatletter
\let\cat@comma@active\@empty
\makeatother

\begin{document}

\title{Building general Langevin models from discrete data sets}
\author{Federica Ferretti}
\affiliation{Dipartimento di Fisica, Universit\a`{a} Sapienza, 00185 Rome, Italy}
\affiliation{Istituto Sistemi Complessi, Consiglio Nazionale delle Ricerche, UOS Sapienza, 00185 Rome, Italy}
\author{Victor Chard\`es}
\affiliation{Laboratoire de Physique de l'\a'{E}cole Normale sup\a'{e}rieure (PSL University), CNRS, Sorbonne Universit\a'{e}, Universit\a'{e} de Paris, 75005 Paris, France}
\author{Thierry Mora}
\affiliation{Laboratoire de Physique de l'\a'{E}cole Normale sup\a'{e}rieure (PSL University), CNRS, Sorbonne Universit\a'{e}, Universit\a'{e} de Paris, 75005 Paris, France}
\author{Aleksandra M Walczak}
\affiliation{Laboratoire de Physique de l'\a'{E}cole Normale sup\a'{e}rieure (PSL University), CNRS, Sorbonne Universit\a'{e}, Universit\a'{e} de Paris, 75005 Paris, France}
\author{Irene Giardina}
\affiliation{Dipartimento di Fisica, Universit\a`{a} Sapienza, 00185 Rome, Italy}
\affiliation{Istituto Sistemi Complessi, Consiglio Nazionale delle Ricerche, UOS Sapienza, 00185 Rome, Italy}
\affiliation{INFN, Unit\a`{a} di Roma 1, 00185 Rome, Italy}

\date{\today}
\begin{abstract}
\FF{Many living and complex systems exhibit second order emergent dynamics. Limited  experimental access to the configurational degrees of freedom results in data that appears to be generated by a non-Markovian  process. This poses a challenge in the quantitative reconstruction of the model from experimental data, even in the simple case of equilibrium Langevin dynamics of Hamiltonian systems. We develop a novel Bayesian inference approach to learn the parameters of such stochastic effective models from discrete finite length trajectories. We first discuss the failure of naive inference approaches based on the estimation of derivatives through finite differences, regardless of the time resolution and the length of the sampled trajectories. We then derive, adopting higher order discretization schemes, maximum likelihood estimators for the model parameters} that provide excellent results even with moderately long trajectories. We apply our method to second order models of collective motion and show that our results also hold in the presence of interactions.
\end{abstract}

\maketitle

\section{Introduction} \label{Intro}

Recent experimental findings on a variety of living systems, from cell migration \cite{bruckner+al_19}, bacterial propulsion \cite{listeria}, worm dynamics \cite{greg_08}, to the larger scale of animal groups on the move \cite{gautrais2009analyzing,gautrais2012deciphering,Attanasi:2014aa,Cavagna:2017aa}, indicate that the observed behavior cannot be explained with a first order dynamical process, but requires a higher order description.  For bird flocks and insect swarms, the case which interests us most, data show that propagating directional information during collective turns in flocks requires rotational inertia, i.e. a reversible dynamical term, to account for the measured dispersion law \cite{Attanasi:2014aa}. The shape of the velocity-velocity correlation function in swarms, which flattens at short times, also points to a second order dynamics for these systems, as suggested  by the value of the dynamical critical exponent \cite{Cavagna:2017aa}. Overall, data indicate that
considering second order dynamics is required to explain how animal groups behave on their natural size and time scales --- 
even though overdamping might theoretically occur for very large systems and on very large time scales.

The emergent dynamics of all the above systems share three fundamental ingredients: an effective inertia, dissipation, and a stochastic contribution. Disentangling such contributions is often crucial to understand the processes at stake and reliable methods are required to extract that information from available data. The example of animal groups, which motivates the present work, is also helpful to discuss the theoretical objectives and experimental constraints of the inference procedure. Ideally, we would like to build the simplest continuous second order model consistent with experimental findings. We seek a continuous time model for \FF{several} reasons: i) it allows computations to be performed; ii) it is a reasonable assumption for systems where microscopic update times are much smaller than observational scales (cognitive processes occur on tenths of milliseconds, whereas behavioral changes on scales of seconds); \FF{iii) it circumvents the inherent arbitrariness of discrete time modelling}. Experimental data, on the other hand, come in the form of discrete time series, where the discretization interval is set by the time resolution of the experimental apparatus. 

In the presence of stochasticity, the nature of the data poses two major problems. First of all, if the dynamics is of second order, all signals (including initial condition and noise) are propagated in time with a memory kernel, making the relation between the coarse grained data that we observe and the underlying process far more complex than in the first order case. 
The memory kernel arises from the contraction of the dynamical description of the second order stochastic process from the full phase space to a lower dimensional subspace ---\,usually that of measurable degrees of freedom \cite{zwanzig_book, Miguel1980, Hanggi1978}. For example, were we able to experimentally measure with the same accuracy a pair of conjugate variables, e.g. positions and velocities of moving individuals, we could seek a model for their joint evolution. But in common experiments that is not the case, as one typically measures one degree of freedom (e.g., positions) and must derive the other. To confront the data, we therefore need to work in a reduced space.  
Secondly, the goal of the inference procedure is to retrieve a continuous stochastic model 
from a collection of discrete sample paths occurring on \FF{finite} observational time scales $\Delta t$. 
In absence of an explicit solution for the stochastic process, the most reasonable thing to do is to transform the stochastic differential equation (SDE) into an approximated difference equation. Such discretization must be performed very carefully, since the resulting equation should correctly represent the underlying stochastic process both at the scales of the sampled data (at which inference works), and in the microscopic limit of vanishing increments.

These two problems are quite general and do not depend on the presence of interactions in the system, but rather on the nature of the dynamics. \FF{Although the issue has been considered before, the literature is sparse and a satisfying Bayesian inference approach is still lacking. Previous attempts to provide systematic inference strategies for second order dynamics consist of building converging estimators for the different terms of the model from proper combinations of measurable quantities  \cite{Ronceray, Lehle2015, Lehle2018}, or in exploiting known relations between model parameters and accessible observables \cite{Pedersen}.

In a more general and refined way, the problem can be reformulated in terms of a dynamical inverse problem, and much work has been done in this field in the last years \cite{MaxCal, Zecchina_review, MacKay}. However most analyses have focused on first order processes in time \cite{Sorensen,Monthus_2011,Otten2010,Dyn-Inf-PRE,Mora:2016aa,frishman2018learning,masson_primer,molecule_traj_masson}. 
Second order processes have been considered within this framework in Refs.~\cite{Gloter, GloterOU}, yet the proposed method differs from a proper maximum likelihood approach, due to the difficulty of deducing a pseudo-likelihood function in the case of non-Markovian processes.

To the best of our knowledge, we present here the first maximum likelihood inference approach for non-Markovian inertial processes. It differs from previous studies in its first principle derivation and absence of a Markovian embedding. We derive explicit formulas for the parameter estimators, and test our approach on synthetic data in a variety of models, including non-linear forces, multiplicative noise and many-body interactions. Results show that the method is accurate and robust, providing an important tool in the analysis and understanding of real systems.} 
The paper is organized in the following way: in Sec.~\ref{ho} we formalize the problem and discuss in detail how to build an appropriate dynamical inference strategy for inertial systems with linear dissipation. \FF{We explain the interplay between the order of convergence of discretized SDEs obtained from Taylor-It\^o expansions and the consistency of the corresponding max-likelihood parameter estimators.} We show that to get accurate results the simplest Euler-like schemes, which work well with first order dynamics, are insufficient, so that one needs to go to the next order of approximation. Theoretical predictions are compared with numerical data to consolidate our results. \FF{ Sec.~\ref{subsec:4} introduces non-Bayesian inference schemes, while in Sec.~\ref{sec:initial} we discuss the problem of eliminating the initial velocity.} In Sec.~\ref{interacting} we address the case of a strongly interacting system: the inference procedure is applied to synthetic data obeying the Inertial Spin Model, a model of self-propelled particles that describes the phenomenology of natural flocks of birds~\cite{Attanasi:2014aa}. \FF{The effect of experimental measurement noise is discussed in Sec.~\ref{noise}.} Finally, in Sec.~\ref{conclusions} we summarize all our results, discuss their conceptual relevance, and outline their potential for applications to real data.

\section{Maximum likelihood inference approach for Langevin dynamics}\label{ho}

\subsection{Problem definition}\label{subsec:1}
Let us assume that the available experimental data are sequences of points $(x_0, x_1, \dots x_L)$ uniformly separated in time by $\Delta t$, and that the underlying dynamics is described by the complete Langevin equation of the form:\FF{
\begin{equation}
	\ddot x = -\eta \dot x + f(x) + \sigma\xi,
	\label{oscillator-1}
\end{equation}
where $f(x)=-V'(x)$ is a conservative force, $\sigma^2 = 2T\eta$, and $\xi$ is a standard white noise: $\langle\xi\rangle=0$, $\langle\xi(t)\xi(t')\rangle=\delta(t-t')$. Without lack of generality, the inertial mass is  set to 1.} Since the noise is additive, it is unnecessary to distinguish between It\^o and Stratonovich integration.

Let us call $\boldsymbol{\lambda}$ the irreducible set of parameters that enter in Eq.~\eqref{oscillator-1}, namely the effective damping coefficient $\eta$, the effective temperature $T$, and the parameters entering in the definition of the potential $V(x)$. The aim of dynamical statistical inference is to provide an estimate of their values.  Following a Bayesian approach, the posterior distribution of parameters given the data reads:
\begin{equation}
	P(\boldsymbol{\lambda}|\{(x_0,\dots, x_L)^{\alpha}\})\propto P(\{(x_0,\dots, x_L)^{\alpha}\}|\boldsymbol{\lambda}) \rho(\boldsymbol{\lambda}), 
	\label{Bayes-lik}
\end{equation}
where each Greek index labels a different experimental sample. By choosing a uniform prior $\rho(\boldsymbol{\lambda})$, the maximum of Eq.~\eqref{Bayes-lik} corresponds to the maximum likelihood estimator. The conceptual and technical difficulty of the whole inference problem is then only about finding a tractable expression for the dynamical likelihood.

The theory of stochastic processes provides us with an explicit but formal expression for the \FF{transition probability $P\left(x(t)|x(0),\dot x(0)\right)$}, involving, in general, integro-differential operators. A closed form solution for the stochastic process may be generally unknown or complicated \footnote{We seek a solution of the stochastic process either as an explicit sample-path solution in position space that does not involve integro-differential operators, or as the time-dependent solution of the associated Fokker-Planck equation \cite{Gardiner}.}, especially for many body or off-equilibrium systems, but finely time-resolved data may be available. What we look for is then an (eventually approximated) expression for the probability of the discrete trajectory, for which a practical connection with the data can be established.

A \FF{first}  general strategy is the following:
\begin{enumerate}
\item{As a preliminary step, Eq.~\eqref{oscillator-1} can be conveniently rewritten as a set of two first order equations:
\begin{equation}
\begin{cases}
	\dot x = v\\
	\dot v = -\eta v + f(x) + \sigma \xi.
	\end{cases}
	\label{SDE-sys}
\end{equation}
} 
\item{Since the dynamics is Markovian when parametrized by the vector variable $\mathbf q = (x,v)$, the probability of a discrete trajectory in this space, given the initial condition $\mathbf q_0 = (x_0,v_0)$, can be split into a product of propagators:
\begin{equation}
	P(\mathbf q_L,\dots,\mathbf q_1|\mathbf q_0) = \prod_{n=1}^L P(\mathbf q_{n}|\mathbf q_{n-1}).
	\label{markov-factor}
\end{equation}
}
\item{Following \cite{Drozdov-PRL}, one can exploit any update rule based on a Taylor-It\^o expansion to approximate, within a certain order of accuracy, the propagator over a small time interval $\Delta t$:
\begin{equation}
	P(\mathbf q_{n}|\mathbf q_{n-1}) = P_{(k)}(\mathbf q_{n}|\mathbf q_{n-1}) + o(\Delta t^k).
	\label{approx-prop}
\end{equation}
Eq.~\eqref{approx-prop} can be replaced into Eq.~\eqref{markov-factor} to get an approximated expression for the probability density of the sequence of points in phase space: 
\begin{equation}
P_{(k)}(\mathbf q_L,\dots\mathbf q_1|\mathbf q_0) = \prod_{n=1}^L P_{(k)}(\mathbf q_{n}|\mathbf q_{n-1}) + o(\Delta t^k).
\end{equation}
}
\item{Marginalizing over the velocity-like degrees of freedom one  gets a probability distribution depending on the $x$'s only. This projection operation on the subspace of $x$ variables is where the original Markovian property of Eq.~\eqref{markov-factor} is generally lost. \FF{A crucial remark, beyond the non-Markovian nature of the resulting dynamics, is that} this procedure does not simply consist of removing the intermediate variables $v_1, \dots, v_L$, but also of eliminating the initial condition $v_0$. \FF{This is at the same time a further technical difficulty and a fundamental conceptual issue in the context of stochastic dynamics. We refer to Sec.~\ref{sec:initial} for a broader discussion.}
}
\end{enumerate}

\FF{When this strategy is adopted,} the first thing we need is then a discrete integration scheme for Eq.~\eqref{oscillator-1} or Eq.~\eqref{SDE-sys}. Although the naive intuition is that any convergent ---\,even if slowly\,--- discretization scheme should work for small $\Delta t$, in fact the order of approximation of the temporal discretization is able to affect the mathematical properties of the discrete path integral measure and, consequently, \FF{the correctness of estimators obtained through a maximum likelihood inference procedure \cite{Drozdov-PRL,Gloter}.}

Alternatively, \FF{one can follow a second strategy, summarized as \emph{`first marginalize, then discretize'}, in contrast to the \emph{`first discretize, then marginalize'} strategy discussed above. The starting point is here the generalized Langevin equation (GLE) corresponding to the desired process, Eq.~\eqref{oscillator-1}, which can be obtained adopting the Mori-Zwanzig formalism \cite{zwanzig_book} (see App.~\ref{app:discretization}):}
\begin{equation}
	\dot x = v_0 e^{-\eta t} + \int_0^t ds K(t-s) f(x(s)) + \zeta(t).
	\label{eqx}
\end{equation}
In this equation, $K(t)=e^{-\eta t}$ and the effective noise is given by $\zeta(t)=\int_0^t ds e^{-\eta (t-s)} \xi(s)$. This formalism shows that, when projected from the full phase space into the $x$ space, the dynamics acquires a memory, described by a friction kernel $K(t)$ and color in the noise. We note that the relation $\langle \zeta(t) \zeta(t')\rangle \propto K(|t-t'|)$ \FF{holds asymptotically in the limit of infinitely long trajectory, and it reduces to the second fluctuation dissipation theorem when $f(x)$ is linear.} Discrete update equations can now be obtained by integrating Eq.~\eqref{eqx} on $\Delta t$ intervals, \FF{and self-consistently removing $v_0$. We notice that, for arbitrary forces $f(x)$, the corresponding term cannot be exactly integrated and it needs to be approximated at small $\Delta t$. The fact that the derivative of the measured coordinate -- position, $x$, -- enters parametrically through $v_0$ in the GLE stems from the second order nature of the process. Its elimination, which is necessary to retrieve a stochastic difference equation where only the $x$ variable appears, is  connected to the problem anticipated in point 4 of the procedure outlined above.

The two strategies must be equivalent: the order of the discretization and marginalization operations should be exchangeable. 
 In the following section we show how the simplest inference schemes derived from Euler-like discretizations of Eq.~\eqref{SDE-sys} do not satisfy this requirement, whereas higher order discretization schemes, strongly convergent as at least $O(\Delta t^{3/2})$, retrieve correct maximum likelihood estimators.
}

\FF{
\subsection{Failure of na\"ive inference schemes}\label{schemes}

Discrete integration approaches for SDEs are well known in the literature in connection to numerical computation methods (see, e.g. \cite{Platen-Kloeden}). Here, we summarize how the order of approximation of these discretization schemes interferes with the non-Markovian character of the observed dynamics. This makes standard claims about the convergence of these integrators not generally valid in cases when only a projection of the original Markovian process is observed. Rigorous results can be found in \cite{Gloter}. We are mainly interested in, from an application point of view, the bias that this fact introduces in na\"ive inference approaches, and possible correction strategies.


Let us start then with the simplest possible construction, i.e. the Euler-Maruyama scheme applied to Eq.~\eqref{oscillator-1} (in this case corresponding to the Milstein scheme) \cite{Platen-Kloeden}. The discrete update equations for the Markov process read: } 
\begin{equation}
	\begin{cases}
	x_{n+1} - x_n = \Delta t\, v_n\\
	v_{n+1} - v_n = -\eta \Delta t\, v_n - \Delta t\, f(x_n) + \sigma\Delta t^{1/2}\,r_n,
	\end{cases}
	\label{Euler-Maruyama}
\end{equation}
with $r_n$ i.i.d. random variables of normal distribution $\mathcal N(0,1)$, for $n=0,\dots, L-1$. We remind that the first neglected terms in Eq.~\eqref{Euler-Maruyama} are $O(\Delta t^{3/2})$. The scheme provides then a deterministic update for the $x$ variables, which manifests itself through $\delta$-functions; a simple change of variables from $r_n$ to $v_{n+1}$ immediately completes the derivation of the discrete propagator in $(x,v)$ space. Finally, in this case one can explicitly marginalize over the velocity degrees of freedom, \FF{and eliminate} the initial condition $v_0$. Indeed, to this order of approximation, information on $v_0$ is fully equivalent to information on $x_1$. From this marginalization, a fully factorized probability distribution for the discrete sequence is obtained:
\begin{equation}
	P_{(1)}(x_L,\dots,x_2|x_0,x_1) ={\prod_{n=1}^{L-1} P_{(1)}(x_{n+1}|x_n, x_{n-1})},
	\label{factorized-x1}
\end{equation}
where transition probabilities are defined as follows:
\begin{equation}
	P_{(1)}(x_{n+1}|x_n, x_{n-1}) = \frac{1}{Z_n}e^{- S_n(x_{n+1},x_n, x_{n-1})}\,,
	\label{transition-master-eq}
\end{equation}
with
\begin{equation}
	Z_n = \sqrt{2\pi\sigma^2\Delta t^3}\,;
	\label{Zn}
\end{equation}
\begin{dmath}
	S_n=\frac{1}{2\sigma^2\Delta t^3} \left[ x_{n+1}-2x_n+x_{n-1} + \eta\Delta t(x_n-x_{n-1}) - \Delta t^2 f(x_n) \right]^2.
	\label{Sn}
\end{dmath}

A factorization of $P(x_L,\dots x_2|x_1,x_0)$ into a product of transition probabilities of this kind is possible because the random variables appearing in the $x$ difference equation, obtained from Eq.~\eqref{Euler-Maruyama} through variable elimination, are independent. This is a crucial \FF{but artificial} feature occurring only at this level of approximation: more accurate discretization procedures produce an effective noise for the $x$ variables which is correlated in time. As a matter of fact, when the description of a Brownian motion is contracted from the full phase space to position space, a colored noise emerges, which is incompatible with the independence of subsequent random variables at any $\Delta t$.

Nonetheless, we find it useful to compute the associated dynamical likelihood, as defined in Eq.~\eqref{Bayes-lik}, \FF{and develop the corresponding inference scheme. For the sake of clarity, we will focus on the  example of the harmonic oscillator, where $f(x) = -\omega_0^2 x$.} Using Eqs.~\eqref{transition-master-eq}--\eqref{Sn}\,, an expression for the likelihood as product of transition probabilities for a second order master equation is recovered. This corresponds  to the discrete path probability one would obtain adopting a maximum caliber approach \cite{MaxCal} when \FF{ certain time-dependent observables are taken as fixed. For the one-dimensional harmonic oscillator, they are the} equal-time correlations, one-time-step correlations and two-time-step correlations of the process. Indeed, rearranging the sum of $S_n$'s in Eq.~\eqref{factorized-x1}, the reduced minus-log-likelihood can be written as: 
\begin{widetext}
\begin{dmath}
	\frac{\mathcal L(\eta,T,\omega_0^2)}{L-1} = \frac{1}{2}\ln(2\pi\sigma^2\Delta t^3) + \frac{1}{2\sigma^2\Delta t^3}\left[C'_s + (2-\eta\Delta t+\omega_0^2\Delta t^2)^2 C_s + (1-\eta\Delta t)^2 C''_s 
	+ 2(1-\eta\Delta t) F_s - 2(2-\eta\Delta t+\omega_0^2\Delta t^2) G_s - 2(1-\eta\Delta t)(2-\eta\Delta t+\omega_0^2\Delta t^2)G'_s\right],
	\label{L-EM}
\end{dmath}
\end{widetext}
where we introduced the following notation for the experimental temporal correlation functions, evaluated at a time distance of 0, $\Delta t$ and $2\Delta t$: 
\begin{eqnarray*}
&C_s = \frac{1}{L-1}\sum_{n=1}^{L-1}x_nx_n; &C'_s = \frac{1}{L-1}\sum_{n=1}^{L-1}x_{n+1}x_{n+1};\\
&C''_s = \frac{1}{L-1}\sum_{n=1}^{L-1}x_{n-1}x_{n-1}; &G_s = \frac{1}{L-1}\sum_{n=1}^{L-1}x_{n}x_{n+1};\\
&G'_s = \frac{1}{L-1}\sum_{n=1}^{L-1}x_{n}x_{n-1}; &F_s = \frac{1}{L-1}\sum_{n=1}^{L-1}x_{n-1}x_{n+1}.
\end{eqnarray*}

Minimization of the quantity in Eq.~\eqref{L-EM} with respect to $\eta$, $T$ and $\omega_0^2$ yields the inference formulas for the parameters of the harmonic oscillator. We express here only the estimator of the damping coefficient $\eta$, while the remaining ones can be found in App.~\ref{app:1}:
\begin{equation}
	\eta^* = \frac{1}{\Delta t}
	\frac{2C_s - G_s - G'_s - \dfrac{G'_s}{C''_s}(2G'_s - C''_s - F_s)}
	{C_s + C''_s - 2G'_s-\dfrac{(C''_s-G'_s)^2}{C''_s}}.
	\label{eta*}
\end{equation}

At this point, having an explicit inference method, it can be both numerically and analytically tested. We simulated discrete trajectories of the stochastic harmonic oscillator in several damping conditions using an exact integrator \cite{Gillespie}, with a numerical time step $\tau^{sim}=0.005$. We applied inference formulas to discrete data sets sampled from synthetic trajectories at time intervals $\Delta t\ge \tau^{sim}$. This choice mimics real experiments, where the time resolution is fixed by the acquisition apparatus, while the true microscopic time-scale of the dynamics is unknown. Filtering the synthetic trajectories in time is a good blind inspection tool to check the robustness of the continuous description given by the inferred parameters, without prior knowledge about the time scales of the process. Moreover, this test on numerical simulations can help us identifying the time window in which any dynamical inference scheme is expected to work: in discretizing the equations of motion, the implicit assumption is that $\Delta t$ must be much smaller than the typical time scales of the process ($\eta^{-1}$ and $\omega_0^{-1}$ in this example).

Results, reported in Fig.~\ref{fig:4panel}, show that a systematic error in the estimation of the damping coefficient emerges, which can be cast into a constant rescaling factor close to 2/3 for the inferred value $\eta^*$ as compared to the true value $\eta^{sim}$. It is worth remarking that this rescaling is {\it independent of }$\Delta t$, as clearly visible in Fig.~\ref{box-eta-all}, so increasing the resolution of the acquisition instruments is of no help in improving the estimation of the damping coefficient. The same problem also occurs when using other variants of the EM scheme obtained from a Taylor-It\^o expansion of the same order, as we illustrate in App.~\ref{app:1}. On the contrary, the estimation of the remaining parameters is in agreement with the parameter values used in the simulations, as shown in Figs.~\ref{box-J-slope} --~\ref{box-T-slope}.

\FF{Numerical evidence for the stochastic harmonic oscillator agrees with the results of Refs.~\cite{Pedersen, Lehle2015}, who pointed out, in a non-Bayesian framework, the failure of the same na\"ive embedding strategy for second order SDEs. We stress that the EM discretization is the simplest and most commonly used extrapolation of the derivative of an observed variable from its finite increment. } This approximated estimation of the velocity works if one observes the system in the overdamped regime, i.e. when $\eta\Delta t \gg 1$ and $\omega_0/\eta < \infty$, and the effective dynamics can be described by a first order equation. In this case, EM-based inference schemes provide in effect excellent results \cite{Dyn-Inf-PRE, Mora:2016aa}. \FF{However, when a non-Markovian signal is observed, such as the partial observation of a higher dimensional Markovian process, these schemes are bound to fail.}

A simple argument can help us to understand what is missing, and why the parameter $\eta$ is the one affected by the approximation. Assuming that experimental averages perfectly reproduce ensemble averages, we can replace into Eq.~\eqref{eta*} the known analytical expression for the self-correlation of the harmonic oscillator in the stationary regime $C(0)$, $C(\Delta t)$ and $C(2\Delta t)$. Since the underlying assumption of the whole procedure is that the time lag $\Delta t$ between subsequent points is small, compared to the typical time scales of the dynamics, we can perform a Taylor expansion around $t=0$,  obtaining from Eq.~\eqref{eta*} an expression for $\eta^*$ depending only on the derivatives of $C(t)$ at $t=0$:
\begin{equation}
	\eta^* \simeq \frac{1}{\Delta t}
	\frac{2\dot C(0) - \frac{2}{3}\dddot C(0)\Delta t^2 -\frac{\ddot C(0) \dot C(0)}{C(0)}\Delta t^3}
	{2\dot C(0)+\ddot C(0)\Delta t + \frac{1}{C(0)}\left[\dot C(0)+\frac{1}{2}\ddot C(0)\Delta t\right]^2\Delta t}.
	\label{eta*-full-exp}
\end{equation}

Knowing explicitly $C(t)$ for the harmonic oscillator \FF{(also in App.~\ref{app:1}, Eq.~\eqref{Ct-harm-osc})}, one can compute the desired derivatives:
\begin{equation}
	C(0) = \frac{T}{\omega_0^2} \;;\quad 
	\dot C(0) = 0 \;;\quad
	\ddot C(0) = -T \;;\quad
	\dddot C(0) = \eta T \;.
	\label{deriv-t0}
\end{equation}
Proper combinations of these quantities allow us to extrapolate all the parameters of the model. The importance of the first derivative as a quantity to discriminate between first and second order dynamics in oscillator-like models has already been stressed in \cite{Cavagna_2016,CAVAGNA20181}, with explicit reference to complex interacting systems. Our point is that we can go beyond the binary answer provided by $\dot C(0)/C(0)$, proportional -- through a time scale factor -- to 1 or to 0 for first or second order dynamics respectively, and give a quantitative estimation of the damping regime in which a system operates, employing all the derivatives at $t=0$ up to the third one.

By replacing Eqs.~\eqref{deriv-t0} into Eq.~\eqref{eta*-full-exp}, we obtain:
\begin{equation}
	\eta^* = -\frac{2}{3}\frac{\dddot C(0)}{\ddot C(0)}\left[1+O(\Delta t)\right] = \frac{2}{3}\eta + O(\Delta t).
	\label{eta*-deriv}
\end{equation}
We find then, at the leading order, a rescaling factor of $2/3$, as observed in numerical tests. No rescaling factors appear for the other inferred parameters: performing the same replacement and expansion of the analytical correlation functions in the inference formulas of $T$ and $\omega_0$, we see that temperature and pulsation are correctly retrieved from proper combinations of $C(0)$ and $\ddot C(0)$.

This result gives us a clue to understand the origin of the $\Delta t$-independent rescaling factor for $\eta$. Looking back at Eq.~\eqref{Euler-Maruyama}, one realizes from simple dimensional analysis that the elimination of the velocity variables makes terms of order $O(\Delta t^{3/2})$ appear, even if the starting accuracy of the expansion is $O(\Delta t)$. This means that Eq.~\eqref{L-EM} has been inconsistently derived retaining only some of the $O(\Delta t^{3/2})$ contributions; in turn this produces missing $O(\Delta t^3)$ contributions to the \FF{fluctuations of $x$.} This explains why Eq.~\eqref{eta*} is incorrect and shows the need of higher order discretization schemes for stochastic second order dynamics.

We finally remark that this $2/3$ rescaling factor is not a specific feature of the stochastic harmonic oscillator, but a recurrent trait in stochastic models of the form of Eq.~\eqref{oscillator-1}. \FF{As rigorously proven by Gloter, the so-called quadratic variation of the discretized velocities (corresponding to an empirical estimate of the squared acceleration) uniformly converges to the expected value for the quadratic variation of the real unobserved velocities rescaled by $2/3$ \cite{Gloter}. These quadratic variations are $O(\Delta t^3)$, and the former one is the only directly measurable quantity containing the necessary dynamical information to disentangle the contribution of dissipation from diffusion and infer $\eta$ in our setting \footnote{The class of models considered by Gloter in \cite{Gloter} isn't exactly the same as the one we consider in Eq.~\eqref{oscillator-1} ($f(x)=0$ is assumed and the presence of nonlinear nonconservative forces and of a multiplicative noise of the form of $\sigma(v)\xi$ is allowed) but we think that the result in \cite{Gloter} may be extended also to the $f(x)\neq0$ case.}. 
}


\subsection{Higher order inference schemes}\label{subsec:3}

The lowest order of convergence required to develop any reasonable dynamical maximum likelihood scheme is $O(\Delta t^{3/2})$. \FF{Since the mean square convergence of the infinitesimal increment of the process is what determines its statistical properties at any time, the minimum requirement for an inference method exploiting only local dynamical information is to reproduce fluctuations correctly at the leading order in $\Delta t$.}

Independently of the details of the discretization, following the procedure outlined in Sec.~\ref{subsec:1}, with $O(\Delta t^{3/2})$ accuracy one reduces to a sequence of intertwined Gaussian integrals for the marginalization of $v_1\dots v_L$, which may be cumbersome to compute for arbitrary length of the trajectory. Therefore, it is convenient to work again with update equations in $x$ space. \FF{They can be obtained either from a temporal discretization of the GLE \eqref{eqx} 
or from the elimination of the velocity variables in the discrete-time equations resulting from a second order Taylor-It\^o expansion of the Markov process in Eq.~\eqref{SDE-sys}. 
In the first case, since the same exponentially decaying kernel propagates both the noise and the initial condition in Eq.~\eqref{eqx}, it is possible to manipulate the integrated GLE to find a stochastic difference equation that does not contain $v_0$ and is driven by a short correlated effective noise:
\begin{dmath}
	x_{n+1}-x_n-e^{-\eta \Delta t} (x_n-x_{n-1})= \frac{1-e^{-\eta\Delta t}}{\eta}\int_{t_{n-1}}^{t_{n+1}} \Psi(t-t_n) f(x(t)) dt + \zeta_n\,
\end{dmath}
where
\begin{flalign}
	\label{zeta-1}\zeta_n=&\int_{t_{n-1}}^{t_{n+1}} \Psi(t-t_n) \xi(t) dt \;;\\
  	\label{psi}\Psi(t)=&
	\begin{cases}\frac{e^{\eta t}-e^{-\eta \Delta t}}{1-e^{-\eta\Delta t}}\quad \text{if}\ -\Delta t <t <0\;; \\
\frac{1-e^{\eta (t-\Delta t})}{1-e^{-\eta\Delta t}}\quad \text{if}\ \ \ \ 0 <t < \Delta t\;. 
	\end{cases}
\end{flalign}
Correspondingly, the \emph{`first discretize, then marginalize'} strategy provides a stochastic difference equation with the same properties. We detail both procedures in App.~\ref{app:discretization}.}

Concentrating on the case of the stochastic harmonic oscillator, any consistent discrete-time description in $x$ space takes the form of a linear stochastic difference equation like: 
\begin{equation}
	x_{n+1} + \alpha x_n + \beta x_{n-1} = \zeta_n,
	\label{x-2ord}
\end{equation}
where the inhomogeneous terms $\zeta_n$ are still Gaussian random variables of null mean, but they are no longer independent. 
This is the crucial difference with the Euler-Maruyama scheme, which takes into account only the diagonal entries of the covariance matrix $C_{nm}=\langle \zeta_n \zeta_m\rangle$.

Eq.~\eqref{x-2ord} defines an affine map:
\begin{equation}
	\boldsymbol \zeta=(\zeta_1,\dots, \zeta_{L-1})^{\top} \mapsto \mathbf x = (x_2,\dots, x_L)^{\top} = \mathbf M^{-1}\boldsymbol\zeta + \mathbf x_0,
	\label{map}
\end{equation}
where $M_{ij}=\delta_{i,j} + \alpha \delta_{i,j-1}+ \beta \delta_{i,j-2}$ and $\mathbf x_0=(x_0,x_1,0,\dots,0)^{\top}$, which can  be generalized to a nonlinear transformation when anharmonic forces are present. This map can be exploited, when the covariance matrix $\mathbf C$ and its inverse are known, to write the new, higher order, dynamical likelihood. For the harmonic oscillator, it reads:
\begin{widetext}
\begin{dmath}
	P_{(2)}(x_L,\dots,x_2|x_1,x_0) = \frac{1}{Z} \exp -\frac{1}{2}\sum_{n,m=1}^{L-1}(x_{n+1}+\alpha x_n +\beta x_{n-1}){C^{-1}}_{nm}(x_{m+1}+\alpha x_m + \beta x_{m-1}),
	\label{likelihood-Toeplitz}
\end{dmath}
\end{widetext}
where $Z$ is the normalization constant:
\begin{equation}
	Z = \left[(2\pi)^{L-1} \det \mathbf C\right]^{1/2} =
	\left[ \prod_{k=1}^{L-1} 2\pi \lambda_k \right]^{1/2}, 
	\label{Z-Jc0}
\end{equation}
with $\lambda_k$ the $k$-th eigenvalue of the covariance matrix $\mathbf C$. The effective parameters $\alpha$ and $\beta$, as well as the entries of the covariance matrix, are known combinations of the parameters of the model, whose details depend on the adopted discretization scheme. \FF{In the following results we adopt $\alpha = -1 -e^{-\eta \Delta t} + \omega_0^2\Delta t\left(1-e^{-\eta\Delta t}\right)/\eta$ and $\beta = e^{-\eta\Delta t}$.}

\FF{For well-chosen  $\alpha$, $\beta$ and $C_{nm}$, Eq.~\eqref{x-2ord} and Eq.~\eqref{likelihood-Toeplitz} are exact, in the limit $L\to \infty$. Thanks to linearity, it is possible to design an exact integration algorithm for the Markov process \eqref{SDE-sys} at any time step increment $\Delta t$ \cite{Gillespie}. For nonlinear generalizations of $f(x)$, the exact Gaussian character of the random increment is lost. However, at leading order in $\Delta t$, a multivariate Gaussian distribution still represents a good approximation for the distribution of the random increments $\zeta_n$ appearing in the $x$ update equation, which takes the form:
\begin{equation}
	x_{n+1}+F(x_n,x_{n-1};\boldsymbol\mu) = \zeta_n,
		\label{gen-update-x}
\end{equation}
with $\boldsymbol\mu$ a set of effective parameters. The corresponding generalization of Eq.~\eqref{likelihood-Toeplitz} can be obtained (see App.~\ref{app:2}).}

To order $O(\Delta t^3)$, for both linear and nonlinear second order processes, one can deduce from Eq.~\eqref{zeta-1} that $\mathbf C$ has a  `nearest-neighbour' structure of the kind:
\begin{equation}
	C_{nm} = \langle\zeta_n\zeta_m\rangle = a\,\delta_{n,m} + b\,\delta_{n,m\pm1}
	\label{Cnm}
\end{equation}
where
\begin{equation}
	a \simeq \frac{2}{3}2T\eta\Delta t^3\;;\quad b \simeq \frac{1}{6}2T\eta\Delta t^3.
	\label{ab-def}
\end{equation}
Hence the covariance matrix has the form of a symmetric tridiagonal Toeplitz matrix of order $L-1$. These mathematical features carry a deep physical meaning: first of all, the presence of non-vanishing off-diagonal elements is the signature of a colored noise. Secondly, the fact that the matrix is banded means that the correlation of the noise variables is short-ranged, i.e. that the associated memory kernel, in a continuous-time description, decays fast \cite{Miguel1980}. Finally, the Toeplitz structure is synonymous with shift invariance.

A more careful derivation of the update equations in $x$ space would require shift invariance not to hold and the first entry of the covariance matrix $C_{11}$ to be different from the other elements of the main diagonal.  Eq.~\eqref{x-2ord} is in fact not valid for the first integration step, where the initial conditions intervene. In this respect the structure of the data also poses the problem of the elimination of the initial condition $v_0$ in favour of $x_0$ and $x_1$. Even if not able to perform it explicitly without stationarity assumptions, we can argue (see App.~\ref{app:discretization}) that it has the effect of modifying the covariance matrix in the following way:
\begin{equation}
	C=\begin{pmatrix}
	\tilde a & b & \dots & 0 \\
	b & a & . & \vdots\\
	\vdots & . & \ddots & b\\
	0 & \dots & b & a
	\end{pmatrix},
	\label{true-Cnm}
\end{equation}
where the shift invariance expressed by the Toeplitz structure of Eq.~\eqref{Cnm} is then broken at the beginning of the time series. 
Despite that, the error we make by replacing $\tilde a$ with $a$ in the quasi-Toeplitz matrix \eqref{true-Cnm} is negligible in the limit of long trajectories, as discussed in Sec.~\ref{sec:initial} and checked in Fig.~\ref{fig:varying_length}. Intuitively, since the breaking of the shift invariance occurs only at the first step, the longer the trajectory, the more similar this is to a truly shift invariant situation. Notice that what matters is not the total length $(L+1)\Delta t$ of the trajectory in units of the physical time scales of the process, but just the number of points $L+1$ of which the trajectory is made up \footnote{This is not surprising if one carefully looks at the expression of the inverse of the tridiagonal Toeplitz matrix Eq.~\eqref{inv-Toeplitz}, which closely resembles Fourier series expansions. Increasing the number of points corresponds to including an increasing number of harmonics; finite size corrections to parameters estimators can be seen as a counterpart of the Gibbs phenomenon.}.

Apart from the difficulty in determining correctly $\tilde a$, the advantage of replacing the true covariance matrix Eq.~\eqref{true-Cnm} with a Toeplitz matrix is that the inverse of the Toeplitz matrix is explicitly known, as well as the eigenvalues \cite{Toeplitz, Meurant}:
\begin{equation}
	{C^{-1}}_{nm} = \frac{2}{L}\sum_{k=1}^{L-1}\dfrac{\sin\left(\frac{nk\pi}{L}\right)\sin\left(\frac{mk\pi}{L}\right)}{a+2b\cos\left(\frac{k\pi}{L}\right)};
	\label{inv-Toeplitz}
\end{equation}
\begin{equation}
	\lambda_k = a + 2b\cos\left(\frac{k\pi}{L}\right).
	\label{eigenvalue-Toeplitz}
\end{equation}
Let us highlight that the inverse of the covariance matrix does not preserve a banded structure. This means that, even if noise correlations are local in time, two-time functions of every pair of points of the trajectory enter into the minus-log-likelihood. Hence Eq.~\eqref{likelihood-Toeplitz} cannot be factorized. Factorization corresponds to a block structure for $\mathbf C^{-1}$, which implies a block structure for $\mathbf C$. This is incompatible with the tridiagonal Toeplitz or quasi-Toeplitz nature of the covariance matrix, where off-diagonal elements are of the same order as the diagonal ones.

Nonetheless, having built an explicit discrete path integral measure, a maximum likelihood approach is practicable, and it reduces to minimizing the quantity $\mathcal L=-\ln P(x_L,\dots,x_2|x_1,x_0)$ with respect to the parameters of the model. Thanks to the regularities of Eq.~\eqref{likelihood-Toeplitz}, the minimization of $\mathcal L$ can be performed analytically in the case of the harmonic oscillator and, in general, of simple single-particle systems. The optimization procedure can be performed semi-analytically also for many-particle systems, like active agent-based microscopic models or spatially discrete counterparts of field theoretical models. In these cases an additional parameter is typically the interaction range of effective pair-wise potentials, which may depend on a different (measurable) variable than the field-like observable $x$. In general, once an expression for $\mathcal L$ is given, a large number of optimization algorithms are available to minimize it with respect to all the extra parameters that do not allow for a full analytical approach.

Complete inference formulas for one-dimensional harmonic \FF{and anharmonic} oscillators and for a system of many coupled harmonic oscillators with parameter-dependent connectivity matrix are reported in Apps.~\ref{app:2}--\ref{app:3}. In all cases, optimal parameter values are given by combinations of all the two-time functions up to the length of the trajectory, and not only those computed at a temporal distance of 0, 1 and 2 time steps.

\FF{For the non-interacting case, we tested the developed schemes numerically by applying the inference formulas to synthetic stochastic trajectories of two reference processes: the Brownian motion in a harmonic potential, and the Brownian motion in a symmetric anharmonic potential $V(x) = \frac{1}{2}kx^2 + \frac{1}{4}\lambda x^4$. The equations of motion corresponding to the latter read:
\begin{equation}
	\begin{cases}
	\dot x = v\\
	\dot v = -\eta v - kx -\lambda x^3 + \sigma\xi\,,
	\end{cases}
	\label{anharmonic-SDE}
\end{equation}
where we chose a unitary mass particle, $\sigma^2 = 2T\eta$ and $\xi(t)$ as a white noise.} We generated synthetic trajectories as in~\cite{CVE} and subsampled them by progressively increasing the time separation $\Delta t$ between subsequent observed points.

\begin{figure*}[t]
	\subfloat{\includegraphics[clip,width=\columnwidth]{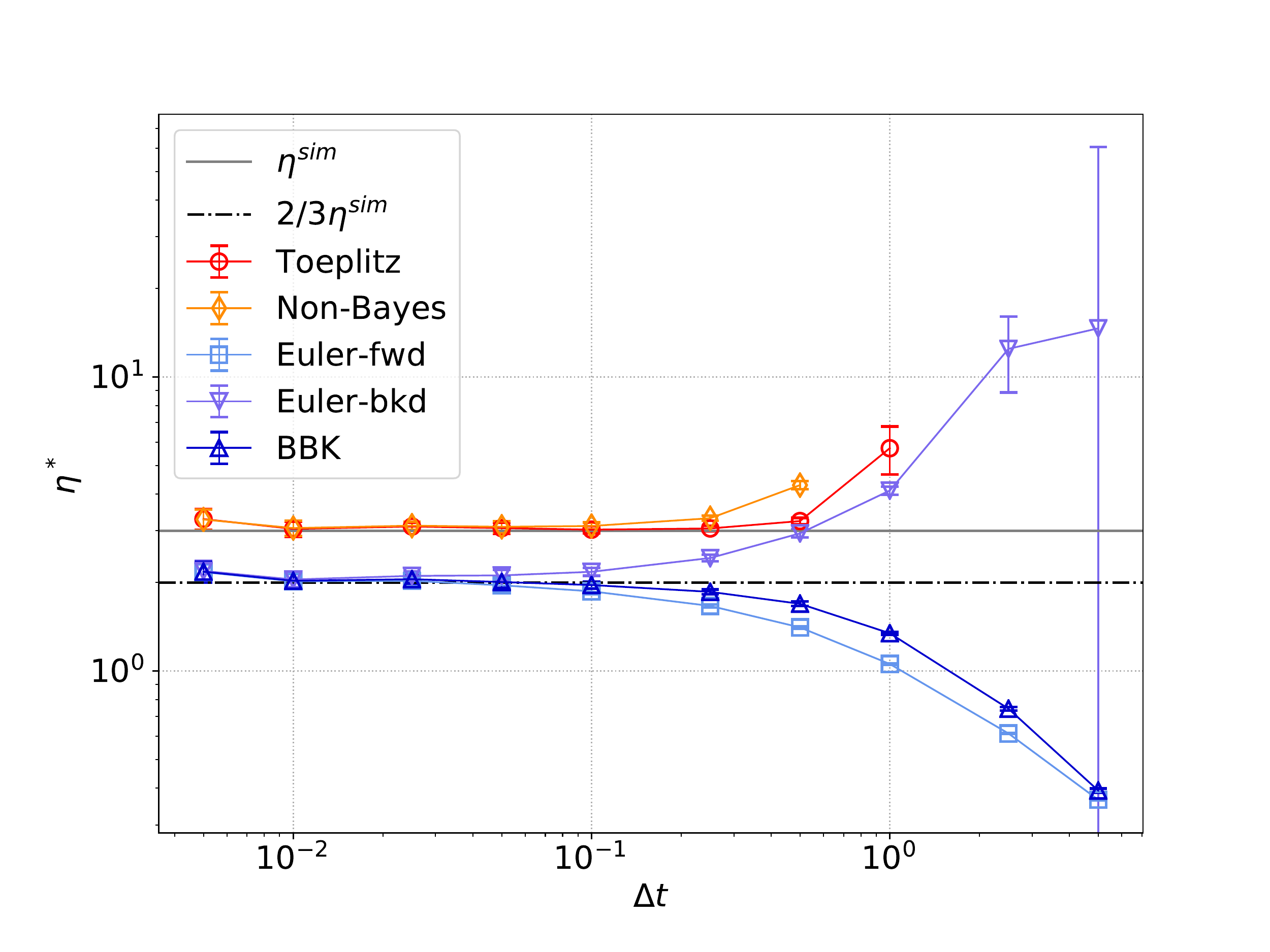}\llap{
  \parbox[b]{16.3cm}{(a)\\\rule{0ex}{2.2in}}}\label{box-eta-all}}	
	\subfloat{\includegraphics[clip,width=\columnwidth]{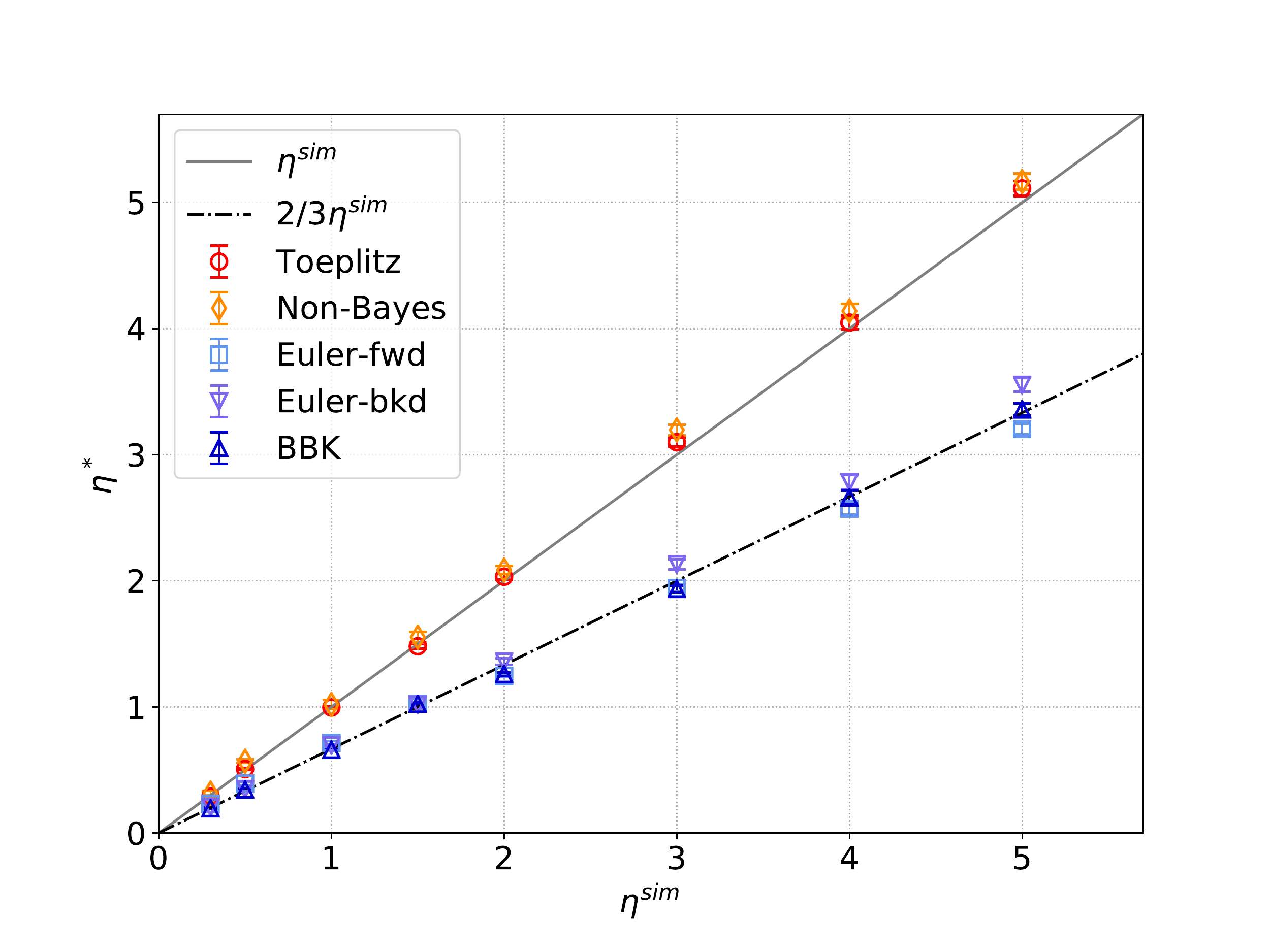}\llap{
  \parbox[b]{16.3cm}{(b)\\\rule{0ex}{2.2in}}}\label{box-slope}}
	
	\subfloat{\includegraphics[clip,width=\columnwidth]{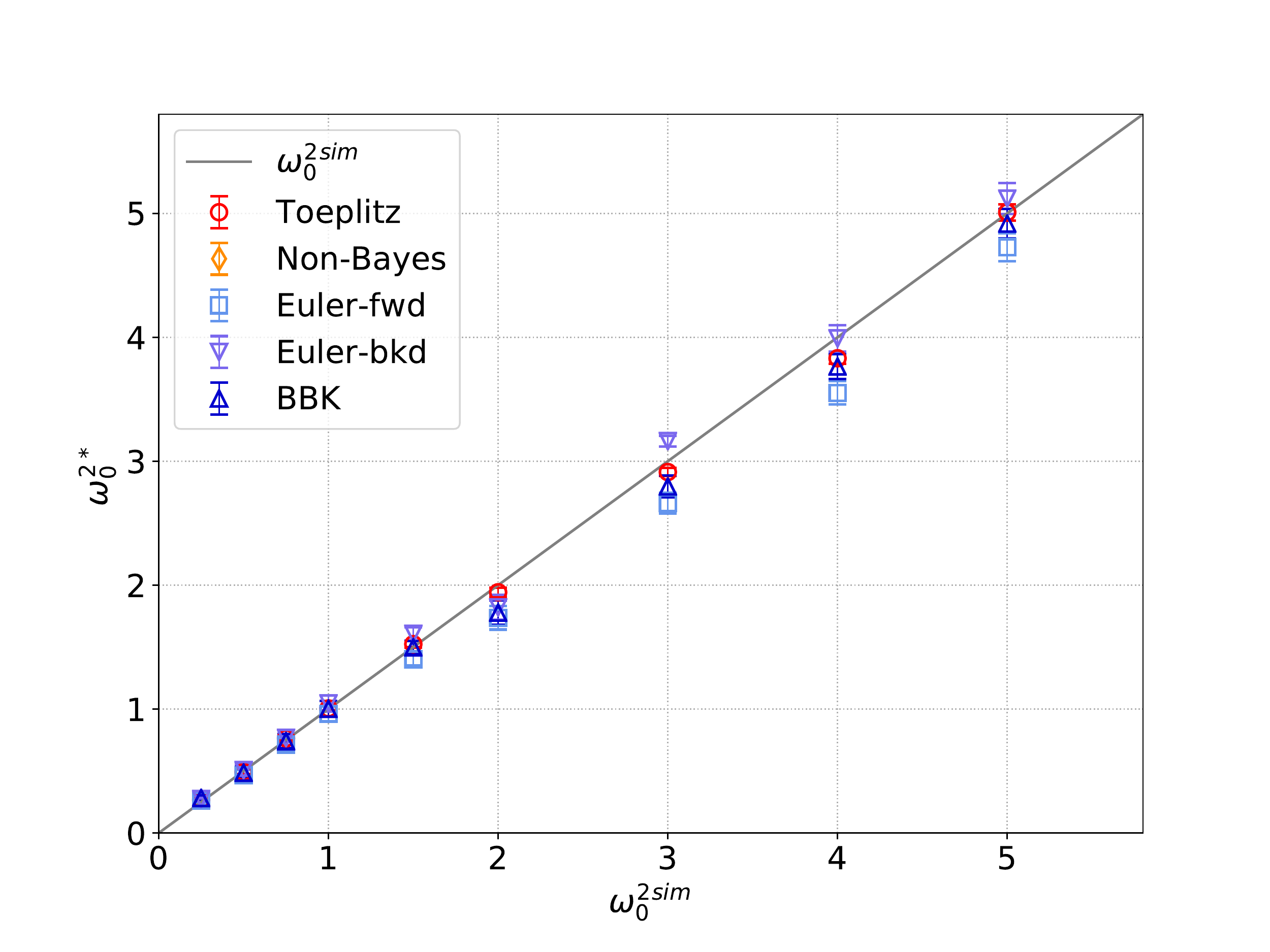}\llap{
  \parbox[b]{16.3cm}{(c)\\\rule{0ex}{2.2in}}}\label{box-J-slope}}	
	\subfloat{\includegraphics[clip,width=\columnwidth]{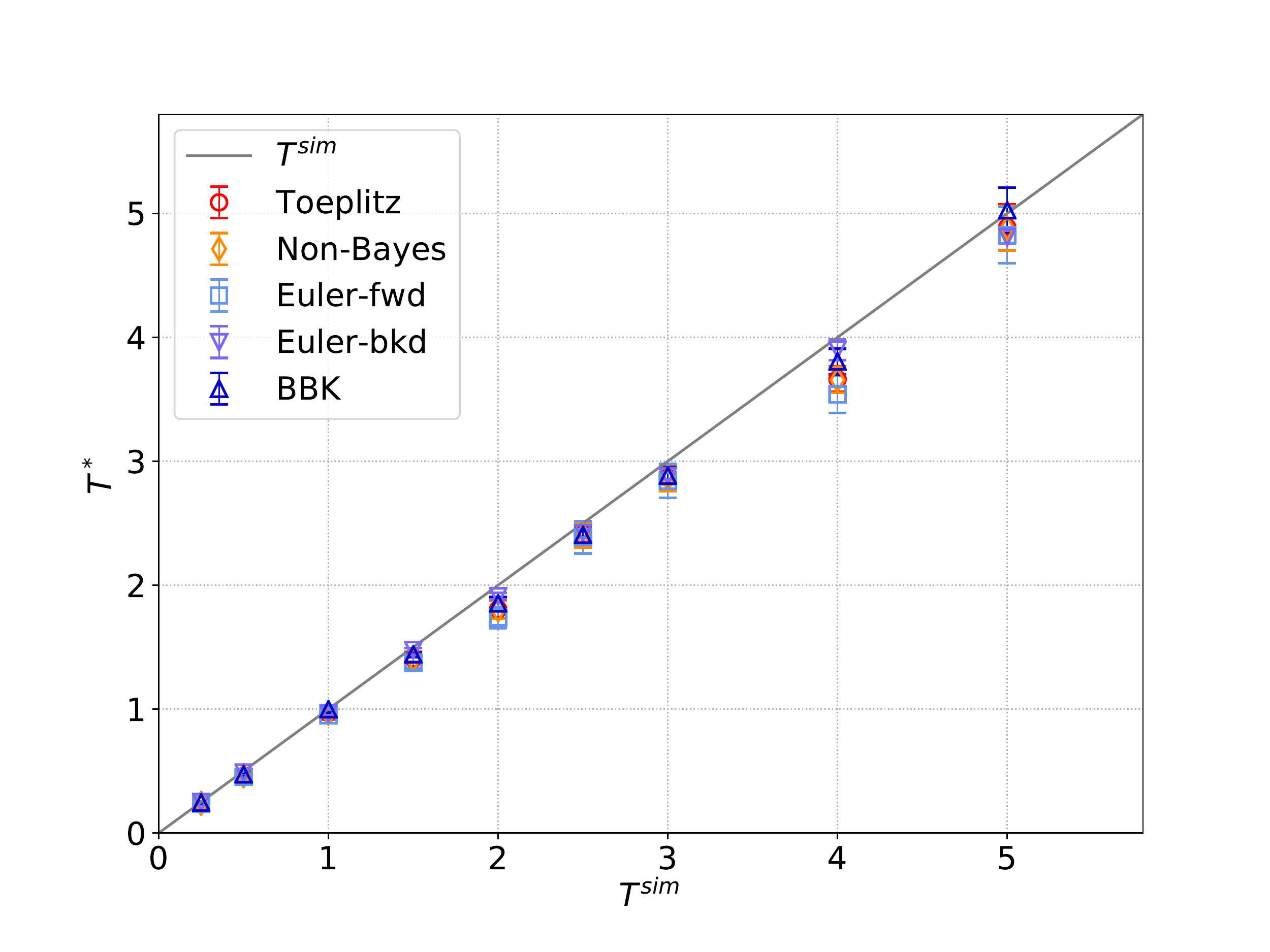}\llap{
  \parbox[b]{16.3cm}{(d)\\\rule{0ex}{2.2in}}}\label{box-T-slope}}
	
	\caption{Inference results for the stochastic harmonic oscillator. Sample trajectories are obtained from \FF{exact numerical integration of the set of first order equations} with parameters $\eta^{sim}$, $\omega_0^{sim}$ and $T^{sim}$. The simulation time step $\tau^{sim}$ is always equal to 0.005, and it corresponds to the minimum displayed value of $\Delta t$ in (a).  Points at higher values of $\Delta t$ are obtained applying the inference procedure to sub-trajectories extracted from the original one. Each of the points displayed in (b) -- (d) is obtained as a weighted average of the inference results for different $\Delta t$ values in the range where the small $\Delta t$ approximation is valid. Weights correspond to the squared inverse of the errorbars, displayed in Fig.~(a) for the $\eta$ parameter. \FF{We compare the accuracy of all the schemes derived in App.~\ref{inf-formulas} from a first order Taylor-It\^o expansion (Euler-fwd, Euler-bkd, BBK) and from a second order expansion (Toeplitz, Non-Bayes).} Fig.~(a)\:: Inferred values for the damping coefficient of the harmonic oscillator, $\eta^*$. Averages over 10 sample trajectories of 5000 points (for any $\Delta t$) are reported with their 0.95 CI. Simulation parameters: $T=1$, $\omega_0=1$, $\eta=3$. Fig.~(b)\:: Inferred damping coefficient $\eta^*$ vs true simulation parameter $\eta^{sim}$: results from higher order methods follow the line of slope 1, whereas numerical results from na\"ive methods fall on the line of slope 2/3. The remaining parameters are fixed: $T=1$, $\omega_0=1$. Fig.~(c)\:: Inferred squared frequency of the harmonic oscillator ${\omega_0^2}^*$ vs true simulation parameter ${\omega_0^2}^{sim}$. All the schemes give correct results in this case in the whole explored range of values. Simulation parameters: $\eta=3$, $T=1$. Fig.~(d)\:: Inferred temperature $T^*$ vs the true value of the simulation parameter $T^{sim}$: again, results from all schemes fall on the line of slope 1 in the whole explored range of values. Remaining simulation parameters: $\eta=1.5$, $J=1$.} 
	\label{fig:4panel}
\end{figure*}

\FF{The comparison with na\"ive inference schemes for the example of the harmonic oscillator confirms the analytical predictions (Fig.~\ref{fig:4panel}). In any damping regime, the higher order inference method outperforms the na\"ive scheme in two ways: perturbatively, since the convergence of the parameter estimators is extended to a larger $\Delta t$ window due to the higher order Taylor-It\^o expansion (an example in Fig.~\ref{box-eta-all}), and non perturbatively in $\Delta t$, since no rescaling factor for the $\eta$ parameter is required (Fig.~\ref{box-slope}). The different behaviour of the various schemes at large $\Delta t$, where the series expansion is non-asymptotic, is probably related to the details of the discretization rules and their stability properties.

\begin{figure*}[t]
\newlength\imageheight
\settoheight{\imageheight}{\includegraphics[clip,height=0.33\textheight,trim={ 0.75cm 0 0.5cm 0}]{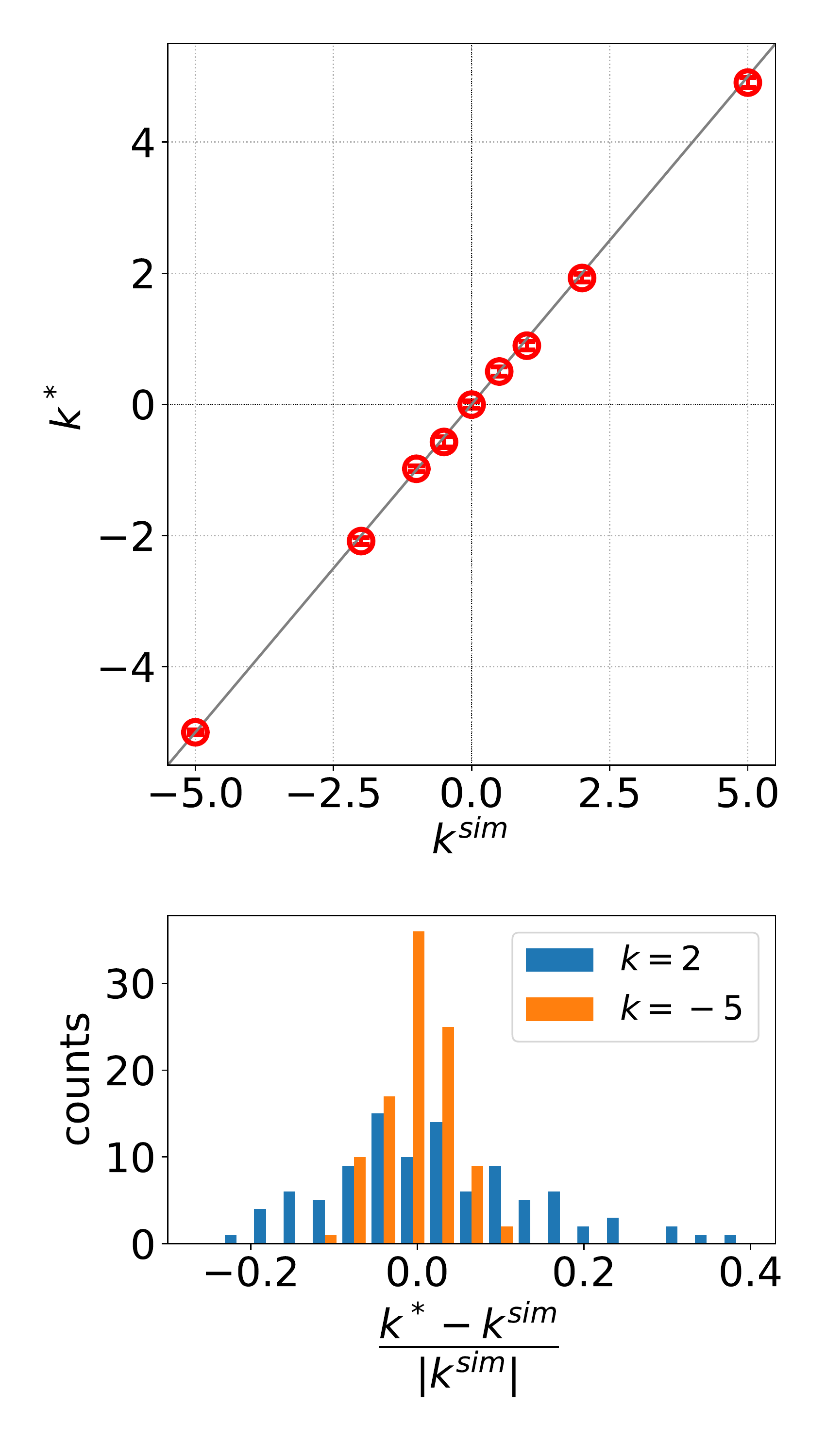}}
	\subfloat{\includegraphics[clip,height=.34\textheight,trim={1.05cm 0 0.2cm 0}]{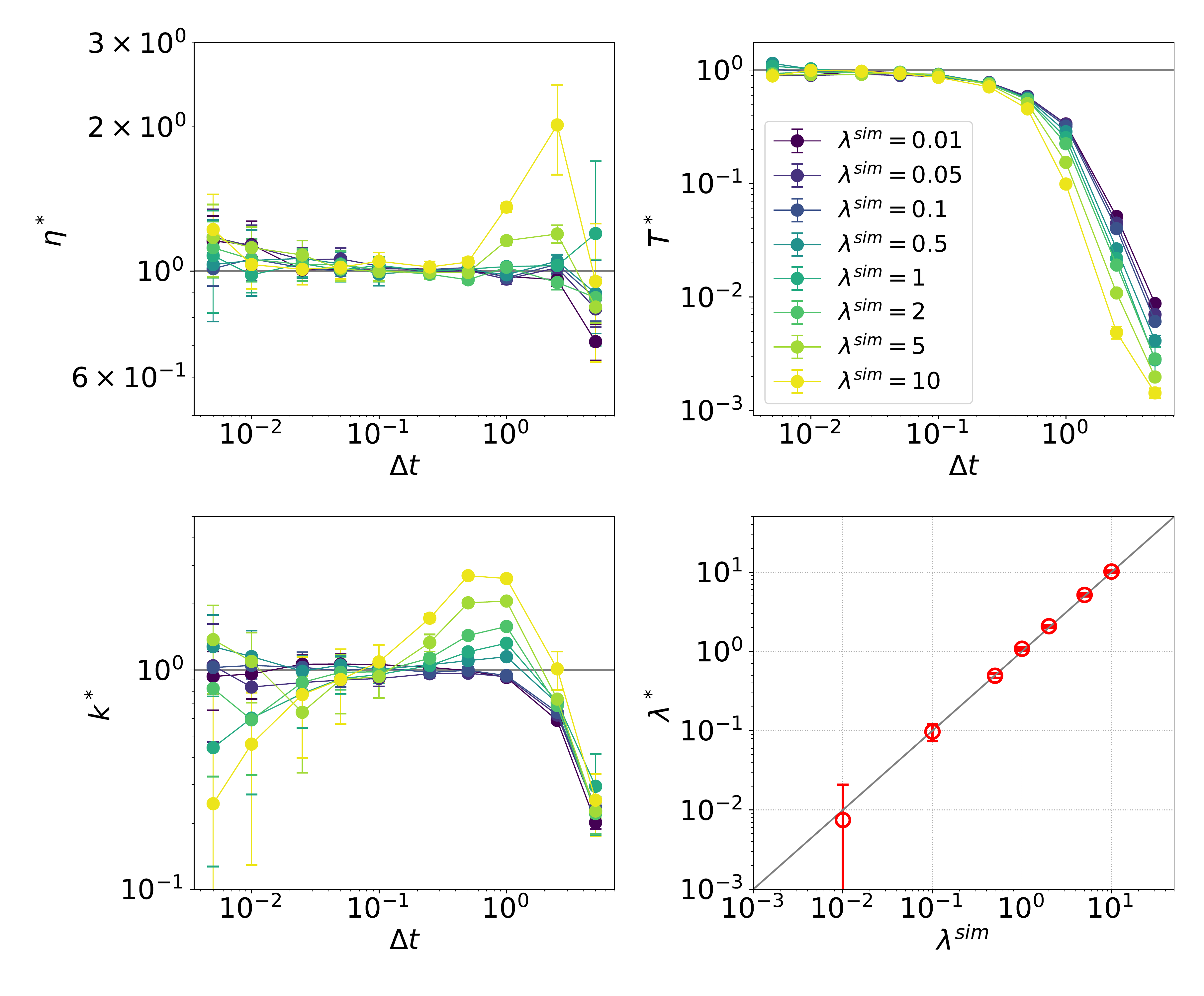}\llap{
  \parbox[b]{18.4cm}{(a)\\\rule{0ex}{2.97in}}}\llap{
  \parbox[b]{8.6cm}{(b)\\\rule{0ex}{2.97in}}}\llap{
  \parbox[b]{18.4cm}{(c)\\\rule{0ex}{1.47in}}}\llap{
  \parbox[b]{8.6cm}{(d)\\\rule{0ex}{1.47in}}}\label{4_anharmonic}}  
  	\subfloat{\tikz\node[minimum height=\imageheight]{\includegraphics[clip,height=0.33\textheight,trim={ 0.5cm 0 1.5cm 0}]{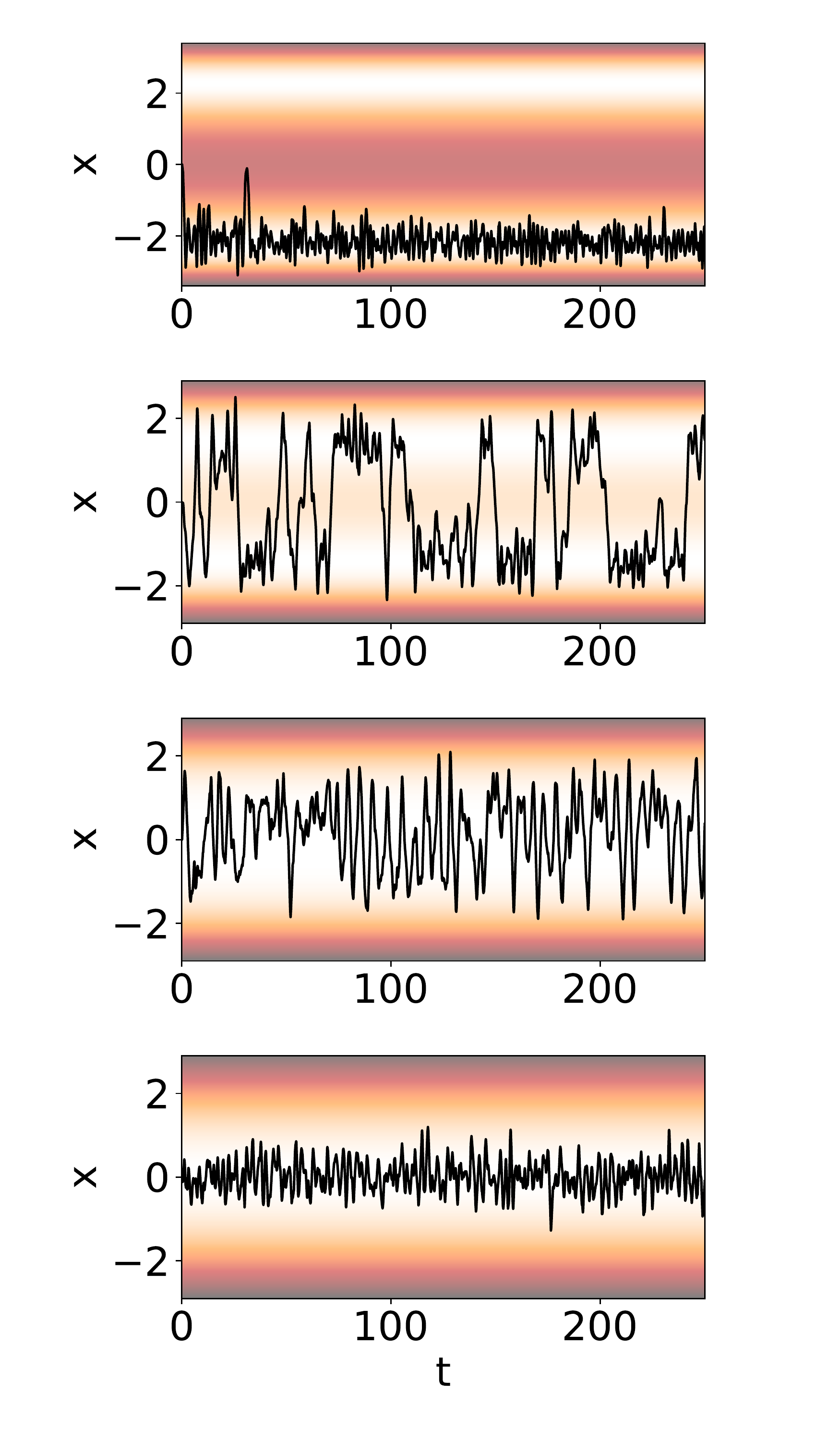}};\llap{
  \parbox[b]{7.8cm}{(e)\\\rule{0ex}{2.97in}}}\label{traj}}
  	\subfloat{\tikz\node[minimum height=\imageheight]{\includegraphics[clip,height=0.33\textheight,trim={ .9cm 0 0.5cm 0}]{k_inf.pdf}};\llap{
  \parbox[b]{8.3cm}{(f)\\\rule{0ex}{2.97in}}\llap{
  \parbox[b]{8.3cm}{(g)\\\rule{0ex}{1.12in}}}\label{k}}}
  
  	\caption{Bayesian inference of the dynamical parameters of a Brownian motion in a force field $f(x)=V'(x)$, with $V(x)=\frac{1}{2}kx^2+\frac{1}{4}\lambda x^4$. Only the Toeplitz method is applied; as for the harmonic oscillator, in Figs.~(a)--(d) and (f) 10 sample trajectories of length 5000 points are considered for each $\Delta t$. Errorbars are 0.95 CI. Figs.~(a)--(c)\:: Inferred model parameters against subsampling parameter $\Delta t$. The true value is equal to 1 in all cases and is marked by the straight grey line. Fig.~(d)\:: Inferred vs true value of the nonlinear coefficient $\lambda$. Fig.~(e)\:: Excerpts of sample trajectories in various landscapes. The strength of the confining potential is qualitatively indicated by the colormap, with light areas corresponding to the minimum of the potential. The following parameters of the simulation are kept fixed: $T=1$, $\eta=1$, $\lambda=1$. By varying the parameter $k$ we realize, from top to bottom: a strong confinement in a double well potential, with long exit times, at $k=-5$; a switching dynamics with relatively short switching times, at $k=-2$; a marginal situation at $k=0$; confined Brownian motion in the vicinity of the origin at positive values of k ($k=5$). Fig.~(f)\:: Inferred vs true value of the parameter of the linear force $k$, assuming both positive and negative values. Fig.~(g)\:: Histogram of counts for the relative distance of the inferred parameter $k^*$ to the simulation parameter $k^{sim}$. With fixed $\lambda^{sim}=1$ and $k^{sim}=\{2,-5\}$, the weight of anharmonicity varies, but the variance of all the estimated parameters seems to be unaffected. As a result, relative errors decrease for larger $|k|$. 100 trajectories are sampled for each $k$ value shown in the histogram, and $\Delta t=0.025$ in all cases.}
	\label{fig:anharmonic}
\end{figure*}

Fig.~\ref{fig:anharmonic} shows numerical results based on the Toeplitz inference scheme for the anharmonic stochastic oscillator for varying values of the parameters $\lambda^{sim}$ (Figs.~\ref{fig:anharmonic}a--d) and $k^{sim}$ (Figs.~\ref{fig:anharmonic}e--g). In all the explored regimes the inference scheme provides excellent results, showing, in particular, that no bias is introduced by the possible imbalance between linear and nonlinear force terms (values close to the origin are correctly estimated in Fig.~\ref{fig:anharmonic}d and Fig.~\ref{fig:anharmonic}.f), even if, for a fixed $\Delta t$, an increase in the relative error or more noisy estimations cannot be prevented in these conditions (Fig.~\ref{fig:anharmonic}g). Moreover, no bias is introduced by the fact that, when $k$ assumes a negative value, the particle may be confined in a single minimum of the double-well potential for all the length of the sampled trajectory (see Fig.~\ref{fig:anharmonic}.e). }


\FF{
\subsection{Generalization to multiplicative noise}\label{multiplicative}

\begin{figure*}[t]
	\settoheight{\imageheight}{\includegraphics[height=.21\textheight]{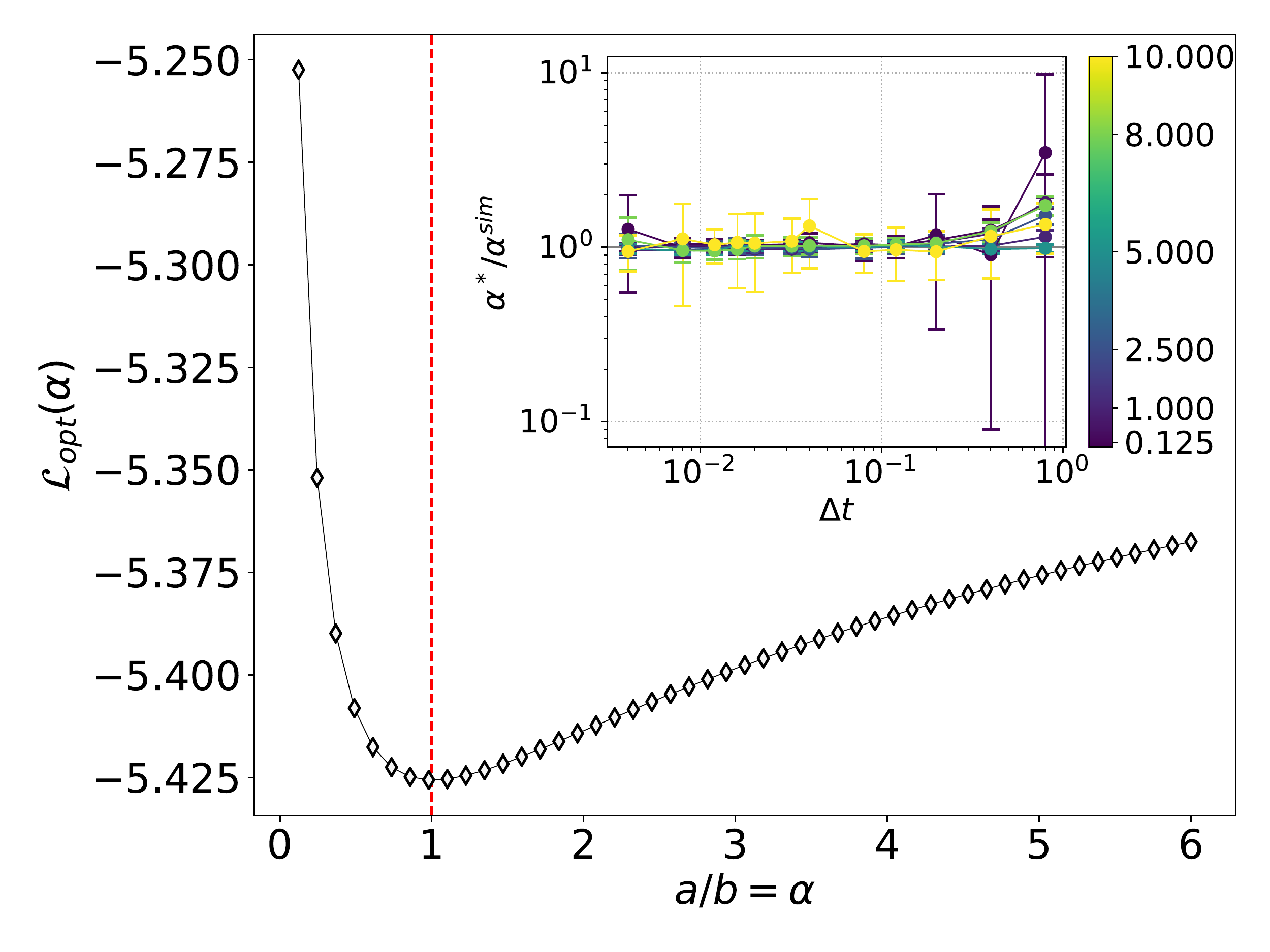}}
	\centering
	  \hspace{.06cm}\subfloat{\tikz\node[minimum height=\imageheight]{\includegraphics[clip,height=.2\textheight]{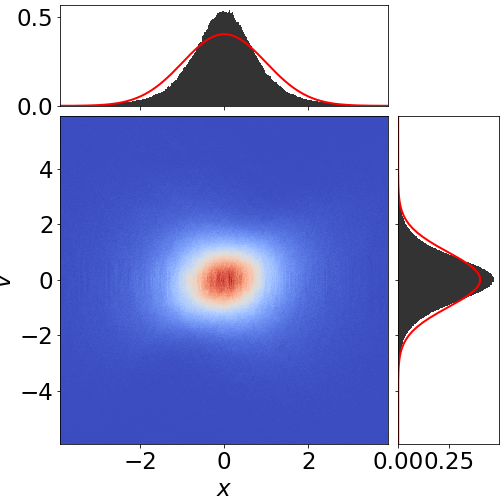}};\llap{  \parbox[b]{9.8cm}{(a)\\\rule{0ex}{1.8in}}}\label{histo-multi}
	}	
	\subfloat{\includegraphics[clip,height=.21\textheight]{lik_inset.pdf}\llap{  \parbox[b]{12.8cm}{(b)\\\rule{0ex}{1.8in}}}\label{lik-inset}
	}	
  \hspace{.1cm}
	\subfloat{\includegraphics[clip,height=.21\textheight]{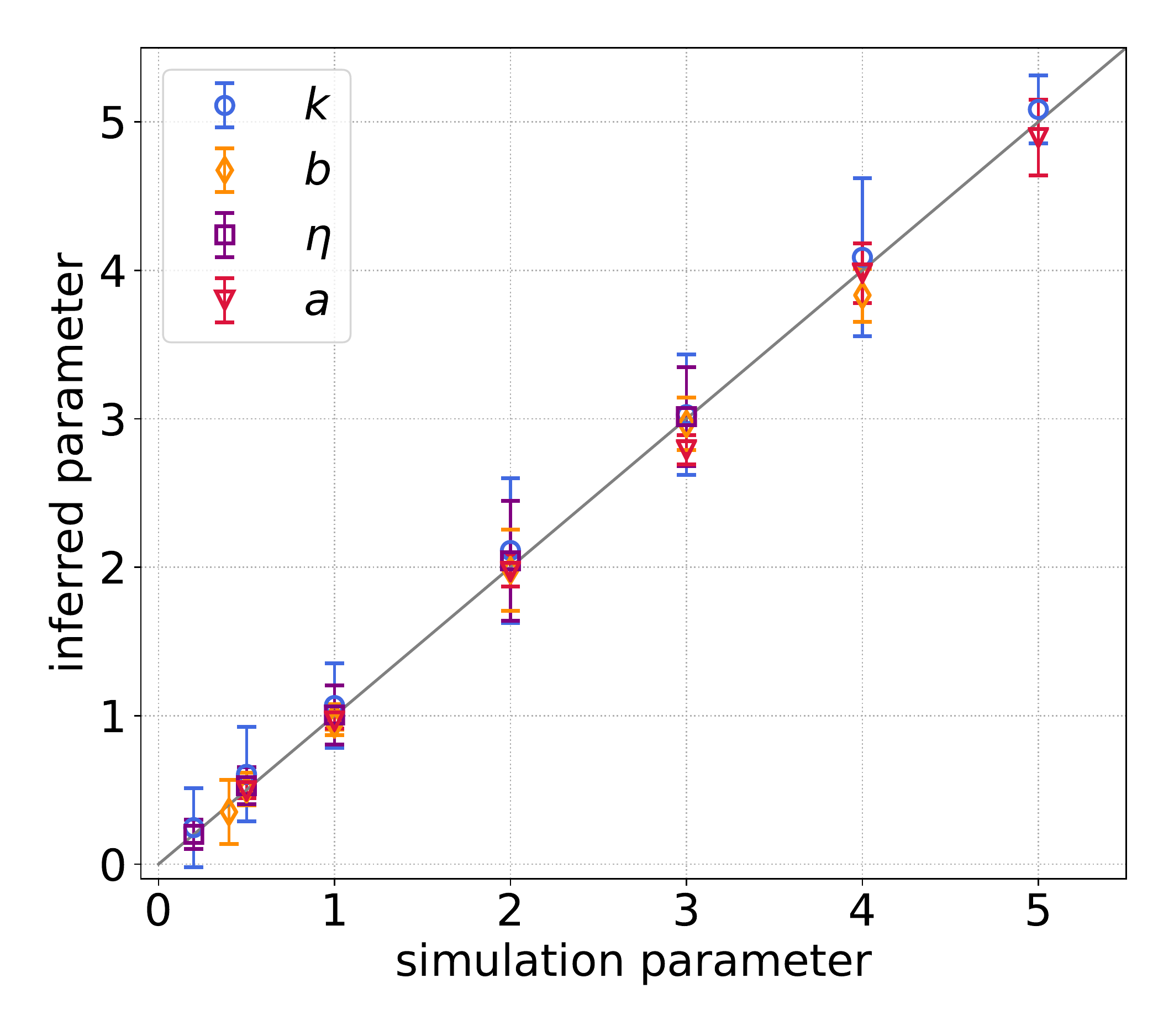}\llap{  \parbox[b]{10.8cm}{(c)\\\rule{0ex}{1.8in}}}\label{par-multi}
	}
	\caption{Inference method applied to a multiplicative process. The process is described by Eq.~\eqref{multi-class} with $f(x)=-kx$ and $\sigma(x)=\sqrt{a+bx^2}$. (a): The fraction of time spent by the system in each region of the phase space for a sample trajectory of length $4\cdot10^4$, with $k=1$, $\eta=1$, $a=1$, $b=1$ and initial condition $(x_0=0,v_0=0)$. There is a clear difference with the Gaussian distribution having the same second moment (red line), showing the effect of the multiplicative noise. (b): Analytically optimized negative log-likelihood as a function of the effective parameter $\alpha=a/b$, computed on a sample sub-trajectory of 5000 points, $\Delta t=0.016$, with the same parameters as in (a). In the inset optimal values of $\alpha$ as a function of $\Delta t$ are reported. Errorbars correspond to 0.95 CI on 10 sample trajectories of 5000 points for each $\Delta t$. The color code refers to the value of $\alpha^{sim}$, measuring the relative contribution of additive and multiplicative part of the noise term. (c): Performance of the method in inferring the whole set of parameters of the model. }
	\label{fig:multi}
\end{figure*}

In order to understand the limits and full potential of the method, we focus in this section on possible generalizations to the case of non-additive noise. An adaptation of our non-Markovian Bayesian inference scheme can be developed for the following class of multiplicative processes: 
\begin{equation}
	\ddot x = -\eta\dot x + f(x) + \sigma(x)\xi,
	\label{multi-class}
\end{equation}
with $\xi(t)$ a standard white noise and initial conditions $x(0)=x_0$, $\dot x(0) = v_0$. This model has two features: linear dissipation, and a velocity-independent diffusion coefficient only proportional to $\sigma^2(x)$. Under these conditions, the memory kernel of the GLE associated to Eq.~\eqref{multi-class} is explicitly known
and, following the same procedure that led to the discretization of the additive process in Sec.~\ref{subsec:3}, we obtain an approximated discrete time update rule of the form:
\begin{dmath}
	x_{n+1} - x_n - e^{-\eta\Delta t}(x_n - x_{n-1}) - \frac{1-e^{-\eta\Delta t}}{\eta}\Delta t f(x_n) = \zeta_n,
	\label{update-multi}
\end{dmath}
where the stochastic term is defined as
\begin{equation}
	\zeta_n = \frac{1-e^{-\eta\Delta t}}{\eta} \int_{t_{n-1}}^{t_{n+1}}dt' \Psi(t'-t_n) \sigma(x(t'))\xi(t').
	\label{zeta-def}
\end{equation}
The function $\Psi(t)$ is defined in the same way as in Eq.~\eqref{psi}.

From now on we will implicitly refer to the It\^o integration prescription. However, due to the fact that $\sigma(x)$ only depends on the configurational degree of freedom, $x$, the mean square convergence of $\zeta_n$ is not affected by a switch to the Stratonovich convention. As a result, one can say that, up to $O(\Delta t^3)$, stochastic terms satisfy
\begin{equation}
	\langle \zeta_n \zeta_m \rangle \simeq \frac{2}{3}\Delta t^3 \sigma^2(x_n)\delta_{n,m} + \frac{1}{6}\Delta t^3 \sigma(x_n)\sigma(x_m) \delta_{n,m\pm1}\,.
	\label{cov-multi}
\end{equation}
This choice of off-diagonal terms ensures the positiveness of the matrix, if $\sigma(x)>0$ \footnote{There exists a similarity transformation that transforms the matrix in Eq.~\eqref{cov-multi} into a strictly diagonally dominant matrix with positive entries. Since the spectrum is unchanged and, in particular,  real, this ensures the non-negativity of all the eigenvalues.}.
The covariance matrix also preserves a tridiagonal symmetric structure. However, the Toeplitz property is lost since, in the presence of multiplicative noise, shift invariance cannot hold. Nevertheless, we can build an efficient maximum likelihood inference routine. Let us rewrite the minus log-likelihood associated to Eq.~\ref{update-multi} as
\begin{widetext}
\begin{dmath}
	\mathcal L = \frac{1}{2}\sum_{k=1}^{L-1}\ln\lambda_k(\bold x;\boldsymbol \nu) + \sum_{n,m=1}^{L-1} \left[x_{n+1} - F(x_n,x_{n-1};\boldsymbol\mu)\right]{C^{-1}}_{nm}(\bold x;\boldsymbol \nu) \left[x_{m+1} - F(x_m,x_{m-1};\boldsymbol\mu)\right],
	\label{lik-multi}
\end{dmath}
\end{widetext}
so that we can distinguish between the subset of parameters $\boldsymbol\mu$, including $\eta$ and the parameters of the conservative potential, and the subset $\boldsymbol \nu$ appearing in the $x$-dependent diffusion coefficient $\sigma(x;\boldsymbol\nu)$. For the parameters in the former set, analytical formulas for their max-likelihood estimators can be found as functions of $\boldsymbol\nu$, while the latter generally requires numerical optimization (unless $\sigma(\bold x;\boldsymbol\nu)$ is univariate and has a purely multiplicative dependence on its single parameter). The effective cost function can be evaluated, also in the case of long trajectories, once the inverse and the spectrum of the symmetric tridiagonal matrix are computed.

To illustrate the method, we applied it to the multiplicative process in Eq.~\eqref{multi-class}, with $f(x) = -k x$ and $\sigma(x)=\sqrt{a+bx^2}$, where $a$ and $b$ are non-negative parameters. In this case the max-likelihood procedure can be reduced to a one-dimensional numerical optimization. Complete inference formulas are reported in App.~\ref{app:multiplicative} and the results are shown in Fig.~\ref{fig:multi}. These confirm that the method provides a reliable inference tool also in the case of a nonequilibrium multiplicative process, independently of the relative strength of the additive and multiplicative contributions to the noise term, and that the procedure does not require equilibrium assumptions to work, nor does it exploit the fluctuation dissipation theorem. 
}


\section{Alternative non-Bayesian approach}\label{subsec:4}

Alternative inference approaches to the maximum likelihood method are also possible. \FF{Several examples are known in the literature: the most general ones, applicable to a vast class of second order stochastic processes, derive the parameters of the assumed model (in the form of a SDE or of a chosen set of projection functions) through a fitting procedure on  measurable quantities, typically involving conditional moments of the increments of the process \cite{Ronceray, Lehle2015, Lehle2018, Pedersen}. Also in this case the relations used for fitting can be found through a Taylor-It\^o expansion even when a nonlocal solution in time is unknown. Other strategies have been proposed with a reformulation of the task -- having relevant application in chemical physics and molecular dynamics -- i.e. not to learn the best model for the measured variables, but to find from higher dimensional data the coarse-grained dynamics of a given system \cite{Dequidt, Harmandaris}.

In this section we put ourselves in a simpler framework than that of Ref.~\cite{Ronceray, Lehle2015, Lehle2018} and derive non-Bayesian parameter estimators just for the stochastic harmonic oscillator, in order to compare on this example the non Bayesian methodology and the maximum likelihood dynamical inference scheme we developed.} From update equations in position space like Eq.~\eqref{x-2ord}, obtained from an $O(\Delta t^{3/2})$ Taylor-It\^o expansion, some relations between experimental correlation functions and model parameters can be found. Let's take the update equation of the Langevin impulse integrator \cite{LI}:
\begin{equation}
	x_{n+1} = x_{n} + e^{-\eta\Delta t} (x_n-x_{n-1}) + \frac{1-e^{-\eta\Delta t}}{\eta}\omega_0^2\Delta tx_n + \zeta_n.
	\label{impulse}
\end{equation}
with $\zeta_n$ the Gaussian random variables characterized by Eq.~\eqref{Cnm}. Multiplying both sides of Eq.~\eqref{impulse} by $x_m$, for $m\in\{n-1,n,n+1\}$, and self-consistently averaging over the noise distribution, yields a set of three independent equations, from which all the parameters of the dynamical model can be extracted (explicit formulas are derived in Appendix~\ref{app:4}).

Notice that, in contrast to the max-likelihood inference method, the obtained relations can involve only three types of temporal correlation functions: equal-time, one-time-step and two-time-step correlations. Even if we are not using all the exploitable information carried by an $N$-point trajectory (the operation outlined above could in principle be performed for all $x_m$), this is the optimal minimal choice. Indeed, the shape of the temporal correlation function at small times contains substantial dynamical information. Moreover, due to the finite length of the trajectories, two-time quantities, like correlation functions, are typically better estimated at small time differences than at large ones.

As expected, parameter estimators provide good values without rescaling. Unfortunately, however, \FFnew{we cannot extend this} approach to interacting systems, where an interaction range is needed to parametrize the potential. As these formulas do not come from the optimization \FF{of any cost function}, there is no \FF{efficient} numerical strategy to find the best parameters of the interaction potential. \FF{The problem is bypassed if no assumption is made about the structure of the interaction, and a different parameter is associated to each element pair in the system. In this framework, however, severe overfitting issues may emerge as well as numerical scaling problems, since the number of parameters grows roughly quadratically with the system size.} \FFnew{We remark that this scaling curse does not afflict all non-Bayesian inference methods \cite{Ronceray}, but only the simple one used here to compare its results with our Bayesian scheme.}

Finally, it is important to specify the probability density function with respect to which we are taking the averages in Eq.~\eqref{impulse}. Since, in order to compute $\langle x_n \xi_n\rangle$ and $\langle x_{n+1}\xi_n\rangle$, we self-consistently used the same update rule and the same shift-invariant noise statistics, we argue that we implicitly introduced a stationarity assumption, overcoming the problem anticipated in Sec.~\ref{subsec:1} and better discussed in Sec.~\ref{sec:initial}. As a result, the inference formulas obtained in this way do not require any rescaling factor, for any length of the trajectory.

\begin{figure}[t]
\includegraphics[width=\columnwidth]{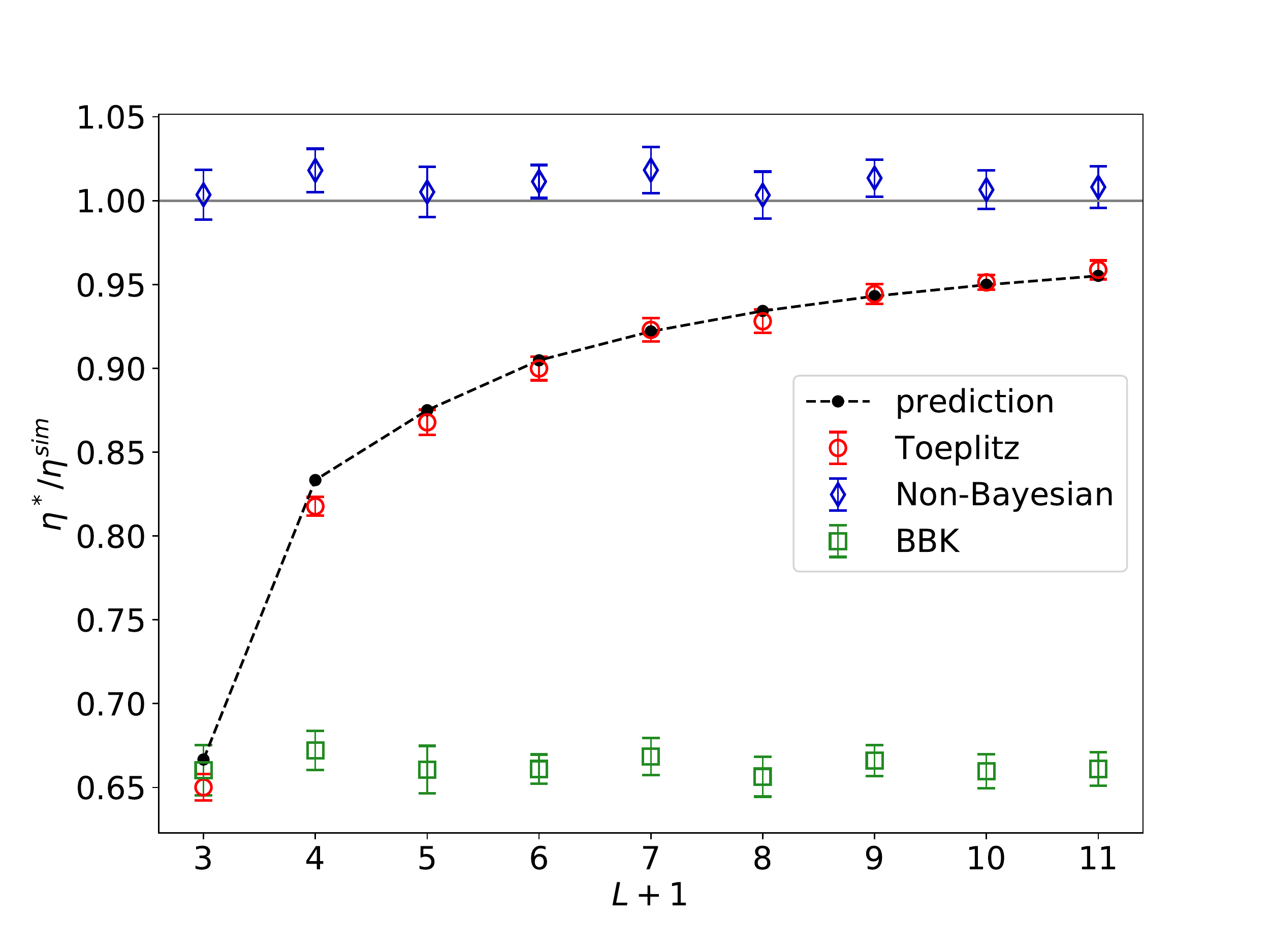}
\caption{Numerical validation of the finite-size distortion introduced by the shift-invariant approximation. Black points connected by dashed lines represent the analytical prediction about the rescaling factor $\varrho(L+1) = \eta^*(L+1)/\eta^{sim}$, with $\varrho(L=3)=2/3$ (shortest possible trajectory) and $\varrho(L+1)\to 1$ monotonically as $L\to\infty$. Numerical results are in agreement with this prediction. As expected, no dependence on trajectory length is found for the non-Bayesian method, nor for Euler-like methods (BBK used here -- see  App.~\ref{app:1} for details).
}
\label{fig:varying_length}
\end{figure}

\FF{
\section{Role of the unobserved initial condition}\label{sec:initial}

Once colored noise is included to take into account the non-Markovian character of the partially observed process, the remaining problem in the application of the Bayesian methodology to second order stochastic models lies in the elimination of the unobserved initial condition. To explain this, let us take a step back. 

In a maximum likelihood setting, the first task is to calculate the probability of observing a given sequence of datapoints, knowing the parameters of the model $\boldsymbol\lambda$. In first order stochastic processes, when all the degrees of freedom allowing for a Markovian description of the dynamics are experimentally accessible, there is no ambiguity on how this likelihood should be computed (see, for example, \cite{Dyn-Inf-PRE}). On the other hand, for second order stochastic processes the inference problem may turn out to be ill-defined. For a first order model
\begin{equation}
	\dot y = g(y) + \xi,
\end{equation}
with initial condition $y(0)=y_0$ the propagator is defined as $P(y_L,\dots,y_1|\boldsymbol\lambda;y_0)$. Here $y_0$ represents quantities that do not change in the inference procedure and the same in the posterior, the likelihood and the prior: the initial condition and the structure of the model. 
We will introduce a semicolon to separate the quantities that do are not updated in the inference. For a second order stochastic process
the initial condition is given by the pair $x(0)=x_0$, $\dot x(0)=v_0$ and the propagator is $P(x_L,\dots,x_1|\boldsymbol\lambda,x_0,v_0)$. However, unlike $x_0$, the initial condition on the velocity is not empirically known, so we the propagator does not result in a  likelihood of the form of $P(x_L,\dots,x_1|\boldsymbol\lambda;x_0,v_0)$.

Let us briefly note that this is strictly connected to the embedding problem in stochastic processes, and that the only consistent way to bypass it is to use the steady state distribution of $v_0$. Nonetheless, in Sec.~\ref{subsec:3} we decided to deal with the initial condition problem in a different way. Firstly, the choice of the LI discretization scheme confined the initial condition problem only to the first timestep, independently of the total number of datapoints and the relation of the decay time of the memory kernel to $\Delta t$. 
Neglecting the breaking of shift invariance, we introduced a Toeplitz approximation for the noise covariance matrix: this approximation works well for long trajectories (with many datapoints), whereas it fails for very short ones. The convergence is however quite fast, as shown in Fig.~\ref{fig:varying_length}. The advantage of this strategy is twofold: it is simpler than exact marginalization, and applies even when a steady state distribution is not available (e.g. in the multiplicative case).
}

Remarkably, the problem of the elimination of the initial condition on the first derivative of the observed variable doesn't affect the non-Bayesian approach. This tells us that non-Bayesian methods apply even to (multiple) disconnected triplets of points or, in general, to disconnected small sequences, if a fragmented observation of the system is the only one achievable. On the contrary, the Toeplitz method is exact only in the infinite trajectory limit, so the smaller the number of subsequent points, the less accurate the inference scheme becomes. In other words, what matters in this case is not only the total number of points for statistical reasons ---\,which is the only thing to worry about in all the other developed schemes\,--- but also their succession in time.

We checked this in numerical simulations of the stochastic harmonic oscillator, keeping constant the total number of points used in the inference procedure, $(L+1)n_S$, and adapting the number of samples $n_S$ as the length $L+1$ of the sample trajectories is varied. A significant deviation of the inferred value from the simulated one is visible in Fig.~\ref{fig:varying_length} for small values of $L$. For small $L$ it is also possible to approximately estimate the distortion introduced by the finite size of the trajectory under the Toeplitz assumption. \FF{Following the same idea that led to the prediction of the 2/3 function for the $\eta$ parameter of the harmonic oscillator, one can expand the two time correlation functions appearing in the Toeplitz inference formulas for small $L$, obtaining 
\begin{equation} 
	\eta^*\simeq -\frac{1}{\Delta t}\ln\left(1+\varrho(L+1)\frac{\dddot C(0)}{\ddot C(0)}\Delta t\right)\left[1+O(\Delta t)\right],
	\label{beta*-deriv}
\end{equation}
from which we deduce that the $\Delta t$-independent rescaling factor of the damping coefficient can be identified with $\varrho(L+1)$ in Eq.~\eqref{beta*-deriv}. The first few values of these rescaling factors are: $\varrho(3)=2/3$, $\varrho(4)=5/6$, $\varrho(5)=7/8$, $\varrho(6)=19/21$, in good agreement with numerical results. 
The exact value is only retrieved in the $L\to\infty$ limit, yet 
time lapse recordings in common motility observation experiments are typically composed by a much larger number of frames than those shown in Fig.~\ref{fig:varying_length}. Although we showed that the wrong marginalization of the initial condition can play a role, in practice this effect can hopefully be neglected in many situations.}

\section{Interacting case}\label{interacting}

\begin{figure*}[t]
	\subfloat{\includegraphics[width=\columnwidth]{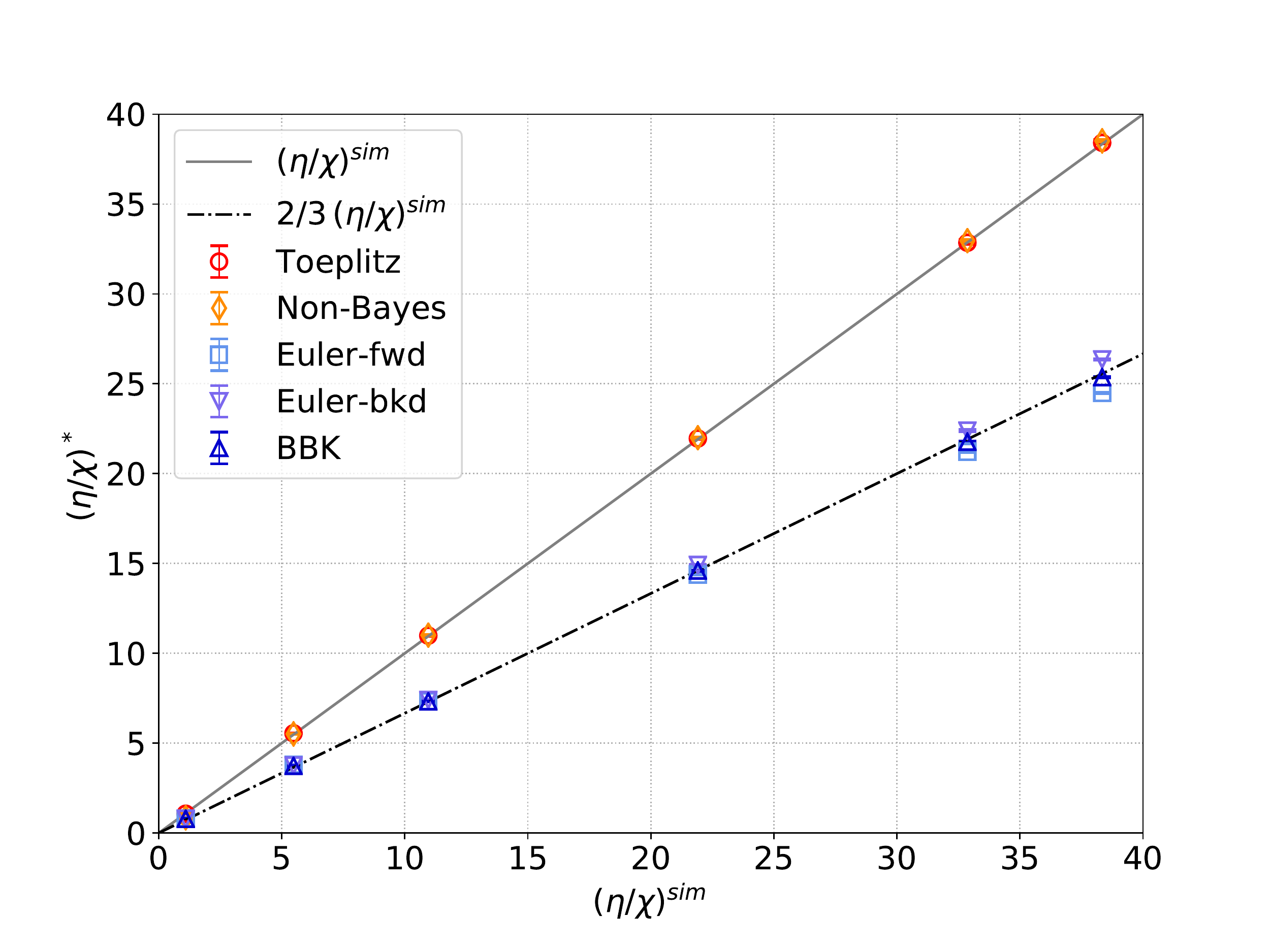}\llap{
  \parbox[b]{17cm}{(a)\\\rule{0ex}{2.2in}}}\label{ISM:eta}}
  	\subfloat{\includegraphics[width=\columnwidth]{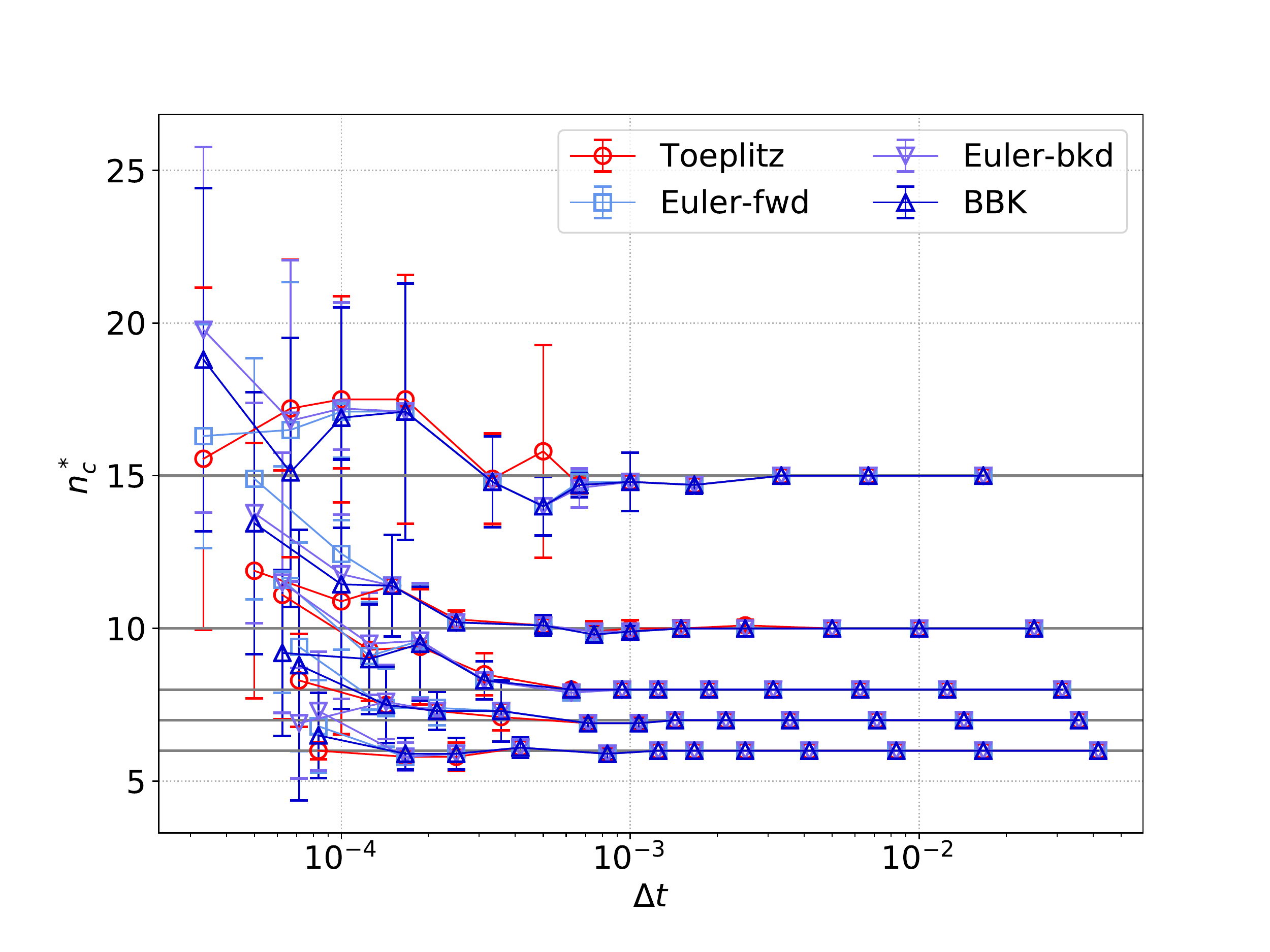}\llap{
  \parbox[b]{17cm}{(b)\\\rule{0ex}{2.2in}}}\label{fig:nc}}
  
	\subfloat{\includegraphics[width=\columnwidth]{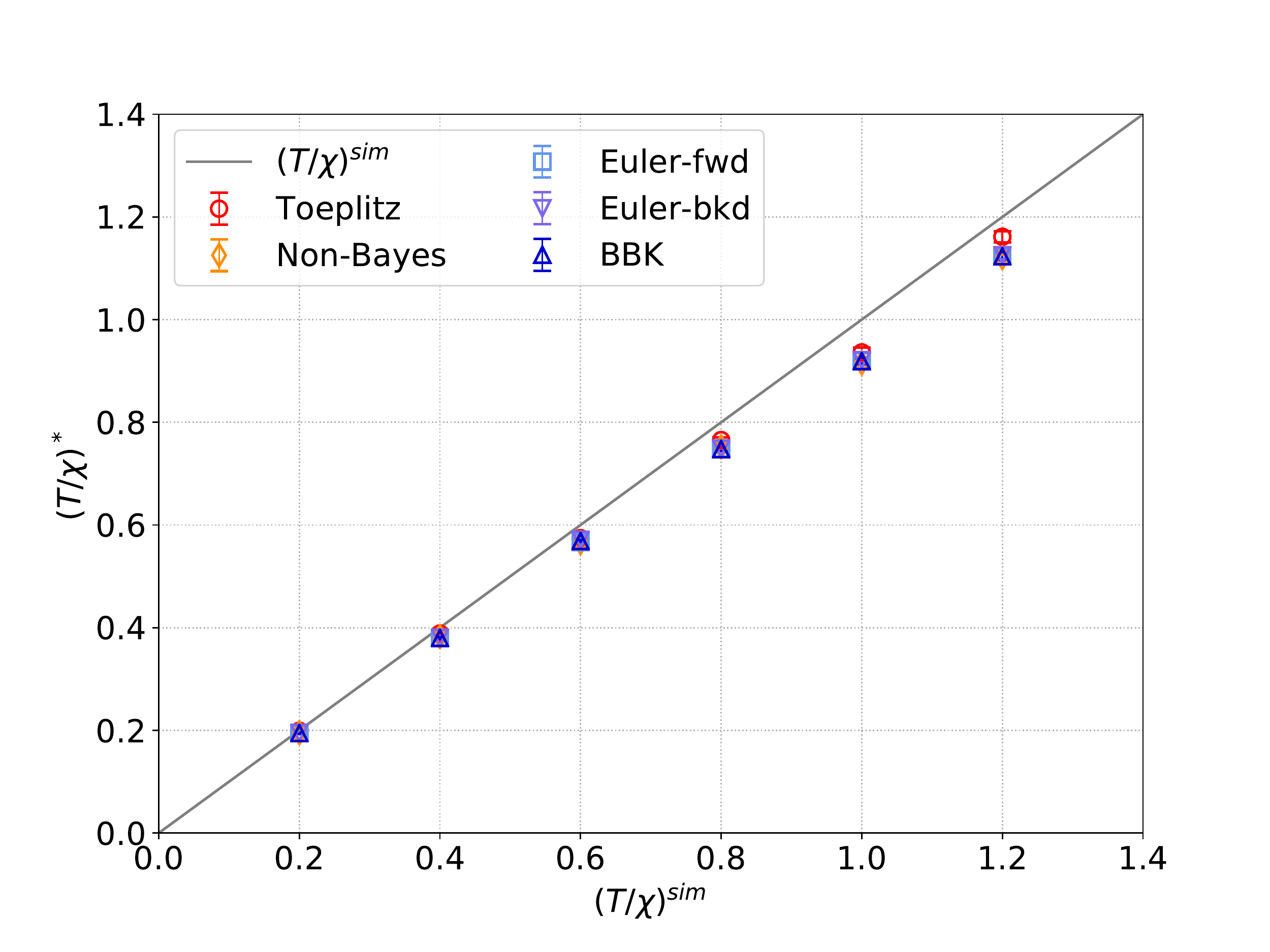}\llap{
  \parbox[b]{17cm}{(c)\\\rule{0ex}{2.2in}}}\label{ISM:T}}
  	\subfloat{\includegraphics[width=\columnwidth]{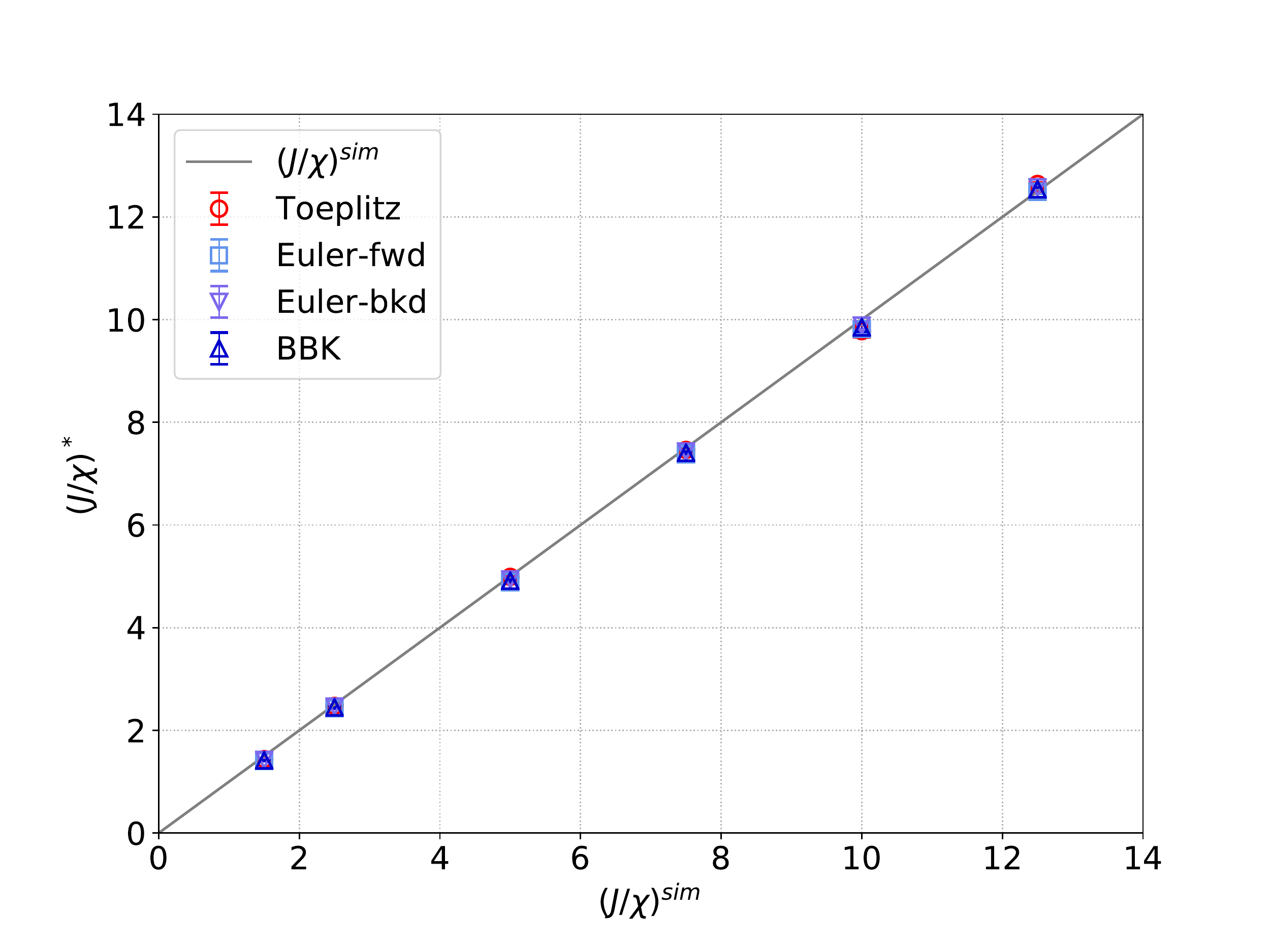}\llap{
  \parbox[b]{17cm}{(d)\\\rule{0ex}{2.2in}}}\label{ISM:J}}

	\caption{Inference results for the Inertial Spin Model. (a)\:: Inferred values for the effective damping coefficient $\eta/\chi$. We notice the emergence of a 2/3 rescaling factor for \FF{na\"ive methods derived from first order Taylor-It\^o expansions}. (b)\:: Inferred topological interaction range from numerical minimization of the reduced minus-log-likelihood, which is properly defined only in the Toeplitz scheme and in the three lower order variants of the Euler scheme. (c)\:: Inferred values for the parameter $T/\chi$, as derived from Eq.~\eqref{ISM-SWA}. One notices a slight divergence from the slope-1 line, which is especially evident at large temperatures. This is due to the spin-wave approximation (SWA), whose first correction only impacts the temperature parameter and can be explicitly evaluated, as explained in App.~\ref{app:SWA}. (d)\:: Inferred values of the interaction strength, $(J/\chi)^*$ vs the parameter value used in simulations, $(J/\chi)^{sim}$. All methods retrieve the correct results.  We remark that only the parameters in the left panels, $\eta/\chi$ and $T/\chi$, can be estimated  by the non-Bayesian method. In all the simulations we took flocks of $N=1000$ birds. Additional information about the choice of the model parameters and numerical methods can be found in App.~\ref{app:7}. Points in (a), (c) and (d) are obtained as in the case of the harmonic oscillator (see Fig.~\ref{fig:4panel}).  For the $O(\Delta t^{1/2})$ methods we consider different integration schemes: standard Euler (Euler-fwd), inverse (Euler-bkd) and BKK defined in section~\ref{app:1}).}
	\label{fig:eta-T}
\end{figure*}

Following our original objective to develop an inference strategy for natural flocks of birds, we generalized the inference equations of Sec.~\ref{ho} and performed numerical simulations of the topological inertial spin model (ISM) on a non-evolving random lattice at low temperature. The model, introduced to account for experimentally observed features that could not be explained within the framework of first order processes \cite{Attanasi:2014aa,Cavagna2015}, represents a second-order generalization of the well-known Vicsek model. The stochastic equations of motion in three dimensions read:
\begin{equation}
	\begin{cases}
	\dot{\mathbf r}_i = \mathbf v_i\\
	\dot{\mathbf v}_i = -\frac{1}{\chi}\mathbf v_i\times \mathbf s_i\\
	\dot {\mathbf s_i} = -\frac{\eta}{\chi}\mathbf s_i + \frac{1}{v_0^2}\sum_{j}J_{ij}\left(\mathbf v_i\times\mathbf v_j\right) + {\boldsymbol \xi_i}_{\perp},
	\end{cases}
	\label{ISM-def}
\end{equation}
where the indexes $i,j=1,\dots, N$ label different individuals, $v_0$ is the constant modulus of each velocity vector $\mathbf v_i$, and ${\boldsymbol \xi_i}_{\perp}$ is the orthogonal projection to $\mathbf v_i$ of a three-dimensional white noise of parameters $T$ and $\eta$: $\langle{\boldsymbol\xi_i}_{\perp}(t)\cdot{\boldsymbol \xi_j}_{\perp}(s)\rangle=2\delta_{ij}2T\eta\delta(t-s)$. Motivated by the  findings of \cite{topo}, we choose to parametrize the coupling constant as $J_{ij}=J\,n_{ij}$, where $n_{ij}=1$ if bird $j$ is  among the first $n_c$ nearest neighbours of bird $i$, whereas it takes a null value otherwise.

In the ordered phase, the spin-wave expansion of the equations of motion of the inertial spin model linearizes the force terms, and Eq.~\eqref{ISM-def} takes the form of a set of SDEs for $N$ coupled harmonic oscillators \cite{Dyn-Inf-PRE}:
\begin{equation}
	\chi{\ddot{\boldsymbol\pi}}_i =-\eta{\dot{\boldsymbol\pi}}_i - J\sum_{j=1}^N \Lambda_{ij}\boldsymbol	\pi_j + {\tilde{\boldsymbol\xi_i}}_\perp\,,\quad i=1,\dots ,N.
	\label{ISM-SWA}
\end{equation}
Here $\boldsymbol \pi_i$ are the birds' normalized velocity fluctuations, lying on the orthogonal plane to the direction of collective motion, $\Lambda_{ij}=n_c\delta_{ij}-n_{ij}$ is the discrete Laplacian of the birds' network, and ${\tilde{\boldsymbol\xi_i}}_\perp$ is now a two-dimensional white noise that lives on the same plane as $\boldsymbol \pi_i$. To  leading order, it is described by the parameters $T$ and $\eta$ appearing in Eq.~\eqref{ISM-def}. For a full derivation of the equations of motion in the spin-wave approximation see App.~\ref{app:SWA}. Thanks to the linearity of Eq.~\eqref{ISM-SWA}, \FF{the same inference strategy one can develop for a system of coupled harmonic oscillators} applies also to the inertial spin model in the highly polarized phase.

For the sake of simplicity, in our simulations we discarded the first equation of Eq.~\eqref{ISM-def} and kept the birds' reciprocal positions fixed. The dynamical maximum likelihood approach, however, should work even when reshuffling birds' reciprocal positions and static approaches fail, since at each time step it is possible to reconstruct the neighborhood of each individual and compute the associated time-dependent observables \cite{Mora:2016aa,Dyn-Inf-PRE,Bialek4786}. \FF{This would introduce an effective nonlinearity which, like in the non-interacting case, is not supposed to modify the leading Gaussian nature of the propagator at small $\Delta t$.}

We applied and compared different inference strategies to the synthetic trajectories. Results are in qualitative agreement to those of the harmonic oscillator. In particular, the expected rescaling factor of $2/3$ for the damping coefficient is retrieved using any EM-like scheme, as shown in Fig~\ref{ISM:eta}. This fact corroborates that the emergence of this 2/3 factor is a universal feature of second order stochastic processes, coming from the interplay between the terms containing second and first order time derivatives, rather than the kind of conservative forces which are applied to the system. Again, \FF{Bayesian and non-Bayesian inference schemes derived from a higher order expansion} do not require any rescaling -- at least for sufficiently long trajectories.

As already mentioned, however, there are some relevant differences with respect to the simple non-interacting case. First of all, the additional difficulty we must face in the case of $N$-body dynamics is that of estimating the interaction range. Since an explicit analytical minimization of the minus-log-likelihood is not operable, 
a numerical approach is needed. The problem is however algorithmically tractable, since it simply consists of a one-dimensional optimization problem. Moreover, if the parametrization of the $n_{ij}$ matrix discussed above is adopted, $n_c$ is a discrete parameter, so the exact minimum value can always be found (see Fig.~\ref{fig:nc}). Wrong estimations of the topological interaction range can be due to a blurred reconstruction of the likelihood from the data. As the number of birds $N$ or the number of trajectory points $L$ is increased, the improved statistics smoothens the rugged reconstructed likelihood and the real minimum becomes easier to detect. To this end, another parameter playing a relevant role is the time lapse $\Delta t$: when the separation between subsequent datapoints is very small compared to the time scales of the system, increments are also very small. Smaller increments correspond to smaller quantities to minimize, which are then subject to bigger relative errors. \FF{This effect is at the origin of what we} observe in Fig.~\ref{fig:nc}.

Once the optimal value of $n_c$ is recovered, it is then used to compute the spatially structured correlation functions which enter into the formulas of the remaining parameters. Non-Bayesian methods are not based on any likelihood definition, and, as a result, do not allow us to infer $n_c$. Despite that, an approximated estimation of the effective temperature $T/\chi$ and of the damping coefficient $\eta/\chi$ is still possible, as shown in Figs.~\ref{ISM:eta} and ~\ref{ISM:T}. On the contrary, the parameters associated to the interaction potential, $n_c$ and $J$, are not evaluated within this framework.

\FF{Applied to large interacting systems, our non-Markovian maximum likelihood method performs well} even for relatively short trajectories. Taking, for instance, trajectories of length $L=200$ for systems of $N=1000$ particles already enables us to achieve good accuracy, with undistinguishable features in the inference of $\eta/\chi$ and $T/\chi$ compared to the non Bayesian method (see Fig.~\ref{fig:eta-T}). As already pointed out, the need for very long trajectories in the max-likelihood scheme, for both single particle and many particle models, stems from two different facts. Firstly, the shift-invariance approximation introduced by enforcing a Toeplitz structure for the noise covariance matrix results in better performance for longer trajectories. Secondly, the empirical reconstruction of two-time correlations, which are the quantities that enter into inference formulas, improves when achieved from longer trajectories as compared to shorter ones. \FF{In other words, the larger the number of datapoints, the higher the amount of available information.} The advantage of moving from the single oscillator to the many-body interacting case is that a restricted number of ``local" quantities turn out to dominate and self-average in sufficiently large systems. So the statistical issue can be at least partially mitigated by averaging over the sample size, rather than relying only on temporal averages as we are compelled to do in the case of the harmonic oscillator.

\FF{
\section{Effect of experimental errors}\label{noise}

So far, we have not included observation errors in the developed inference scheme, but we assumed that stochastic trajectories are sampled with infinite accuracy. However, data are typically affected by accuracy limitations and other sources of experimental errors. In the current section we show the effects of an additional source of noise on the estimation of the model parameters.

The simplest (still realistic, in many practical cases) way to model experimental errors is through a superposition of the discretely sampled trajectory with a sequence of i.i.d. Gaussian random variables $\mathcal N(0,\sigma^2)$.

\begin{figure*}[t]
	\includegraphics[clip,width=\textwidth]{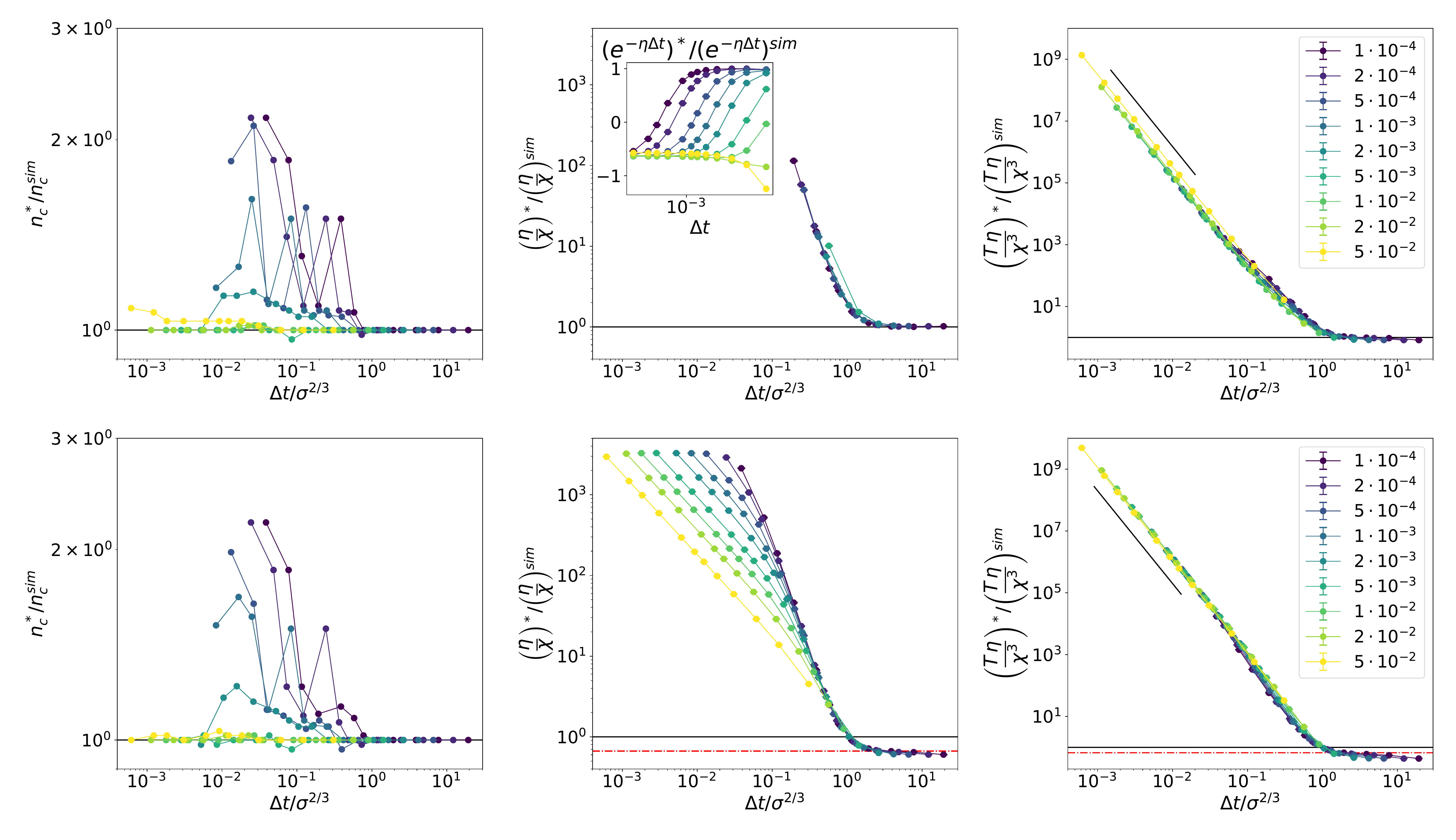}\llap{\parbox[b]{24.6cm}{\small{(a)}\\\rule{0ex}{9.5cm}}}\llap{\parbox[b]{12.9cm}{\small{(c)}\\\rule{0ex}{9.5cm}}}\llap{\parbox[b]{9cm}{\small{(e)}\\\rule{0ex}{9.5cm}}}\llap{\parbox[b]{24.6cm}{\small{(b)}\\\rule{0ex}{4.5cm}}}\llap{\parbox[b]{12.9cm}{\small{(d)}\\\rule{0ex}{4.5cm}}}\llap{\parbox[b]{9cm}{\small{(f)}\\\rule{0ex}{4.5cm}}}
	\caption{Effect of measurement error on some selected parameter estimators for the ISM. In the top row results from the Toeplitz inference scheme are reported; in the bottom row results from the BBK inference scheme are reported. The rescaling of the inverse sampling rate in the abscissa make the curves in (c\,--\,f) to depart at the same point ($\Delta t/\sigma^{2/3}\sim1$) from the expected value in absence of experimental errors (1 for the Toeplitz method, 2/3, marked by the red dot-dashed line, for the BBK method). The collapse of the curves shown in (e\,--\,f) proves that the control parameter is the ratio between stochastic and experimental noise: $T\eta\Delta t^3/\sigma^2$. The black lines, having a slope -3, are a guide for the eye. We notice that for large noise-to-signal ratio the estimate of $\eta$ with the Toeplitz method may be problematic since estimators of positive definite quantities built with noisy data can become negative, as visible in the inset of (c). We refer to App.~\ref{app:2} for details on the inference formulas. Errorbars on Figs.~(a\,--\,b) are not shown, for sake of clarity, whereas in Figs.~(c\,--\,f) the 0.95 CI is smaller than the markersize.}
	\label{fig:noise}
\end{figure*}
As  pointed out by several authors \cite{Pedersen, Lehle2015, Ronceray}, even when $\sigma^2$ is very small measurement noise can impact dynamical inference. A large modification of the high-frequency region of the power spectrum of reconstructed velocities is introduced \cite{Pedersen}, which in turns results in a diverging bias in parameter estimation as $\Delta t\to 0$ \cite{Lehle2015}. This bias and its trend with $\Delta t$ appear also in our inference method (see Fig.~\ref{fig:noise}). Intuitively, the inference procedure relies on the increments of the measured degree of freedom, $\Delta x$, whose average absolute value has a monotonic dependence on $\Delta t$, and need to be compared with the amplitude of measurement errors $\sigma$, which we assume to be independent of the data acquisition sampling rate. At very high sampling rates experimental errors will dominate over the effective dynamics, resulting into an artificial trend $\sim \Delta t^{-1}$ for the parameter $\eta$, and $\sim \Delta t^{-2}$ for the effective temperature and pulsation of the harmonic oscillator (the same dimensional analysis argument can be extended to the parameters of the inertial spin model). 


Since noise cannot be ignored, we include it in the model in the form of a hidden (non) Markov model. Suppose we measure noisy discrete datapoints $\{(\hat x_0,\hat x_1,\dots,\hat x_L)^{\alpha}\}$ corresponding to trajectory points $\{(x_0,x_1,\dots,x_L)^{\alpha}\}$. Following a maximum likelihood argument, we estimate the parameters $\boldsymbol\lambda$ of the dynamical hidden model as
\begin{dmath}
\boldsymbol\lambda^*_{H} = \arg \max_{\boldsymbol\lambda} P(\boldsymbol\lambda|\hat x_0,\dots,\hat x_L) = \arg\max_{\boldsymbol\lambda} P(\hat x_0,\dots,\hat x_L|\boldsymbol\lambda),
\end{dmath}
where 
\begin{dmath}
	P(\hat x_0,\dots,\hat x_L|\boldsymbol\lambda) = \int dx_0\dots dx_LP(x_0,\dots,x_L|\boldsymbol\lambda)\cdot  \prod_{n=0}^{L} P(\hat x_n|x_n).
	\label{lik-HMM}
\end{dmath} 
We assume $P(\hat x_n|x_n) = \frac{1}{\sqrt{2\pi\sigma^2}}\exp-\frac{(\hat x_n-x_n)^2}{2\sigma^2}$ and $P(x_0,\dots,x_L|\boldsymbol\lambda)$ is determined by the hypothesized dynamical model. As long as we deal with linear models, as in the interacting and non-interacting cases considered above, $P(\hat x_0,\dots,\hat x_L|\boldsymbol\lambda)$ reduces to Gaussian integrals and the marginalization over the hidden variables can be performed explicitly. A full treatment at any noise-to-signal ratio is then possible, but not easily generalizable beyond the harmonic case. For this reason here we limit ourselves to showing the predicted effect of experimental uncorrelated noise on numerical simulations. Explicit rewriting of the likelihood in Eq.~\eqref{lik-HMM} allows us to identify the combination of parameters that control the transition from the small to large noise regime. If $T\eta\Delta t^3/\sigma^2\ll1$, noise dominates and, to lowest order, $P(\hat x_0,\dots,\hat x_L|\boldsymbol\lambda) \simeq \prod_{n=0}^{L}\frac{1}{\sqrt{2\pi\sigma^2}}e^{-\frac{1}{2\sigma^2}{\hat x_n}^2}$. If $T\eta\Delta t^3/\sigma^2\gg1$, the effect of noise will be small, and the likelihood will converge to the one we found in absence of experimental errors.

We conclude, in agreement with Ref.~\cite{Ronceray,Lehle2015}, that whenever the experimental apparatus and the observed process are such that the chain of conditions $\sigma^2\ll T\eta\Delta t^3\ll1$ holds, the developed inference strategy still provides a reliable methodology to infer the parameters of the dynamics. When that condition is not fulfilled, controlled denoising procedures or inference strategies based on hidden modelling must be employed.
}

\section{Conclusions}\label{conclusions}

We proposed \FF{a maximum likelihood inference} strategy to tackle the problem of learning the best continuous inertial stochastic model from time lapse recordings of an observed process. The problems arising in this context are general, as they stem from the combination of the following three ingredients: the second (or higher) order nature of the process, when described in terms of the directly measurable degrees of freedom, stochasticity, and the use of discrete sequences of datapoints. Because of that, contrary to first order processes, reconstructing the continuous-time dynamical model from the data is not a straightforward task in the case of second order dynamics. Careful attention must be paid to the mathematical peculiarities of Brownian motion, and in particular to the \FF{minimum order of convergence of the Taylor-It\^o expansion allowing for a correct description of infinitesimal fluctuations.

We want a robust inference methodology which could be applied to a wide class of inertial processes, without knowing their exact time-dependent solution. Such a method must then exploit only the local dynamical information carried by the differential equation. Locally in time, the statistical properties of a Markovian or non-Markovian process are determined by the random variable appearing in the discretized Langevin or generalized Langevin equation respectively. It is then crucial to evaluate correctly the incremental fluctuations, at least to leading order in $\Delta t$.

In the considered non-Markovian scenario, the minimum order of convergence required for the Taylor-It\^o expansion is $O(\Delta t^{3/2})$. We showed that lower order approximations lead to the emergence of a 2/3 rescaling factor for the inferred damping coefficient, as already pointed out in Ref.~\cite{Gloter} and in Ref.~\cite{Lehle2015, Pedersen} in non-Bayesian settings. Employing known numerical integration schemes \cite{CVE,LI}, we developed, to the best of our knowledge, the first max-likelihood inference approach for non-Markovian dynamics (or, equivalently, partially observed Markovian dynamics, since the Markovian embedding is not exploited). We demonstrated the robustness and wide applicability of the method by applying it to different processes: an exactly solvable stochastic oscillator with additive noise with a Gaussian propagator; the Brownian motion of a particle in an anharmonic potential in thermal contact with a heat bath at constant temperature $T$; a stochastic harmonic oscillator driven by multiplicative noise. While the first two examples are described in equilibrium by Gibbs-Boltzmann distribution, the latter is intrinsically out of equilibrium. In all these cases our maximum likelihood estimators for the model parameters are in good agreement with the values used in simulations.

The method can also be successfully and efficiently applied to large interacting systems,} \FFnew{with prior modelling of the interaction mechanism. }
\FF{It is in this aspect that the most promising applications of our max-likelihood method possibly lie. The class of processes for which the method has been developed correspond to the simplest way of incorporating memory effects in the equilibrium dynamics of complex Hamiltonian systems. Its fundamental ingredients are linear dissipation and additive noise. With these conditions fulfilled, the problem is computationally efficient and tractable. For non-interacting systems we showed it is possible to generalize the Bayesian inference approach to non equilibrium processes driven by multiplicative noise. This generalization should work also for interacting ones.

An important remark is that in this setting only single valued parameters can be inferred. Heterogeneities in time and space are not taken into account. The proposed method  is able to cope with slow time dependence of the parameters compared to the available experimental frame rate, by assuming effectively constant parameters along long sub-trajectories. For fast varying parameters, a better approach is to describe the parameter as a random variable drawn from an unknown distribution and infer the parameters of this distribution. For the spatial heterogeneity in very large systems, unless it is modelled using a small number of parameters, a brute force maximum likelihood approach is not feasible and more sophisticated strategies must be developed, as for static inverse problems \cite{Zecchina_review}.

Another possible extension of the method is to include a position dependent dissipation coefficient. This modification would not alter the Gaussian nature of the propagator at short times, even if the noise covariance matrix will no longer be tridiagonal.  Nonlinearities in the first derivative of the measured degree of freedom $x$ and $v$-dependent multiplicative noise could also be considered.} Finally, one could try to generalize the approach to higher order processes, provided that this is motivated by some experimental evidence. 

\FF{Relating the exact maximum likelihood procedure to alternative effective inference schemes, like Gloter's minimum contrast strategy \cite{Gloter}, also remains an open question. Specifically, is it possible to associate to these non-Markovian processes an effective Markovian description with uncorrelated noise (corresponding to factorized dynamical likelihood) and rescaled parameters? Our analysis suggests that} it should be possible to adopt, even if incorrectly, one of the na\"ive methods discussed in Sec.~\ref{schemes} and introduce an \emph{a posteriori} correction of the wrongly estimated parameter, to take into account the effect of the lowest order discretization.


\FF{Another interesting development would be to provide a reliable inference method even in the presence of strong measurement errors. The maximum likelihood framework provides a natural formulation for the problem in terms of hidden Markov models.}

The natural use of the developed framework is application to real data. Technical specifications of acquisition systems have remarkably improved in the last \FF{decades}, and it is now possible to collect well resolved trajectories for long enough time windows. This is also true for animal groups on the move, where experiments are performed in the field and strong limitations are usually set on the acquisition length due to global motion. We know from previous work that the emergent dynamics of groups of birds is dominated by an effective rotational inertia~\cite{Attanasi:2014aa}. This inertia allows information to propagate linearly and in an almost undamped way allowing flocks to turn coherently. Retrieving the effective damping coefficient in this case will allow us to predict the scales where damping becomes relevant,  setting a size limit for groups able to collectively change direction. In the context of swarm dynamics, recent theoretical findings \cite{cavagna2019renormalization, cavagna2019dynamical} suggest that the value of the damping coefficient sets ---\,again\,--- a size crossover for groups displaying different critical behavior on the large scale. Understanding the interplay between size, information propagation and response is a key issue in collective behavior and a reliable inference approach is crucial to provide well grounded answers to these questions.

\begin{acknowledgments}
FF thanks M. Baldovin for helpful discussions and suggestions. IG and FF also thank A. Cavagna, A. Vulpiani and M. Viale. IG thanks E. Aurell, S. Bo and R. Eichhorn. This work was partially supported by the European
Research Council Consolidator Grant n. 724208, the European Research Council Advanced Grant n. 785932, and by the Italian Ministry of Foreign Affairs and International Cooperation through the Adinmat project.
\end{acknowledgments}

\appendix

\section{Discretization procedure}\label{app:discretization}

Let us briefly summarize two possible systematic strategies to obtain a discretized equation in the space of the $x$ variables up to the desired $O(\Delta t^{3/2})$ order. 
Following \cite{zwanzig_book}, we can derive from Eq.~\eqref{SDE-sys} the associated GLE by formally solving the second equation of the system: 
\begin{equation}
	v(t)=\int_0^t ds\, e^{-\eta (t-s)} \left [f(x(s)) +\xi(s)\right] +v_0 e^{-\eta t}.
\end{equation}
Plugging this expression back into the equation for $x$, we get a closed equation in $x$ space:
\begin{equation}
	\dot x = \int_0^t ds K(t-s) f(x(s)) + \zeta(t) + v_0 e^{-\eta t},
	\label{eqx-app}
\end{equation}
where $K(t)=e^{-\eta t}$ and $\zeta(t)=\int_0^t ds e^{-\eta (t-s)} \xi(s)$. Discrete update equations on the scale $\Delta t$ can now be obtained by integrating Eq.~\eqref{eqx-app} between $t_n$ and $t_{n+1}$ and between $t_{n-1}$ and $t_n$. An exponentially decaying memory kernel propagates both the noise and the initial condition $v_0$ in Eq.~\eqref{eqx-app}; it is then possible to identify an appropriate reweighing of its integrated counterparts in order to get rid of both effects. Indeed the combination $x_{n+1}-x_n - e^{-\eta \Delta t} (x_n-x_{n-1})$ does not contain $v_0$ and has a short correlated effective noise:
\begin{widetext}
\begin{equation}
	x_{n+1}-x_n-e^{-\eta \Delta t} (x_n-x_{n-1})= \frac{1-e^{-\eta\Delta t}}{\eta}\int_{t_{n-1}}^{t_{n+1}} \Psi(t-t_n) f(x(t)) dt + \zeta_n\,
\end{equation}
with
\begin{equation}
 	\zeta_n=\int_{t_{n-1}}^{t_{n+1}} \Psi(t-t_n) \xi(t) dt \;;\quad 
	\Psi(t)=\frac{e^{\eta t}-e^{-\eta \Delta t}}{1-e^{-\eta\Delta t}} \left[\theta(t+\Delta t) - \theta(t)\right] + \frac{1-e^{\eta (t-\Delta t})}{1-e^{-\eta\Delta t}} \left[\theta(t) - \theta(\Delta t-t)\right]\,,
	\label{noise-li}
\end{equation}
\end{widetext}
$\theta(t)$ being the Heaviside function. We can check that $\langle \zeta_n \zeta_m\rangle$ has the nearest neighbor structure of Eq.~\eqref{Cnm}:
$$\langle \zeta_n \zeta_m\rangle=C_{nm}=a\delta_{n,m}+b\delta_{n,m\pm1}.$$
From Eq.~\eqref{noise-li} one deduces that, to order $O(\Delta t^{3})$, the coefficients $a$ and $b$  of the covariance matrix assume the expression reported in Eq.~\eqref{ab-def}.

So far, these equations are exact. Some approximation is needed at this stage to evaluate the integral of the force. Various methods have been investigated in the literature; among the simplest is the Langevin Impulse method \cite{LI}, which approximates the integral with the function at the midpoint, leading to
 \begin{equation}
	x_{n+1} = x_{n} + e^{-\eta\Delta t} (x_n-x_{n-1}) + \frac{1-e^{-\eta\Delta t}}{\eta}\Delta t f(x_n) + \zeta_n.
	\label{impulse-app}
\end{equation}

An alternative approach, followed in \cite{CVE} (see also \cite{Mannella}), is to consider the full system of equations in the $(x,v)$ phase space in integral form:
\begin{equation}
	\begin{cases}
	x_{n+1} = x_n + \int_{t_n}^{t_{n+1}} v(t) dt\\[1em]
	v_{n+1} =  v_n + \int_{t_n}^{t_{n+1}} f(x(t)) dt+  \int_{t_n}^{t_{n+1}} \xi(t) dt 
	\end{cases}
	\label{CV}
\end{equation}
and perform a second order Taylor-It\^o expansion around the point $t_n$:
\begin{widetext}
\begin{equation}
	\begin{cases}
	x_{n+1} = x_n + v_n \Delta t + D_n\\
	v_{n+1} = (1-\eta\Delta t) v_n +\frac{1}{2}\Delta t\left[f(x_{n+1})+f(x_n)\right] + 
	      \sigma\Delta t^{1/2}\xi_n -\eta D_n,
	\end{cases}
	\label{CV}
\end{equation}
\end{widetext}
where $D_n$ is defined as follows:
\begin{equation}
D_n = \frac{1}{2}\Delta t^2 \left[f(x_n)-\eta v_n\right] + \sigma\Delta t^{3/2}\left[\frac{1}{2}\xi_n + \frac{1}{2\sqrt{3}}\theta_n\right]
\label{Dn}
\end{equation}
and $\xi_n$ and $\theta_n$ are i.i.d. Gaussian variables sampled from $\mathcal N(0,1)$.
Eliminating the velocity variables $v_n$ and $v_{n-1}$, we find a difference equation of the form of Eq.~\eqref{x-2ord}:
\begin{dmath}
	x_{n+1} = 2 x_n - x_{n-1} - \eta\Delta t\left(1-\frac{\eta\Delta t}{2}\right)(x_n-x_{n-1})-\Delta t^2f(x_n)+\frac{\eta\Delta t^3}{4}\left[f(x_n)-3f(x_{n-1})\right]+\Delta t^{3/2} \zeta_n,
	\label{CV-x}
\end{dmath}
with $\alpha$ and $\beta$ coinciding, up to $O(\Delta t^2)$, to the Taylor expansion of the coefficients in Eq.~\eqref{impulse-app}. The noise variable $\zeta_n$ is defined from Eq.~\eqref{CV} as a linear combination of $\xi_n,\ \xi_{n-1},\ \theta_n,\ \theta_{n-1}$. As a result, due to overlapping Wiener processes, correlations between subsequent noise extractions emerge, which are still described by Eq.~\eqref{Cnm}.

This second derivation is helpful in justifying the quasi-Toeplitz structure of the covariance matrix discussed in the main text. Indeed, fixing $x_1$ implies fixing the first random increment which is responsible for position update in the integration scheme Eq.~\eqref{CV}, when the known initial conditions are $(x_0,v_0)$. Since this stochastic increment enters into the definition of $\zeta_1$ but not in that of $\zeta_2$, the true covariance matrix must have a different entry $C_{11}$ than the other elements on the main diagonal, as in Eq.~\eqref{true-Cnm}.


\section{Inference formulas}\label{inf-formulas}

\subsection{\FF{Na\"ive} max-likelihood approaches for the harmonic oscillator}\label{app:1}

Several Euler-like schemes for the numerical integration of second order stochastic differential equations can be defined. From each of them, inconsistently retaining only the diagonal $O(\Delta t^{3/2})$ stochastic terms when we write the update equations in $x$ space, we can extract a factorized expression for the dynamical likelihood, such as Eq.~\eqref{transition-master-eq}. 

Let us focus on three particular examples: the standard explicit Euler-Maruyama scheme (EM-fwd), its implicit variant (EM-bkd), and the symmetric BBK scheme \cite{BBK}. The three of them may be obtained from the second order SDE Eq.~\eqref{oscillator-1} by approximating first and second time derivatives adopting a forward, backward or symmetric prescription respectively.
The resulting update equations in the three cases read:
\begin{widetext}
\begin{eqnarray}
	\label{x-update-fwd}\text{[EM-fwd]}\qquad &x_{n+1} - (2-\eta\Delta t) x_n + (1-\eta\Delta t + \omega_0^2\Delta t^2)x_{n-1} &= \sigma\Delta t^{3/2} r_{n-1}\\
	\label{x-update-bkd}\text{[EM-bkd]}\qquad &(1+\eta\Delta t)x_{n+1} - (2+\eta\Delta t-\omega_0^2\Delta t^2) x_n + x_{n-1} &=\sigma\Delta t^{3/2} r_{n+1}\\
	\label{x-update-sym}\text{[BBK]}\qquad & \left(1+\frac{\eta\Delta t}{2}\right)x_{n+1}-(2-\omega_0^2\Delta t^2)x_n + \left(1-\frac{\eta\Delta t}{2}\right)x_{n-1}&=\sigma\Delta t^{3/2} r_n
\end{eqnarray}
\end{widetext}
with $\sigma = \sqrt{2T\eta}$ and $\{r_n\}$ a sequence of $L-1$ i.i.d. Gaussian random variables of null mean and unit variance. 

Thanks to the independence of the random variables appearing in Eqs.\eqref{x-update-fwd}--\eqref{x-update-sym}, the discrete propagator takes an approximate factorized form, which we can generally write as:
\begin{equation}
	P_{(1)}(x_2,\dots,x_L|x_0,x_1) = \prod_{n=1}^{L-1}\frac{1}{Z_n}e^{-S_n(x_{n+1},x_n,x_{n-1})}.
\label{factorized}
\end{equation}
The reduced minus-log likelihood, defined as
\begin{equation}
	\frac{\mathcal L}{L-1}:=\frac{-\ln P(x_2,\dots,x_L|x_0,x_1)}{L-1}\,,
	\label{del-minus-log-lik}
\end{equation} 
corresponds in the factorized case to the temporal average of the quantity $(S _n+\ln Z_n)$. This quantity is defined in a slightly different way in the three cases above; consequently, in each of these cases the reduced minus-log-likelihood will be slightly different, as reads in the following. We recall the notation used  in the main text to indicate the equal-time, one-step and two-step experimental correlation functions:
\begin{equation*}
\begin{split}
C_s = \frac{1}{L-1}\sum_{n=1}^{L-1}x_nx_n&;\ C'_s = \frac{1}{L-1}\sum_{n=1}^{L-1}x_{n+1}x_{n+1};\\
C''_s = \frac{1}{L-1}\sum_{n=1}^{L-1}x_{n-1}x_{n-1}&;\ G_s = \frac{1}{L-1}\sum_{n=1}^{L-1}x_{n}x_{n+1};\\
G'_s = \frac{1}{L-1}\sum_{n=1}^{L-1}x_{n}x_{n-1}&;\ F_s = \frac{1}{L-1}\sum_{n=1}^{L-1}x_{n-1}x_{n+1}.
\end{split}
\end{equation*}

\begin{widetext}
\begin{flalign}
          \notag\text{[EM-fwd]}\qquad \frac{\mathcal L}{L-1}=&\frac{1}{2}\ln\left(4\pi T\eta\Delta t^3\right)+\frac{1}{4T\eta\Delta t^3}\big[C'_s + (2-\eta\Delta t^2)^2C_s + (1-\eta\Delta t+\omega_0^2\Delta t^2)^2C''_s - 2(2-\eta\Delta t)G_s + \\
	\label{log-lik-fwd} &2(1-\eta\Delta t+\omega_0^2\Delta t^2)F_s - 2(2-\eta\Delta t)(1-\eta\Delta t+\omega_0^2\Delta t^2)G'_s\big]; \\
	\notag\text{[EM-bkd]}\qquad \frac{\mathcal L}{L-1}=&\frac{1}{2}\ln\left(4\pi T\eta\Delta t^3\right)-\ln \left(1+\eta\Delta t\right) + \frac{1}{4T\eta\Delta t^3}\big[(1+\eta\Delta t)^2C'_s+(2+\eta\Delta t-\omega_0^2\Delta t^2)^2C_s+C''_s-\\
	\label{log-lik-bkd}&2(1+\eta\Delta t)(2+\eta\Delta t-\omega_0^2\Delta t^2)G_s+2(1+\eta\Delta t)F_s-2(2+\eta\Delta t-\omega_0^2\Delta t^2)G'_s\big]\;;\\
	\notag\text{[BBK]}\qquad \frac{\mathcal L}{L-1}=&\frac{1}{2}\ln\left(4\pi T\eta\Delta t^3\right)-\ln \left(1+\frac{\eta\Delta t}{2}\right)+\frac{1}{4T\eta\Delta t^3} \big[\left(1+\eta\Delta t/2\right)^2C'_s+(2-\omega_0^2\Delta t^2)^2C_s + \\
	\notag&\left(1-\eta\Delta t/2\right)^2C''_s -2(2-\omega_0^2\Delta t^2)\left(1+\eta\Delta t/2\right)G_s+2\left(1+\eta\Delta t/2\right)\left(1-\eta\Delta t/2\right)F_s-\\
	\label{log-lik-sym}&2(2-\omega_0^2\Delta t^2)\left(1-\eta\Delta t/2\right)G'_s\big]\:.
\end{flalign}
\end{widetext}

Minimization of Eqs.~\eqref{log-lik-fwd}--\eqref{log-lik-sym} with respect to the parameters of the model yields the following optimal values, according to the adopted scheme:
\begin{itemize}
\item Euler-forward:
\begin{flalign}
	\eta^*_{fwd}=&\frac{1}{\Delta t} \frac{G_s+G'_s-2C_s+\frac{G'_s}{C''_s}(2G'_s-C''_s-F_s)}{-C_s+\frac{{G'_s}^2}{C''_s}}\;;\\
	{\omega_0^2}^*_{fwd}=&\frac{1}{\Delta t^2} \frac{(2-\eta\Delta t)G'_s-(1-\eta\Delta t)C''_s-F_s}{C''_s}\;;\\
	\notag T^*_{fwd}=&\frac{1}{2\eta\Delta t^3}\big[C'_s + (2-\eta\Delta t^2)^2C_s +\\
	\notag &(1-\eta\Delta t+\omega_0^2\Delta t^2)^2C''_s - 2(2-\eta\Delta t)G_s +\\
	\notag &2(1-\eta\Delta t+\omega_0^2\Delta t^2)F_s -\\
	\label{T*-fwd} &2(2-\eta\Delta t)(1-\eta\Delta t+\omega_0^2\Delta t^2)G'_s\big]\;;
\end{flalign}
\item Euler-backward:
\begin{flalign}
	\eta^*_{bkd}=&\frac{1}{\Delta t}\frac{C''_s+F_s-\frac{G'_s}{C_s}(G_s+G'_s)}{\frac{G_sG'_s}{C_s}-F_s}\;;\\
	{\omega_0^2}^*_{bkd} =&\frac{1}{\Delta t^2}\frac{(2+\eta\Delta t)C_s-G'_s-(1+\eta\Delta t)G_s}{C_s}\;;\\
	\notag T^*_{bkd} =&\frac{1}{2\eta\Delta t^3}\big[(1+\eta\Delta t)^2C'_s+(2+\eta\Delta t-\omega_0^2\Delta t^2)^2C_s+\\
	\notag &C''_s-2(1+\eta\Delta t)(2+\eta\Delta t-\omega_0^2\Delta t^2)G_s+\\
	\label{T*-bkd}&2(1+\eta\Delta t)F_s-2(2+\eta\Delta t-\omega_0^2\Delta t^2)G'_s\big]\;;
\end{flalign}
\item BBK:
\begin{flalign}
	\eta^*_{_{BBK}}=&\frac{2}{\Delta t}\frac{C''_s+F_s-\frac{G'_s}{C_s}(G_s+G'_s)}{C''_s-F_s-\frac{G'_s}{C_s}(G'_s-G_s)}\;;\\
	{\omega_0^2}^*_{_{BBK}} =&\frac{1}{\Delta t^2}\frac{2C_s-\left(1+\frac{\eta\Delta t}{2}\right)G_s-\left(1-\frac{\eta\Delta t}{2}\right)G'_s}{C_s}\;;\\
	\notag T^*_{_{BBK}} =&\frac{1}{2\eta\Delta t^3} \big[\left(1+\eta\Delta t/2\right)^2C'_s+(2-\omega_0^2\Delta t^2)^2C_s +\\
	\notag &\left(1-\eta\Delta t/2\right)^2C''_s -\\
	\notag &2(2-\omega_0^2\Delta t^2)\left(1+\eta\Delta t/2\right)G_s+\\
	\notag &2\left(1+\eta\Delta t/2\right)\left(1-\eta\Delta t/2\right)F_s-\\
	&2(2-\omega_0^2\Delta t^2)\left(1-\eta\Delta t/2\right)G'_s\big]\;.
\end{flalign}
\end{itemize}

All of the schemes above are derived from numerical integrators \FF{with weak and strong convergence order $O(\Delta t)$}, and consequently give a 2/3 rescaling factor for the inferred damping coefficient $\eta^*$. This can be checked using the procedure outlined to derive Eq.~\eqref{eta*-full-exp}, which consists of replacing the experimental two-time quantities with the known correlation functions for the harmonic oscillator:
\begin{equation}
	C(t) = \frac{T}{\omega_0^2}e^{-\gamma t}\left[\cos\left(\sqrt{\omega_0^2-\gamma^2}t\right) + \gamma \frac{\sin(\sqrt{\omega_0^2-\gamma^2}t)}{\sqrt{\omega_0^2-\gamma^2}} \right],
	\label{Ct-harm-osc}
\end{equation} 
where $\gamma = \eta/2$, and performing a Taylor expansion around the zero temporal distance. In the same way, the exactness of the inference formulas for $T^*$ and ${\omega_0^2}^*$ can be checked for the three methods.\\

\begin{figure*}
	\includegraphics[clip,width=\textwidth,trim={0.2cm 0 8cm 0}]{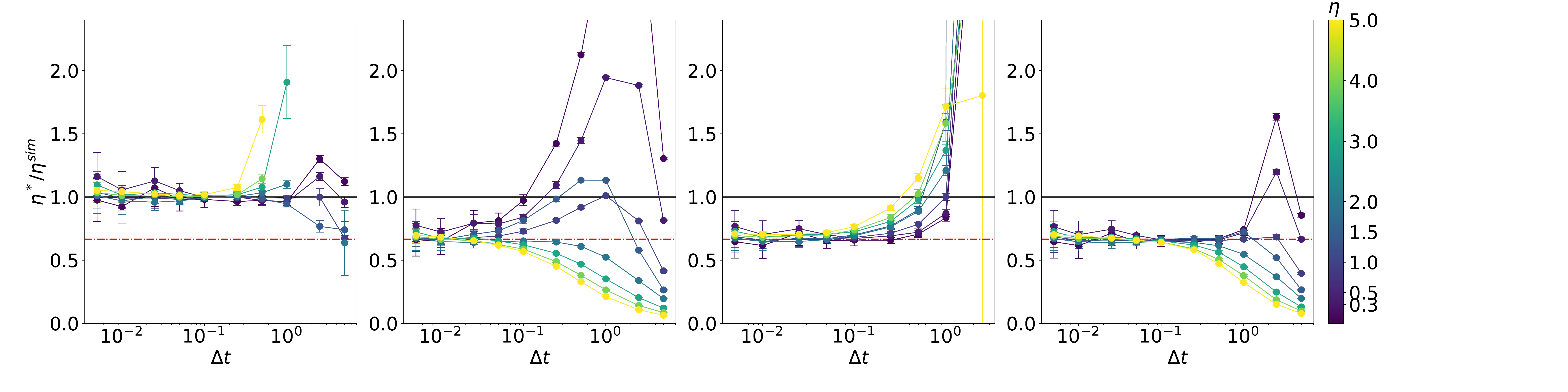}\llap{\parbox[b]{31.9cm}{\small{(a) Toeplitz}\\\rule{0ex}{4.3cm}}}\llap{\parbox[b]{23.45cm}{\small{(b) Euler-fwd}\\\rule{0ex}{4.3cm}}}\llap{\parbox[b]{15.2cm}{\small{(c) Euler-bkd}\\\rule{0ex}{4.3cm}}}\llap{\parbox[b]{7.6cm}{\small{(d) BBK}\\\rule{0ex}{4.3cm}}}
	\caption{Accuracy of the different likelihood-based methods in inferring the damping coefficient of the harmonic oscillator, in varying damping regimes: (a) shows the $O(\Delta t^{3/2})$ Toeplitz method; (b) -- (d) show the three $O(\Delta t^{1/2})$ variants corresponding, respectively, to the Euler forward, Euler backward and BBK schemes. The main features to highlight are the appearance of the  2/3 rescaling factor for the $O(\Delta t^{1/2})$ scheme (red dot-dashed line), compared to the absence of any rescaling for the $O(\Delta t^{3/2})$ scheme, and the higher stability of the latter with respect to $\Delta t$ filtering. Different damping regimes are explored: the sampled values of $\eta^{sim}$ are indicated in the colorbar. The remaining parameters are: $T=1$, $\omega_0=1$. Each point is the average of the inference results of 10 different trajectories of 5000 points (for any $\Delta t$). Errorbars are taken as 0.95 CI.}
	\label{fig:ho-dt}
\end{figure*}


\subsection{Shift-invariant $O(\Delta t^{3/2})$ Bayesian approach}\label{app:2}

We argued that the joint probability of sequences of points in real space is not factorized into a chain of conditional probabilities. This happens because the dynamics of the harmonic oscillator, when projected into the $x$ space, is governed by evolution equations containing a colored noise. The right scheme to adopt is then of the kind of Eq.~\eqref{x-2ord}: as discussed in the main text, this requires correlations between subsequently extracted random variables to be taken into account, resulting, in the case of additive noise, in a covariance matrix with a (quasi-)Toeplitz symmetric tridiagonal structure (cfr. Eqs. \eqref{true-Cnm} and \eqref{Cnm}). We pursue a maximum likelihood approach taking as the function of the parameters of the model to maximize:
\begin{widetext}
\begin{equation}
	P_{(2)}(x_L,\dots,x_2|x_1,x_0) = \frac{1}{Z} \exp -\frac{1}{2}\sum_{n,m=1}^{L-1}(x_{n+1}+F(x_n,x_{n-1};\boldsymbol\mu)){C^{-1}}_{nm}(x_{m+1}+F(x_m,x_{m-1};\boldsymbol\mu)).
	\label{likelihood-Toeplitz-app}
\end{equation}
\end{widetext}
The partition function is specified by Eq.~\eqref{Z-Jc0} and Eq.~\eqref{eigenvalue-Toeplitz}, whereas the relation between $\boldsymbol\mu$ and the physical parameters of the dynamical model depends on the details of the discretization scheme which is adopted.

Thanks to the peculiar structure of this likelihood, one can go pretty far with simple algebra in the optimization problem. First of all, it is convenient to reformulate the issue as a minimization problem for the minus log-likelihood:
\begin{widetext}
\begin{dmath}
	\mathcal L = \frac{L-1}{2}\ln \left(2\pi\frac{2}{3}T\eta\Delta t^3\right) + \frac{1}{2}\sum_{k=1}^{L-1} \ln \left(2+\cos\left(\frac{k\pi}{L}\right)\right) + \frac{3/2}{LT\eta\Delta t^3} \sum_{n,m=1}^{L-1}\left[(x_{n+1}+F(x_n,x_{n-1};\boldsymbol\mu))\tilde A_{nm}(x_{m+1}+F(x_m,x_{m-1};\boldsymbol\mu))\right],
	\label{minus-log-lik}
\end{dmath}
\end{widetext}
being
\begin{equation}
	\tilde A_{nm} = \sum_{k=1}^{L-1}\frac{\sin\left(\frac{n k \pi}{L}\right)\sin\left(\frac{m k \pi}{L}\right)}{2+\cos\left(\frac{k\pi}{L}\right)}.
	\label{A-tilde}
\end{equation}

As usual, the temperature just appears as a prefactor for the effective action, without affecting its actual dynamical structure. The optimal value is given by:
\begin{widetext}
\begin{equation}
	T^{*} = \frac{3}{L(L-1)\eta\Delta t^3} \sum_{n,m=1}^{L-1}\left[(x_{n+1}+F(x_n,x_{n-1};\boldsymbol\mu))\tilde A_{nm}(x_{m+1}+F(x_m,x_{m-1};\boldsymbol\mu))\right].
 	\label{T*}
\end{equation}
Replacing it into Eq.~\eqref{minus-log-lik} and getting rid of additional constants, we obtain a reduced minus-log-likelihood:
\begin{equation}
	\mathcal L \propto \frac{1}{L-1}\sum_{n,m=1}^{L-1}\sum_{k=1}^{L-1}\frac{\sin\left(\frac{nk\pi}{L}\right)\sin\left(\frac{mk\pi}{L}\right)}{2+\cos\left(\frac{k\pi}{L}\right)}(x_{n+1}+F(x_n,x_{n-1};\boldsymbol\mu))(x_{m+1}+F(x_m,x_{m-1};\boldsymbol\mu)).
	\label{L-T*}
\end{equation}
\end{widetext}
One can now split all the terms appearing in the sum and derive with respect to the effective parameters $\boldsymbol\mu$. \FF{Focusing on the case of the simple stochastic harmonic oscillator, $F(x_n,x_{n-1};\boldsymbol\mu) = \alpha x_n + \beta x_{n-1}$,  the set of effective parameter corresponds to $\boldsymbol\mu=(\alpha,\beta)$.} By adopting the Langevin Impulse integrator (see App.~\ref{app:discretization}), they correspond to:
\begin{equation}
	\begin{cases}
	\alpha = -1 -e^{-\eta \Delta t} + \omega_0^2\Delta t\left(1-e^{-\eta\Delta t}\right)/\eta\\
	\beta = e^{-\eta\Delta t}.
	\end{cases}
	\label{impulse-parameters}
\end{equation}
By adopting  a second order Taylor expansion around the prepoint, they correspond to:
\begin{equation}
	\begin{cases}
	\alpha = -2 + \eta\Delta t\left(1-\frac{\eta\Delta t}{2}\right) + \omega_0^2\Delta t^2\\
	\beta = 1-\eta\Delta t\left(1-\frac{\eta\Delta t}{2}\right).
	\end{cases}
	\label{Taylor2-parameters}
\end{equation}
As required for them to be consistent, the two variants are equivalent up to $O(\Delta t^3)$. The numerical results shown in this paper are obtained using Eq.~\eqref{impulse-parameters}.

Imposing that the derivatives of $\mathcal L$ w.r.t. $\alpha$ and $\beta$ are zero leads to:
\begin{equation}
	\alpha^* = -\frac{T_1+\beta^* T_3}{2T_4}\;;\quad
	\beta^* = \frac{T_1T_3 - 2T_2T_4}{-T_3^2 + 4T_4T_5},
	\label{alpha-beta}
\end{equation}
where, with implicit sum over the indexes $n,m$ form 1 to $L-1$,
\begin{flalign}
	\notag T_1 &= \frac{2}{L} \tilde A_{nm}\, x_{n}x_{m+1}\;;\ T_2 = \frac{2}{L} \tilde A_{nm}\, x_{n-1}x_{m+1}\;;\\
	\notag T_3 &= \frac{2}{L} \tilde A_{nm}\, x_{n-1}x_{m}\;;\ T_4 = \frac{1}{L} \tilde A_{nm}\, x_{n}x_{m}\;;\\
	\label{T5} T_5 &= \frac{1}{L} \tilde A_{nm} \, x_{n-1}x_{m-1}\;.
\end{flalign}

\FF{This procedure can be applied to find explicit formulas for any non-interacting system described by a Kramers process with velocity-independent forces $f(x)$, as in Eq.~\eqref{oscillator-1}. We report here those we derived and used for the anharmonic model with force $f(x)=-kx-\lambda x^3$. Referring again to the Langevin Impulse integrator, one possible set of independent parameters is given by $\boldsymbol\mu = (\beta, K, \Lambda)$, where $\beta=e^{-\eta\Delta t}$, $K = k\Delta t/\eta$, $\Lambda= \lambda\Delta t/\eta$. The Toeplitz inference formulas for those parameters read:
\begin{flalign}
	\notag \beta^* =& \left[P_5 - \frac{P_6 P_8 }{P_2} - \frac{(P_2 P_9 - P_3 P_8)(P_2 P_7 - P_3 P_6)}{P_2(P_2 P_4 - P_3^2)}\right]\\
	& \cdot \left[P_1 - \frac{P_6^2}{P_2} - \frac{(P_2 P_7 - P_3 P_6)^2}{P_2(P_2 P_4 - P_3^2)}\right]^{-1}\;;\\
	\Lambda^* =& \frac{\beta^*(P_2 P_7 - P_3 P_6) - (P_2 P_9 - P_3 P_8)}{(1-\beta^*)(P_2 P_4 -P_3^2)}\;;\\
	K^* =& \frac{\beta^* P_6 - P_8}{(1-\beta^*)P_2} - \Lambda^*\frac{P_3}{P_2}\;;
\end{flalign}
where
\begin{flalign}
	\notag P_1 &= (x_{n} - x_{n-1}) \tilde A_{nm}(x_{m} - x_{m-1}) \;;\\
	\notag P_2 &= x_n \tilde A_{nm} x_m \;;\ P_3 = x_n \tilde A_{nm} x_m^3  \;;\ P_4 = x_n^3 \tilde A_{nm} x_m^3 \;;\\
	\notag P_5 &= (x_{n} - x_{n-1}) \tilde A_{nm}(x_{m+1} - x_{m}) \;;\\
	\notag P_6 &= (x_{n} - x_{n-1}) \tilde A_{nm}x_{m} \;;\ P_7 = (x_{n} - x_{n-1}) \tilde A_{nm}x_{m}^3 \;;\\
	P_8 &= (x_{n+1} - x_{n}) \tilde A_{nm} x_{m} \;;\ P_9 = (x_{n+1} - x_{n}) \tilde A_{nm} x_{m}^3 \;.
\end{flalign}
From these equations, the max-likelihood estimators for the physical parameters $\lambda^*$, $k^*$ and $\eta^*$ can be found.
}


\subsection{Generalization to the interacting case (ISM)}\label{app:3}

\begin{figure*}
	\includegraphics[clip,width=\textwidth,trim={0.5cm 0 8cm 0}]{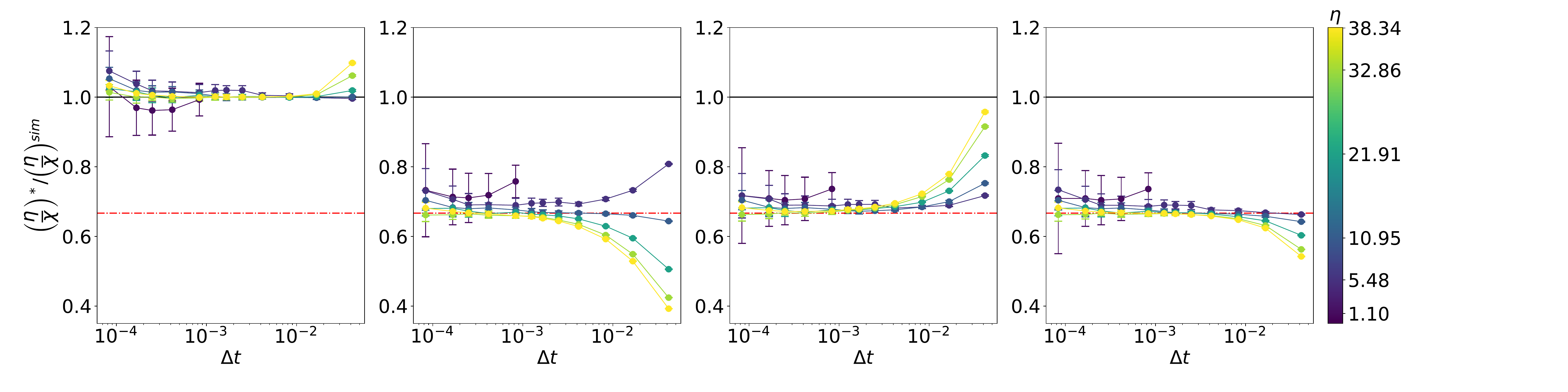}\llap{\parbox[b]{31.75cm}{\small{(a) Toeplitz}\\\rule{0ex}{1cm}}}\llap{\parbox[b]{23.3cm}{\small{(b) Euler-fwd}\\\rule{0ex}{1cm}}}\llap{\parbox[b]{15.1cm}{\small{(c) Euler-bkd}\\\rule{0ex}{1cm}}}\llap{\parbox[b]{7.5cm}{\small{(d) BBK}\\\rule{0ex}{1cm}}}
	\caption{Accuracy of the different likelihood-based methods in inferring the effective parameter $\eta/\chi$ of the inertial spin model: (a) shows the $O(\Delta t^{3/2})$ Toeplitz method; (b) -- (d) show the three $O(\Delta t^{1/2})$ variants corresponding, respectively, to the Euler forward, Euler backward and BBK schemes. We see the 2/3 factor for the $O(\Delta t^{1/2})$ schemes (red dot-dashed line), and no rescaling for the $O(\Delta t^{3/2})$ scheme. ISM simulations are performed in different damping regimes: the sampled values for the parameter $\eta^{sim}$ are indicated along the colorbar. The remaining parameters are: $\chi=1$, $T=0.4$, $J=5.0$, $n_c=6$, $N=1000$. Each point is the average of the inference results of 10 different trajectories of 200 points (for any $\Delta t$). Errorbars correspond to 0.95 CI.}
	\label{fig:ISM-dt}
\end{figure*}

As one moves from single to many particle systems, extra parameters are needed: position and velocity variables are conveniently represented as $N$-component vectors, $N$ being the number of constituents of the group, and model parameters become matrices. For the equations of motion of the three-dimensional ISM on a fixed lattice in the spin-wave approximation Eq.~\eqref{ISM-SWA}, the update rule becomes:
\begin{equation}
	\boldsymbol\zeta_n^i = \boldsymbol\pi^i_{n+1} + \alpha^{ij}{\boldsymbol\pi_n}_j + \beta^{ij} {\boldsymbol\pi_{n-1}}_j,
	\label{vec-update-rule}
\end{equation}
with 

$$\alpha^{ij} = \alpha^0 \delta^{ij} + \alpha^1\Lambda^{ij}\ \text{and}\ \beta^{ij} = \beta\delta^{ij},$$ 
where $\Lambda^{ij}$ is the discrete Laplacian, and sums over the $j$ index are implicit. 

The definitions of $\alpha^0$, $\alpha^1$ and $\beta$ depend on the details of the discretization. Using, for instance, the LI integrator for Langevin equations, 
\begin{equation}
	\left(\frac{\eta}{\chi}\right)^*_{LI} = -\frac{\ln\beta^*}{\Delta t}\;;\quad
	\left(\frac{J}{\chi}\right)^*_{LI} = -\frac{\ln\beta^*}{1-\beta^*}\frac{\alpha^*}{\Delta t^2}\;.
	\label{par-imp}
\end{equation}
Using instead a second order Taylor expansion, we get:
\begin{equation}
	\left(\frac{\eta}{\chi}\right)^*_{IIT} = \frac{1-\sqrt{2\beta^*-1}}{\Delta t}\;;\quad
	\left(\frac{J}{\chi}\right)^*_{IIT} = \frac{\alpha^*}{\Delta t^2}\;.
	\label{par-2T}
\end{equation}

The three parameters $\alpha^0$, $\alpha^1$ and $\beta$ are not independent, since the extra independent parameters of the interacting problem are hidden in the adjacency matrix. In both of the cases considered above (Eq.~\eqref{par-2T} and Eq.~\eqref{par-imp}), $\alpha^0$ and $\beta$ are linked by the same relation: $\alpha^0 = -\beta - 1.$ Renaming $\alpha^1 = \alpha$, the minus-log-likelihood reads:
\begin{widetext}
	\begin{dmath}
	\mathcal L = \frac{(L-1)(d-1)}{2}\ln\left(\frac{T\eta}{\chi^2}\Delta t^3\right) + \text{const} + \frac{3/2}{L \frac{T \eta}{\chi^2} \Delta t^3} \sum_{n,m=1}^{L-1}\frac{1}{N}\sum_{i=1}^N \tilde A_{nm} \left[\boldsymbol\pi^i_{n+1}-\boldsymbol\pi_n^i - \beta(\boldsymbol\pi^i_n - \boldsymbol\pi^i_{n-1}) + \alpha\Lambda^{ij}{\boldsymbol\pi_n}_j \right] \cdot \left[\boldsymbol\pi^i_{m+1}-\boldsymbol\pi_i^m - \beta(\boldsymbol\pi^i_m -\boldsymbol\pi^i_{m-1}) + \alpha\Lambda^{il}{\boldsymbol\pi_m}_l \right].
	\label{L-vector}
\end{dmath}
\end{widetext}

Again, one can proceed with an analytic minimization with respect to $T$, $\alpha$ and $\beta$, giving:

\begin{flalign}
	\alpha^* &= \frac{\beta K_4 - K_3}{2K_5}\;; \\
	\beta^* &= \frac{-K_3 K_4 + 2K_1K_5}{-K_4^2 + 4 K_2 K_5}\;;
	\label{ab*} 
\end{flalign}

\begin{dmath}
T^* = \frac{3}{(d-1)\left(\eta/\chi\right)^*\Delta t^3}\left[K_0 - \beta^*K_1 + {\beta^*}^2K_2 + \alpha^* K_3 -\alpha^*\beta^* K_4 +{\alpha^*}^2 K_5\right]\;,
\label{T*}
\end{dmath}
with $K_0\dots K_5$ the generalization to the many-particle case of the combinations of experimental observables $T_1,\dots, T_5$ defined above (again with implicit sums over $n,m$):

\begin{flalign}
	\notag K_0 &= \frac{1}{N} \sum_{i=1}^N \frac{ \tilde A_{nm}}{L(L-1)} (\boldsymbol\pi^i_{n+1} - \boldsymbol\pi^i_n) \cdot (\boldsymbol\pi^i_{m+1} - \boldsymbol\pi^i_m) \;;\\
	\notag K_1 &= \frac{2}{N} \sum_{i=1}^N \frac{ \tilde A_{nm}}{L(L-1)} (\boldsymbol\pi^i_{n+1} - \boldsymbol\pi^i_n) \cdot (\boldsymbol\pi^i_m - \boldsymbol\pi^i_{m-1}) \;;\\
	\notag K_2 &= \frac{1}{N} \sum_{i=1}^N \frac{ \tilde A_{nm}}{L(L-1)} (\boldsymbol\pi^i_n - \boldsymbol\pi^i_{n-1}) \cdot (\boldsymbol\pi^i_m - \boldsymbol\pi^i_{m-1}) \;;\\
	\notag K_3 &= \frac{2}{N} \sum_{i,j=1}^N \frac{ \tilde A_{nm}}{L(L-1)}  \Lambda^{ij} (\boldsymbol\pi^i_{n+1} - \boldsymbol\pi^i_n) \cdot\boldsymbol\pi_m^j \;;\\
	\notag K_4 &= \frac{2}{N} \sum_{i,j=1}^N \frac{ \tilde A_{nm}}{L(L-1)}  \Lambda^{ij} (\boldsymbol\pi^i_n - \boldsymbol\pi^i_{n-1}) \cdot \boldsymbol\pi^j_m \;;\\
	\notag K_5 &= \frac{1}{N} \sum_{i=1}^N \frac{ \tilde A_{nm}}{L(L-1)}  \Lambda^{ij}\Lambda^{il} \boldsymbol\pi^j_n \cdot \boldsymbol\pi^l_m \;.
\end{flalign}


\FF{\subsection{Generalization to the multiplicative case}\label{app:multiplicative}

From Eqs.~\eqref{update-multi} -- \eqref{zeta-def}, which define the discrete update rule for the multiplicative process described by Eq.~\eqref{multi-class}, one can derive max-likelihood estimators for the parameters of the model by minimizing the effective cost function in Eq.~\eqref{lik-multi}. Let us carry on this explicitly for the following reference example:
\begin{equation}
	\ddot x = -\eta v - k x + \sqrt{a+bx^2}\xi,
\end{equation}
such that the quantities appearing in Eq.~\eqref{lik-multi} read:
\begin{dmath} 
	F(x_n,x_{n-1};\boldsymbol\mu) = x_n - e^{-\eta\Delta t}(x_n - x_{n-1}) + (1-e^{-\eta\Delta t})\frac{k \Delta t}{\eta}x_n,
\end{dmath} 
with $\boldsymbol \mu = \left(e^{-\eta\Delta t},k\Delta t/\eta\right)$ and 
\begin{equation}
	C_{nm} = \left(a + bx_n^2\right) \delta_{n,m} + \sqrt{\left(a + bx_n^2\right)\left(a + bx_m^2\right)} \delta_{n,m\pm1} .
\end{equation}
Simple manipulations allow us to reduce to the minimization problem to a one-dimensional numerical optimization, since analytical formulas for the optimal values of the effective parameters $\beta= e^{-\eta\Delta t}$, $K = k\Delta t/\eta$ and $b$ can easily be found:
\begin{flalign}
	\notag b^* &= \frac{3}{L-1}\left[P_0 - \frac{P_4^2}{P_2} - \frac{\left(P_3P_2 - P_4P_5\right)\left(P_3P_2 - P_4P_5\right)}{P_1P_2 - P_5^2}\right];\\
	\beta^* &= \frac{P_3^*P_2^* - P_4^*P_5^*}{P_1^*P_2^* - (P_5^*)^2};\qquad K^* = \frac{\beta^* P_5^* - P_4^*}{(1-\beta^*)P_2^*}\;.
	\label{multi-analytical}
\end{flalign}
where we renamed $\alpha = a/b$ and $P_i^* = P_i(\alpha^*)$. The optimal value of the new effective parameter $\alpha^*$ is the minimizer of the following function of $\alpha$:
\begin{dmath}
	\mathcal L = \frac{1}{L-1}\sum_{k=1}^{L-1}\ln\tilde\lambda_k + \ln\left[P_0 - \frac{P_4^2}{P_2} - \frac{\left(P_3P_2 - P_4P_5\right)\left(P_3P_2 - P_4P_5\right)}{P_1P_2 - P_5^2}\right],
	\label{cost}
\end{dmath}
where $\{\tilde \lambda_k\}$ is the set of eigevnvalues of the reduced covariance matrix
\begin{dmath}
	{A^{-1}}_{nm} = C_{nm}/b = \left(\alpha + x_n^2\right) \delta_{n,m} + \sqrt{\left(\alpha + x_n^2\right)\left(\alpha + x_m^2\right)} \delta_{n,m\pm1}\,,
	\label{Anm-multi}
\end{dmath}
and
\begin{flalign}
	\notag P_0 &=  \frac{1}{L-1}\sum_{n,m=1}^{L-1} (x_{n+1} - x_n) A_{nm} (x_{m+1} - x_m)\;;\\
	\notag P_1 &=  \frac{1}{L-1}\sum_{n,m=1}^{L-1} (x_{n} - x_{n-1}) A_{nm} (x_{m} - x_{m-1})\;;\\
	\notag P_2 &=  \frac{1}{L-1}\sum_{n,m=1}^{L-1} x_n A_{nm} x_m\;;\\
	\notag P_3 &= \frac{1}{L-1}\sum_{n,m=1}^{L-1} (x_{n+1} - x_n) A_{nm} (x_{m} - x_{m-1})\;;\\
	\notag P_4 &=  \frac{1}{L-1}\sum_{n,m=1}^{L-1} (x_{n+1} - x_n) A_{nm} x_m\;;\\
	\notag P_5 &=  \frac{1}{L-1}\sum_{n,m=1}^{L-1} (x_{n} - x_{n-1}) A_{nm} x_m\;.
\end{flalign}
}


\subsection{Non-Bayesian approach: inference formulas without a likelihood}\label{app:4}

We build in this section an alternative approach to the Bayesian one, as outlined in Section~\ref{subsec:4} of the main text. To be explicit, we need to choose a discrete update equation in $x$ space: let us choose again the one corresponding to the usual continuation rule of the LI:
\begin{equation}
x_{n+1} = x_n +e^{-\eta\Delta t}(x_n-x_{n-1}) + \frac{1-e^{-\eta\Delta t}}{\eta}\omega_0^2\Delta t\, x_n + \zeta_n,
\label{start_eq}
\end{equation}
and multiply its r.h.s. and l.h.s. by $x_n$, $x_{n+1}$ and $x_{n-1}$ and take the average over the noise distribution. The resulting equations are:
\begin{widetext}
\begin{flalign}
	\label{xn}    \langle x_{n+1} x_n\rangle &= \langle{x_n}^2\rangle + e^{-\eta\Delta t}(\langle{x_n}^2\rangle - \langle x_n x_{n-1}\rangle) + \frac{1-e^{-\eta\Delta t}}{\eta}\omega_0^2\Delta t \langle{x_n}^2\rangle + \langle x_n\zeta_n\rangle \;;\\
	\label{xn+1}    \langle x_{n+1}x_{n+1}\rangle &= \langle x_n x_{n+1}\rangle + e^{-\eta\Delta t}(\langle x_n x_{n+1}\rangle - \langle x_{n-1} x_{n+1}\rangle) + \frac{1-e^{-\eta\Delta t}}{\eta}\omega_0^2\Delta t \langle x_n x_{n+1}\rangle + \langle\zeta_n x_{n+1}\rangle\;;\\
	\label{xn-1}    \langle x_{n+1}x_{n-1}\rangle &= \langle x_n x_{n-1}\rangle + e^{-\eta\Delta t}(\langle x_n x_{n-1}\rangle - \langle x_{n-1}^2\rangle) + \frac{1-e^{-\eta\Delta t}}{\eta}\omega_0^2\Delta t\langle x_n x_{n-1}\rangle\;.
\end{flalign} 

Using again Eq.~\eqref{start_eq} -- combined with the covariance matrix of the Gaussian variables -- to compute $\langle \zeta_n x_n\rangle$ and $\langle\zeta_n x_{n+1}\rangle$, the relations we find are:
\begin{flalign}
	\label{box-1}	G_s &= C_s + e^{-\eta\Delta t}(C_s-G'_s) + \frac{1-e^{-\eta\Delta t}}{\eta}\omega_0^2\Delta t\, C_s + b\;;\\
	\label{box-2}	C'_s &= G_s + b + a + e^{-\eta\Delta t}(G_s-F_s+b) + \frac{1-e^{-\eta\Delta t}}{\eta}\omega_0^2\Delta t(G_s+b)\;;\\
	\label{box-3}	F_s &= G'_s + e^{-\eta\Delta t}(G'_s - C''_s) + \frac{1-e^{-\eta\Delta t}}{\eta}\omega_0^2\Delta t\, G'_s\;.
\end{flalign}
\end{widetext}
In order to find Eqs. \eqref{box-1}--\eqref{box-3}, we identified the actual correlation functions with the empirical ones, denoted with $C$, $G$ and $F$ symbols, and we hypothesized a stationarity assumption to hold to explicitly compute them. 
After proper manipulation, one can extract ``inference relations'' for $b$, $e^{-\eta\Delta t}$ and $\omega_0^2\Delta t$, and derive from them the physical parameters of the model. In order, $e^{-\eta\Delta t}$ is given as the solution of the second-degree polynomial equation:
\begin{dmath}
(2G'_s-C_s-C''_s)e^{-2\eta\Delta t} + \left[ 2G_s + C''_s - C_s - 2F_s	+ 5(2G'_s-C_s-C''_s) \right]e^{-\eta\Delta t} + \left[G_s - G'_s + F_s - C'_s + 5(G'_s-C_s-F_s+G_s)\right]=0;
\end{dmath}
then $b$ and $\omega_0^2\Delta t$ are computed as follows:
\begin{flalign}
	\label{B} b &= G'_s - F_s + G_s - C_s + e^{-\eta\Delta t}\left(2G'_s-C''_s-C_s\right)\;;\\
	\label{omega} \omega_0^2\Delta t &= \frac{-\eta}{1-e^{-\eta\Delta t}}\left[\frac{G_s-C_s-b}{C_s}-e^{-\eta\Delta t}\frac{C_s-G'_s}{C_s}\right]\;.
\end{flalign}

Notice that these inference equations are not unique. Combining the starting equations in a different way would result into slightly different inference formulas, which, however, should provide the same result if the experimental correlation functions faithfully reproduce ensemble averages at the steady state.

This strategy cannot be adapted to interacting problems, outside of the mean field approximation. The obstacle comes from the parametrization of the interaction matrix, which is the discrete counterpart of the introducing an interaction range in the corresponding field theory. Without a priori parametrization, the issue of sufficient statistics arises: one can think about repeating the same procedure in the multi-particle case for each particle pair and look for independent inference formulas for any matrix element $J\Lambda^{ij}$. Bypassing the technical difficulties related to solving the resulting system of $N^2+2$ second degree equations for the unknowns $b$, $e^{-\eta\Delta t}$ and $\{J\Lambda^{ij}\}_{i,j=1\dots N}$, we have a much greater number of parameters to infer than of points in each frame. This problem becomes totally untractable if one also allows $\Lambda^{ij}$ to evolve in time, as in active animal groups \cite{diffusion,Mora:2016aa}.

Assumptions about the structure of the matrix $\Lambda_{ij}$ dramatically diminish the number of parameters and help us deal with the worry of insufficient statistics, but require an alternative strategy to estimate the interaction range, since this physically motivated parametrization does not allow us to find closed-form equations.

It is possible yet to approximately estimate the damping coefficient and the effective temperature of the system of interacting particles, assuming that they are all immersed in the same uniform thermal bath. Under this assumption, Eqs. \eqref{xn}--\eqref{xn-1} can be adapted to the interacting case and properly manipulated to find the following relations:
\begin{widetext}
\begin{equation}
	F_s - G'_s - G_s + C_s = e^{-\frac{\eta}{\chi}\Delta t} \left(2G'_s-C''_s-C_s\right)+\frac{G'_{int}-C_{int}}{C_{int}}\left[G_s-C_s-b-e^{-\frac{\eta}{\chi}\Delta t}(C_s-G'_s)\right] - b\;;
	\label{box-1-ISM}
\end{equation}
\begin{dmath}
C'_s-2G_s+C_s = e^{-\frac{\eta}{\chi}\Delta t}\left(G_s-F_s-C_s+G'_s\right) + b\left\{4+e^{-\frac{\eta}{\chi}\Delta t} + \frac{n_c}{C_{int}}\left[G_s-C_s-b-e^{-\frac{\eta}{\chi}\Delta t}(C_s-G'_s)\right]\right\} + \frac{G_{int}-C_{int}}{C_{int}}\left[G_s-C_s-b-e^{-\frac{\eta}{\chi}\Delta t}(C_s-G'_s)\right]\;;
	\label{box-2-ISM}
\end{dmath}
\end{widetext}
where we have used the third independent equation to eliminate $J/\chi$ and exploited the fact that $a =4b$, with $b=\frac{1}{6}2\frac{T\eta}{\chi^2}\Delta t^3$. Let us define the empirical spatio-temporal correlation functions involved in these inference formulas:
\begin{itemize}
\item Equal-time correlations:
	\begin{flalign}
	\label{C-start}C_{ij} &= \frac{1}{L-1}\sum_{n=1}^{L-1} \boldsymbol\pi^i_{n} \cdot \boldsymbol\pi^j_{n}\;;\\
	C'_{ij} &= \frac{1}{L-1}\sum_{n=1}^{L-1} \boldsymbol\pi^i_{n+1} \cdot \boldsymbol\pi^j_{n+1}\;;\\
	C''_{ij} &= \frac{1}{L-1}\sum_{n=1}^{L-1} \boldsymbol\pi^i_{n-1} \cdot \boldsymbol\pi^j_{n-1}\;;
	\end{flalign}
\item One-step correlations:
	\begin{flalign}
	G_{ij} &= \frac{1}{L-1}\sum_{n=1}^{L-1} \boldsymbol\pi^i_{n+1} \cdot \boldsymbol\pi^j_{n}\;;\\
	G'_{ij} &= \frac{1}{L-1}\sum_{n=1}^{L-1} \boldsymbol\pi^i_{n} \cdot \boldsymbol\pi^j_{n-1}\;;
	\end{flalign}
\item Two-step correlations:
	\begin{equation}
	\label{C-end}F_{ij} = \frac{1}{L-1}\sum_{n=1}^{L-1} \boldsymbol\pi^i_{n+1} \cdot \boldsymbol\pi^j_{n-1}\;.
	\end{equation}
\end{itemize}
The observables appearing in Eqs. \eqref{box-1-ISM}--\eqref{box-2-ISM} are defined from \eqref{C-start}--\eqref{C-end} as in the following. We can distinguish the contribution of self-correlations, encoded by: 
\begin{eqnarray}
	\notag C_s = \frac{1}{N}\Tr \mathbf C\;&;\ C'_s = \frac{1}{N}\Tr \mathbf C'\;&;\ C''_s = \frac{1}{N}\Tr \mathbf C''\;;\\
	\notag G_s = \frac{1}{N}\Tr \mathbf G\;&;\ G'_s = \frac{1}{N}\Tr \mathbf G'\;&;\ F_s = \frac{1}{N}\Tr \mathbf F\;;
\end{eqnarray} 
and that of correlations between directly interacting birds, encoded by the quantities:
\begin{equation}
	\notag C_{int} = \frac{\Tr(\mathbf\Lambda\mathbf C)}{N}\;;\ G_{int} =  \frac{\Tr(\mathbf\Lambda^{\top}\mathbf G)}{N}\;;\ G'_{int} = \frac{\Tr(\mathbf\Lambda\mathbf G')}{N}\;;
\end{equation}
where $\Lambda_{ij}=n_c\delta_{ij}-n_{ij}$. Notice that all of them are by definition self-averaging quantities, which obviously tend to be more and more stable as the size of the system increases.

As already stressed, in absence of a proper likelihood, an unattainable task is that of dealing with functions denoted with an {\em int} subscript; however, the manipulation we carried out to derive Eqs. \eqref{box-1-ISM}--\eqref{box-2-ISM} confines them into sub-leading terms. This can be checked by looking at the combinations:
\begin{equation}
	\frac{G_{int}-C_{int}}{C_{int}}\left[\left(1-e^{-\frac{\eta}{\chi}\Delta t}\right)(G_s-C_s)-b\right]\simeq O(\Delta t^5)\,,
	\label{comb-1}
\end{equation}
the one obtained replacing $G_{int}$ with $G'_{int}$, and
\begin{equation}
	b \cdot \frac{n_c}{C_{int}}\left[(1-e^{-\frac{\eta}{\chi}\Delta t})(G_s-C_s)-b\right] \simeq O(\Delta t^6).
	\label{comb-2}
\end{equation}

Under the working hypothesis that $\Delta t$ is sufficiently small, we can neglect these terms and find usable relations to extract the effective parameters of the thermal bath ($\eta/\chi$, $T/\chi$) from the experimental self-correlations only. Precisely, $\eta/\chi$ is found as a solution of the equation:
\begin{dmath}
	(C''_s+C_s-2G_s)e^{-2\frac{\eta}{\chi}\Delta t}+2(F_s-5G'_s-G_s+3C_s+2C''_s)e^{-\frac{\eta}{\chi}\Delta t} + 4F_s-4G'_s-6G_s+5C_s+C'_s=0\,,
\end{dmath}
whereas the effective temperature is extracted from $b$, being:
\begin{equation}
	b = G'_s+G_s-F_s-C_s+e^{-\frac{\eta}{\chi}\Delta t}\left(2G'_s-C_s-C''_s\right)\,.
	\label{b-NBayes-ISM}
\end{equation}
Notice that this formula is exactly equivalent to Eq.~\eqref{B}, since we defined the effective damping coefficient of the harmonic oscillator as $\eta=\mu/m$, whereas the corresponding quantity, having the dimension of an inverse time scale, is $\eta/\chi$ for the ISM. These formulas have been applied to find the results shown in Fig.~\ref{fig:eta-T}.

\section{Equations of motion of the ISM in the spin wave approximation (SWA)}\label{app:SWA}

\begin{figure*}[t]
	\subfloat{\includegraphics[width=\columnwidth]{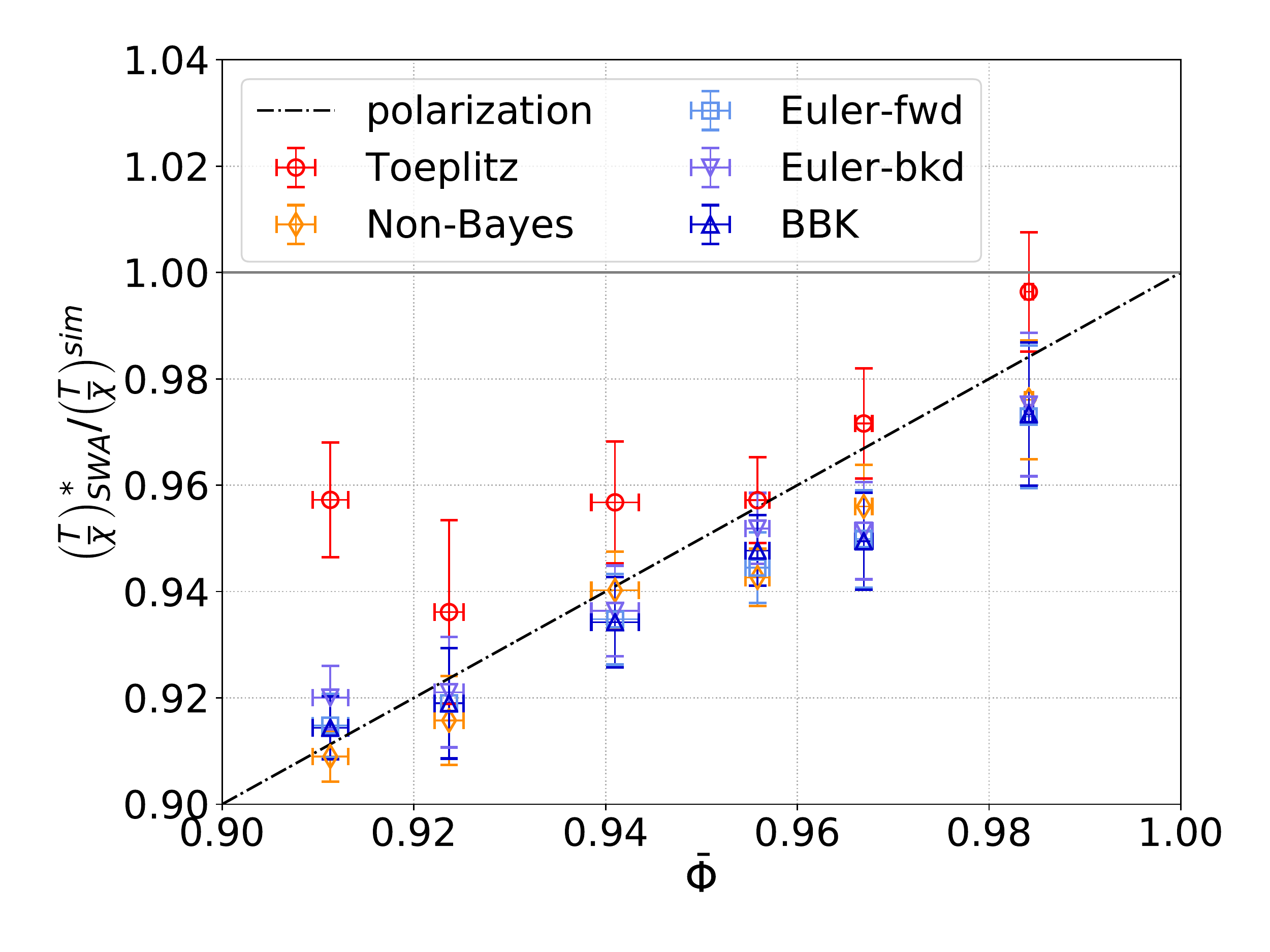}\llap{
  \parbox[b]{17cm}{(a)\\\rule{0ex}{2.2in}}}\label{SWA:a}}
	\subfloat
	{\includegraphics[width=\columnwidth]{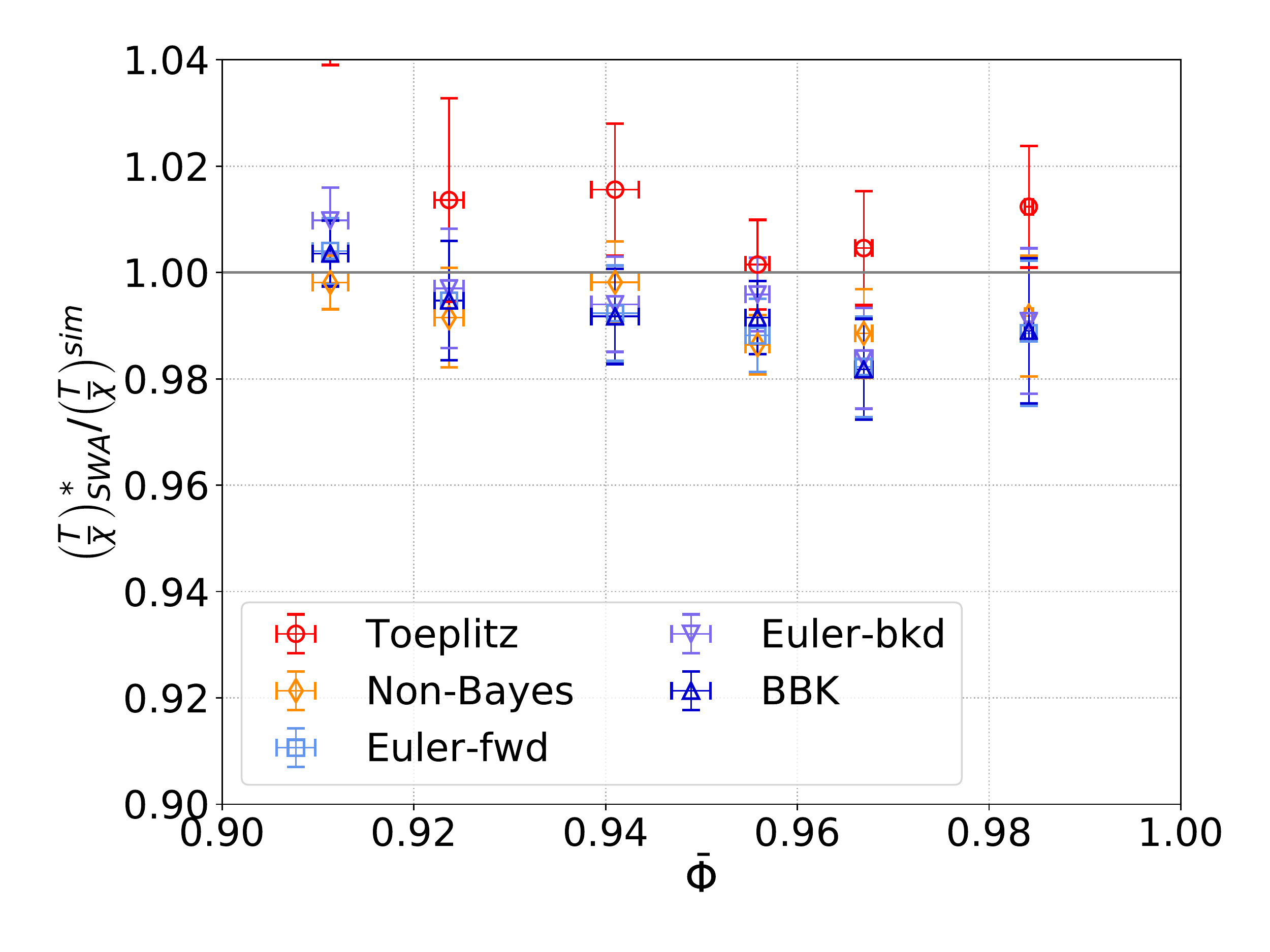}\llap{
  \parbox[b]{17cm}{(b)\\\rule{0ex}{2.2in}}}\label{SWA:b}}
	\caption{First correction to the SWA.	The comparison between the plots shows the effect of the SWA: in (a) the raw inferred values of $T/\chi$, obtained using the inference formulas derived from Eq.~\eqref{ISM-SWA}, are reported.  In (b) we included the first correction by rescaling the output with the time-averaged polarization, for each sample trajectory. $\bar \Phi$ is the average of the averaged polarizations among different simulated flocks, at any given temperature. Errorbars for $\bar \Phi$ correspond to standard errors, whereas vertical bars represent, as in the other figures, 0.95 CI.}
	\label{fig:SWA}
\end{figure*}

We derive in this appendix the equations of motion of the inertial spin model (ISM) in the so-called \emph{spin wave} approximation (SWA). The name comes from the analogy with ideal Heisenberg ferromagnets which, at very low temperatures, can be studied using an approximate theory, whose basic idea is that the lowest energy excitations in a ferromagnet are those produced by a single reversed spin over a large number of otherwise aligned spins in a crystal lattice. Dyson showed that an interaction between spin-wave states arises from this approximation and it should be taken into account to consistently work out the spin wave expansion \cite{Dyson}. In a similar way, since natural flocks of starling are in a deeply ordered phase, we can perform an expansion around the perfectly ordered state of the flock, where all of the birds' velocities are aligned along the same direction.

Let us denote by $\mathbf n$ the collective direction of motion of the flock. Each vector $\mathbf v_i$ can be decomposed into its longitudinal and transverse components with respect to $\mathbf n$:
\begin{equation}
	\mathbf v_i = v_i^L\mathbf n + \boldsymbol\pi_i.
	\label{v-decomposition}
\end{equation}
In the case of bird flocks, the spin-wave approximation reduces to approximating the longitudinal components as follows:
\begin{equation}
	v_i^L = \sqrt{1-|\boldsymbol\pi_i|^2} \simeq 1-\frac{1}{2}|\boldsymbol\pi_i|^2,
	\label{viL-approx}
\end{equation}
having $\mathbf v_i$ a unit length. The equations of motion of the ISM (with fixed interaction network) can be written in the form of a set of second order SDEs for the velocity variables:
\begin{equation}
	\frac{d^2\mathbf v_i}{dt^2} = \left(-\eta\frac{d\mathbf v_i}{dt} - J \sum_{j=1}^Nn_{ij} \mathbf v_j + \boldsymbol \xi_i \right)_{\perp} - \left\lvert\frac{d\mathbf v_i}{dt}\right\rvert^2 \mathbf v_i,
	\label{ISM-2ord}
\end{equation}
where the $\perp$ symbol indicates the projection onto the orthogonal plane to the direction of motion of the $i$-th bird, $\mathbf v_i$. This projection operator and the last term of Eq.~\eqref{ISM-2ord} are the required ingredients to ensure individual speed conservation: $|\mathbf v_i(t)|=v_0 = 1\ \forall i,t$. Thanks to this property, Eq.~\eqref{ISM-2ord} further simplifies:
\begin{equation}
	\frac{d^2\mathbf v_i}{dt^2} = -\eta\frac{d\mathbf v_i}{dt} - J \sum_{j=1}^Nn_{ij} {\mathbf v_j}_{\perp} + {\boldsymbol \xi_i }_{\perp} - \left\lvert\frac{d\mathbf v_i}{dt}\right\rvert^2 \mathbf v_i.
	\label{ISM-2ord-bis}
\end{equation}
Using Eqs. \eqref{v-decomposition} -- \eqref{viL-approx}, and exploiting the fact that, for any vector $\mathbf a$,
\begin{equation}
	\mathbf a_{\perp} = -\mathbf v_i \times \left(\mathbf v_i \times \mathbf a\right),
	\label{a-perp}
\end{equation}
one can evaluate all the terms appearing in Eq.~\eqref{ISM-2ord-bis}, at the desired order of approximation.

Let us focus firstly on time derivatives: we notice that, in principle, they also produce terms containing $\frac{d\mathbf n}{dt}$ and $\frac{d^2\mathbf n}{dt^2}$. In the following we will assume that the direction of collective motion $\mathbf n$ is constant. This is legitimate in the limit $N\to\infty$, when the wandering of the order parameter is suppressed, or at least when it is very slow compared to the relaxational dynamics of the degrees of freedom. If, on the contrary, one wants to take this effect into account, apparent forces emerge because the chosen reference frame is non-inertial.

Neglecting apparent forces enables to segregate  \emph{on-plane} (i.e. perpendicular to $\mathbf n$) and \emph{off-plane} (i.e. parallel to $\mathbf n$) contributions, and completely disentangle the corresponding equations. One can then consider the equations in the $\boldsymbol\pi$-plane only:
\begin{equation}
	\frac{d^2\boldsymbol\pi_i}{dt^2} + \eta\frac{d\boldsymbol\pi_i}{dt} + J\Lambda_{ij}\boldsymbol \pi_j =  \hat P {\boldsymbol\xi_i}_{\perp} + O(|\boldsymbol\pi|^3),
	\label{approx-ISM}
\end{equation}
where $\Lambda_{ij} = n_{ij}-n_c\delta_{ij}$ and $\hat P$ is the projection operator onto the plane perpendicular to the collective velocity $\mathbf V = \frac{1}{N}\sum_{i=1}^N \mathbf v_i \equiv \Phi \mathbf n$. The velocity fluctuations $\boldsymbol\pi_i$ play in this case the same role as spin excitations in Dyson's SWA, both becoming the new degrees of freedom and displaying a linear interaction.

At this stage, what remains to explicitly evaluate is only $\hat P{\boldsymbol\xi_i}_{\perp}$. We know that ${\boldsymbol\xi_i}_{\perp}$ lives in the plane perpendicular to $\mathbf v_i$, so that the perpendicular component to the plane spanned by $\mathbf V$ and $\mathbf v_i$ is left unchanged by this projection operator, while the other one is contracted with a factor $\cos\theta_i$, with $\theta_i$ the angle between $\mathbf v_i$ and $\mathbf n$. As a result:
\begin{equation}
	\langle\hat P \boldsymbol\xi_i(t) \cdot\hat P\boldsymbol\xi_i(s)\rangle = 2(1+\cos^2\theta_i) \frac{T\eta}{\chi^2}\delta(t-s). 
\end{equation}
The second moment of each noise term is then rescaled, with respect to the original one, by a factor:
\begin{equation}
	\frac{1}{2}(1+\cos^2\theta_i)=\frac{1}{2}\left(1+\left(v_i^L\right)^2\right) = 1-\frac{1}{2}|\boldsymbol\pi_i|^2\simeq v_i^L.
	\label{rescaling-T}
\end{equation}
In order to let the fluctuation-dissipation theorem hold, this rescaling can be re-adsorbed by the temperature parameter $T/\chi$, which is in principle different for each bird. At an averaged level, we can define a new spin wave temperature that differs form the original temperature of the inertial spin model by a factor $\frac{1}{N}\sum_{i=1}^N v_i^L$, which is by definition equivalent to the polarization of the flock $\Phi = \lvert\frac{1}{N}\sum_{i=1}^N\mathbf v_i\rvert$. In the low temperature case, where $|\boldsymbol\pi|\ll1$, $\Phi=1+O(|\boldsymbol\pi|^2)$; the first correction to the temperature parameter is then of a lower order with respect to the terms which have been neglected in the deterministic part of Eq.~\eqref{approx-ISM} and shall correctly be included through this simple effective rescaling.

As long as the experimental or statistical errors are wide enough and the system pretty ordered, this SWA-related correction is negligible. Thanks to the large statistics and high accuracy we managed to have with our simulations and inference machinery, we are able to detect it in Fig.~\ref{ISM:T}, where points are systematically placed below the line of slope 1, especially for higher values of the temperature, which in turn correspond to lower polarization values. A comparison between the two panels of Fig.~\ref{fig:SWA} confirms that this is truly the origin of the observed trend and not an intrinsic defect of the inference procedure.


\section{ISM simulations}\label{app:7}





We implemented a numerical integrator for the ISM in $d=3$ that combines the leapfrog method with Boris's trick to ensure speed conservation \cite{Boris1970}. We performed simulations on fixed Poisson random lattices (i.e. sites are randomly chosen points with uniform distribution), discarding the update of particle positions and consequent reshuffling effects. As a result, the adjacency matrix of the graph associated to the interacting particle system is time-independent and the constant speed $v_0$ of each bird does not play any role. Thus the numerical integrator we used consists of the following set of update equations:
\begin{widetext}
\begin{equation}
	\begin{cases}
	\mathbf v_i^{\ n+1} = \mathbf v_i^{\ n} + \left(\mathbf v_i^{\ n} + \mathbf v_i^{\ n}\times \mathbf t^n\right) \times \mathbf u^n\\
	\mathbf s_i^{\ n+1/2} = \left(1+\frac{\eta \Delta t}{2\chi}\right)^{-1}\left\{\left(1-\frac{\eta \Delta t}{2\chi}\right) \mathbf s_i^{\ n-1/2} + \mathbf v_i^{\ n} \times \left[ \frac{J\Delta t}{\chi} \sum_{j} n_{ij} \mathbf v_j^{\ n} + \boldsymbol \Xi_i^{\ n}\right]\right\}.
	\end{cases}
	\label{ISM-integrator}
\end{equation}
\end{widetext}
with $\mathbf t^n = -\frac{1}{2\chi}\Delta t\mathbf s^{n+1/2}$ and $\mathbf u^n = 2\mathbf t^n/(1+\lvert \mathbf t^n \rvert^2)$.  $\boldsymbol \Xi_i^{\ n}$ is a three-dimensional isotropic Gaussian variable of zero mean and of variance: 
\begin{equation}
\langle\boldsymbol \Xi_i^{\ n} \cdot \boldsymbol \Xi_j^{\ m}\rangle=\delta_{ij}\delta_{mn}\,2\cdot3\cdot T\eta\,\Delta t.
\end{equation}

The adjacency matrix explicitly reads:
\begin{equation}
	n_{ij} = \begin{cases}
	1 \quad \text{if}\ r_{ij}\leq n_c\\
	0 \quad \text{if}\ r_{ij}>n_c 
	\end{cases}
\end{equation}
with $r_{ij}$ the rank of bird $j$ as a neighbour of bird $i$ (excluding the bird itself, to which we conventionally associate rank $r_{ii}=0$). In all of our simulations we worked with periodic boundary conditions.

We tried to ensure that the system was sampled in a stationary regime by starting from microscopic configurations corresponding to polarization values close to the equilibrium ones. The polarization is the macroscopic order parameter of the system and it is defined, in perfect analogy to the magnetization in a 3-dimensional Heisenberg model, as $\Phi = \frac{1}{N v_0}\left\lvert\sum_{i=1}^N\mathbf v_i\right\rvert$.

Flocks of $N=1000$ birds are simulated to obtain the results shown in this paper, with topological range of interaction $n_c=6$ (except for the data in Fig.~\ref{fig:nc}), alignment strength $J/\chi=5$ and effective temperature $T/\chi$ in the range $[0.2,1.2]$. When not explictly indicated, we took $T/\chi=0.4$, approximately corresponding to a polarization of 0.97 (for $n_c =6$). We chose an integration time step of  $\tau_{sim}=0.0005/(J n_c)$ for all the simulations. Different damping regimes have been explored, and the performance of the inference method was tested in each of them,  and for various choices of the time lag $\Delta t$. In order to disentagle the effects of the discrete nature of the simulation from proper malfunctioning of the inference schemes, the minimum inference time step $\Delta t$ displayed in Figs.~\ref{fig:nc} and~\ref{fig:ISM-dt} is $5\tau_{sim}$. 

\bibliography{bibliography}

\end{document}